\documentclass[journal=jacsat,manuscript=article]{achemso}
\setkeys{acs}{articletitle = true}

\graphicspath{ {./images/} }

\usepackage[version=3]{mhchem} % Formula subscripts using \ce{}

\usepackage{multicol}
\usepackage{graphicx} 
\usepackage{changepage}
\usepackage{amsmath}

\usepackage{soul}
\soulregister\cite7
\usepackage{xcolor}
\sethlcolor{yellow}

\usepackage{tabularx}

\newcolumntype{C}{>{\centering\arraybackslash}X}
\usepackage{array}

\usepackage{array}
\usepackage{booktabs}
\usepackage{makecell}

\usepackage{longtable}
\usepackage{booktabs}
\usepackage{array}
\usepackage{makecell}

\author{Alexandre Peuch}
\affiliation{Yusuf Hamied Department of Chemistry, University of Cambridge, Lensfield Rd, Cambridge, CB2 1EW, UK}
\alsoaffiliation{Lennard--Jones Centre, University of Cambridge, Trinity Ln, Cambridge, CB2 1TN, UK}

\author{Angus Whiteside}
\affiliation{Yusuf Hamied Department of Chemistry, University of Cambridge, Lensfield Rd, Cambridge, CB2 1EW, UK}
\alsoaffiliation{Lennard--Jones Centre, University of Cambridge, Trinity Ln, Cambridge, CB2 1TN, UK}

\author{Giaan Kler-Young}
\affiliation{Yusuf Hamied Department of Chemistry, University of Cambridge, Lensfield Rd, Cambridge, CB2 1EW, UK}
\alsoaffiliation{Lennard--Jones Centre, University of Cambridge, Trinity Ln, Cambridge, CB2 1TN, UK}

\author{Kaifeng Niu}
\affiliation{Yusuf Hamied Department of Chemistry, University of Cambridge, Lensfield Rd, Cambridge, CB2 1EW, UK}
\alsoaffiliation{Lennard--Jones Centre, University of Cambridge, Trinity Ln, Cambridge, CB2 1TN, UK}
\alsoaffiliation{Materials Design Division, Department of Physics, Chemistry and Biology, IFM, Linköping University, 581 83 Linköping, Sweden}

\author{Jinwoo Hwang}
\affiliation{Department of Chemical and Biological Engineering, University of Wisconsin-Madison, Madison, Wisconsin 53706, United States}

\author{Manos Mavrikakis}
\affiliation{Department of Chemical and Biological Engineering, University of Wisconsin-Madison, Madison, Wisconsin 53706, United States}

\author{Angelos Michaelides}
\affiliation{Yusuf Hamied Department of Chemistry, University of Cambridge, Lensfield Rd, Cambridge, CB2 1EW, UK}
\alsoaffiliation{Lennard--Jones Centre, University of Cambridge, Trinity Ln, Cambridge, CB2 1TN, UK}
\email{am452@cam.ac.uk}

\author{Fabian Berger}
\affiliation{Yusuf Hamied Department of Chemistry, University of Cambridge, Lensfield Rd, Cambridge, CB2 1EW, UK}
\alsoaffiliation{Lennard--Jones Centre, University of Cambridge, Trinity Ln, Cambridge, CB2 1TN, UK}
\alsoaffiliation{Max Planck Institute for Polymer Research, Ackermannweg 10, 55128 Mainz, Germany}
\email{fb593@cam.ac.uk}

\title{Assessing the Transferability of General-Purpose Machine Learning Interatomic Potentials for Heterogeneous Catalysis with HetCat26
}

\begin{document}

\clearpage

\begin{abstract}

Foundation machine learning interatomic potentials (MLIPs) promise near-density functional theory (DFT) accuracy across broad areas of chemistry and materials science.
However, their performance in describing systems and processes relevant to heterogeneous catalysis remains underexplored.
Here, we introduce \textit{HetCat26}, a collection of benchmark tests designed to assess pre-trained MLIPs across key aspects of catalytic modeling, including surface energetics, metal--metal oxide interactions, adsorption, and catalytic reaction networks.
Evaluating eighteen foundation models, we find that performance on existing general materials benchmarks is only weakly predictive of performance on \textit{HetCat26}; transferability to heterogeneous catalysis cannot be inferred from these general benchmarks.
Across the benchmark tests, current models describe surface energetics and, perhaps surprisingly, reaction barriers well, whereas larger errors are observed for adsorption, and DFT site preferences are often not reproduced.
Three models, eSEN-30M-OAM, SevenNet-Omni-i12 and MACE-MH-1-OMAT, nevertheless achieve high accuracy across the properties evaluated.
Beyond model benchmarking, \textit{HetCat26} highlights the importance of training data consistency: PBE and PBE+U calculations should not be mixed within a training set.
Overall, we identify challenges limiting the transferability of current foundation MLIPs to heterogeneous catalysis and, more broadly, to chemical reactions at interfaces, providing guidance for the development of the next generation of foundation models.

\end{abstract}

\clearpage

\section{Introduction}
Many of the catalysts currently employed in industrial processes were developed during the 20$^\text{th}$ century, before climate change emerged as a major global concern.\cite{baiker2000,marcilly2003,armor2011,liu2014,bowker2022}
Accordingly, there is a growing need for more efficient catalysts across numerous industrially important reactions.
Catalyst performance is governed by a complex interplay between surface structure, adsorbate binding, and reaction kinetics and thermodynamics.\cite{norskov2009,schlogl2015,swetlana2015}
Understanding these phenomena at the atomistic scale is therefore central to the rational design of next-generation catalysts.

Over the past decades, density functional theory (DFT) has become indispensable for computational catalysis, providing atomistic insights into reaction mechanisms, active sites, and catalyst design principles.\cite{honkala2005,kandoi2006,norskov2009,chen2021,chen2025}
DFT has enabled the widespread use of adsorption energies as catalytic descriptors and facilitated the establishment of relationships linking thermodynamics and kinetics, such as Brønsted--Evans--Polanyi (BEP) scaling relations.\cite{norskov2002,michaelides2003,bligaard2004,liu2011}
Yet its computational cost has largely restricted simulations to simplified catalyst models.

Machine learning has become an important tool in heterogeneous catalysis.\cite{bozal-ginesta2025}
Specifically, machine learning interatomic potentials (MLIPs)\cite{deringer2019,ko2023,wang2024,jacobs2025} are opening the door to atomistic simulations of increasing complexity, including reactive events involving bond breaking and bond formation, while extending the length and time scales accessible at near-\textit{ab initio} accuracy by several orders of magnitude.\cite{chen2023,olajide2025,omranpour2025,xie2026}
A major bottleneck, however, is that substantial effort is required to generate training data and to develop and validate individual, highly system-specific models.\cite{deringer2019}

To overcome this limitation, foundation MLIPs (fMLIPs), pre-trained on large and chemically diverse datasets, have emerged as a promising alternative.\cite{yuan2026}
Their broad transferability enables stable atomistic simulations for systems that were not explicitly represented during training, making them attractive for wide-ranging applications across materials chemistry and beyond.\cite{batatia2025}
Although fMLIPs do not yet consistently match the accuracy of carefully developed bespoke MLIPs, they offer substantial practical advantages.
Domain-specific reactive potentials occupy a middle ground, sacrificing generality across chemical space for accuracy within a targeted class of systems.\cite{yang2025}
In practice, foundation models can be used out-of-the-box for exploratory simulations, accelerate active-learning workflows, provide good initial configurations, and serve as efficient starting points for developing bespoke potentials by leveraging the physical knowledge already encoded in the pre-trained models. 
Compelling examples of their use are emerging from the catalysis community.\cite{perego2024,ma2026,cheula2026}

Benchmark suites have been developed in conjunction with the emergence of fMLIPs to understand the models' capabilities.
For example, Matbench Discovery\cite{riebesell2025} and LAMBench\cite{peng2026} effectively evaluate fMLIP performance across general materials science, providing valuable insights into model accuracy, transferability, and physical robustness.
Heterogeneous catalysis presents a particularly stringent test of transferability, owing to low-coordination surface sites, bond-breaking and bond-forming events, reaction pathways far from equilibrium, and complex interfaces.
Besides this, many existing benchmarks have largely emphasized energy and force accuracy, focusing less on the application-relevant metrics that underpin the complexity of catalytic systems.\cite{wu2026}
It thus remains unclear whether fMLIP performance on existing general benchmarks is predictive of that in catalytic modeling.

Efforts directed specifically at heterogeneous catalysis have recently begun to emerge, each addressing certain aspects of catalytic modeling.
The CatBench framework focuses on the prediction of adsorption energies by fMLIPs,\cite{moon2025} while an assessment of their performance in describing perovskite oxide surfaces, reactions on alloyed metal and oxide surfaces, and metal--oxide interfaces was recently published.\cite{kempen2026}
The ability of fMLIPs to accurately capture the reverse water--gas shift reaction has also been investigated.\cite{loveday2026}
Here, we introduce \textit{HetCat26}, a benchmark designed to cover the important aspects of heterogeneous catalysis, complementing and extending existing efforts and bringing this research into focus for model development.
Curated literature data together with newly generated reference calculations are included to probe increasingly complex catalytic phenomena, ranging from surface energetics and atom migration to adsorption site preferences and catalytic reaction networks.

We assess the performance of a representative selection of eighteen fMLIPs on \textit{HetCat26}, providing a snapshot of their current capabilities and limitations.
Performance on existing general materials benchmarks correlates only weakly with performance on \textit{HetCat26}, demonstrating that transferability to heterogeneous catalysis cannot be reliably inferred from conventional benchmark results alone.
At the same time, three fMLIPs, eSEN-30M-OAM, SevenNet-Omni-i12 and MACE-MH-1-OMAT, achieve strong performance across a broad range of properties.
\textit{HetCat26} further reveals classes of catalytic environments that remain challenging for current foundation models, providing insight into the factors that limit transferability and identifying priorities for future dataset and model development.
Beyond these findings, we anticipate that \textit{HetCat26} will provide a useful compilation of datasets for the community to evaluate future generations of foundation models for application in heterogeneous catalysis.

\section{Methods and Materials}

\subsection{The \textit{HetCat26} Benchmark Framework}

\begin{figure}[htb]
    \centering \includegraphics[width=14.9cm,height=\textheight,keepaspectratio]{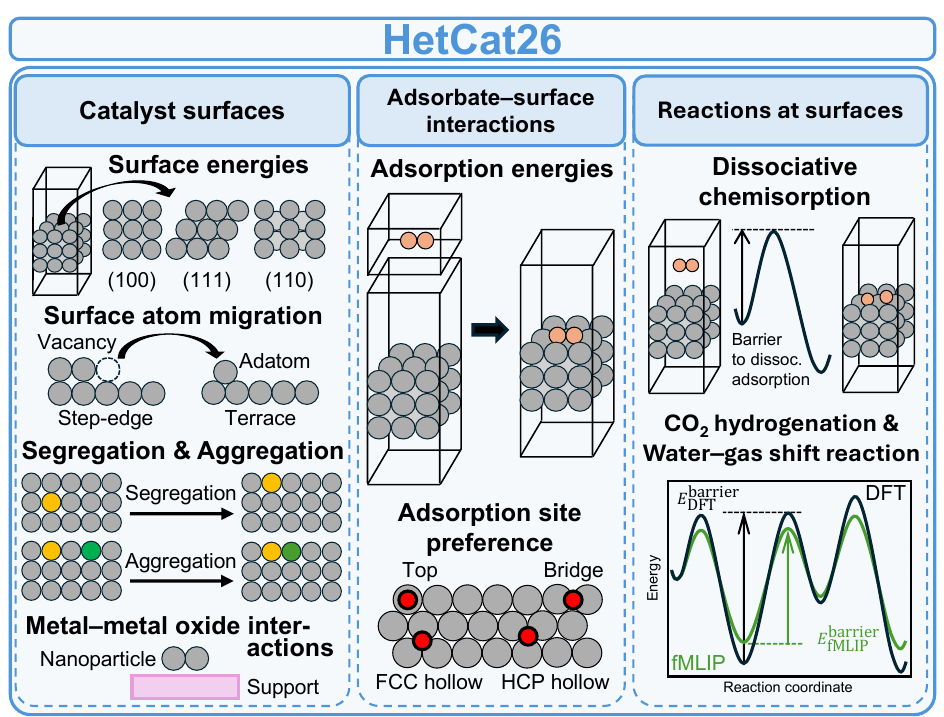}
    \caption{
    Composition of the \textit{HetCat26} benchmarking framework.
    The properties included can be broadly separated into three categories.
    The first covers catalytic surfaces: surface energies, energetics for the migration of atoms on surfaces, segregation and aggregation energies and metal--metal oxide support interactions.
    The second covers adsorbate--surface interactions, including adsorption energies across a wide range of systems and site preferences for the adsorption of CO on Cu(111) and Pt(111).
    The third assesses reactivity through reaction energies and barriers in dissociative chemisorption and in catalytic reaction networks.
    }
    \label{fig:hetcat26}
\end{figure}

Catalytic performance is underpinned by an intricate combination of a catalyst's structure, its dynamic evolution, adsorbate binding at the surface, and reaction kinetics and thermodynamics.\cite{norskov2009,schlogl2015,swetlana2015}
A meaningful assessment of transferability to heterogeneous catalysis consequently requires benchmarking that captures these key physical processes.
For this reason, the \textit{HetCat26} benchmark framework introduced here is broadly organized into three categories: (i) catalyst surface properties, (ii) adsorbate--surface interactions, and (iii) reactions at surfaces.
Figure~\ref{fig:hetcat26} summarizes the metrics included in \textit{HetCat26}.

Specifically, the first category comprises surface energies of transition metals, which set the equilibrium morphology of catalytic surfaces and the shape of nanoparticles,\cite{balluffi2005,tran2016,zhuang2016} together with the energetics of adatom--vacancy creation at step edge and kink sites.\cite{xu2023}
In dilute alloys, an emerging class of catalysts, dopant atoms serve as the active sites and differ from the host metal in their electronic structure,\cite{greiner2018free,berger2025dopant} stability,\cite{ruban1999surface,berger2026atlas} formation energetics,\cite{wang2020surface,karageorgiou2026mechanisms} and reactivity.\cite{hannagan2020single,berger2024bringing}
We therefore include segregation energies of 4$d$ transition metal dopants in Cu, which determine whether the dopants are stable at the surface, where they are accessible to reactants, and dopant aggregation energies, which determine whether the active sites consist of isolated atoms or larger ensembles.
Metal--support interactions are equally important: supported nanoparticles are widely used as catalysts owing to their high surface-area-to-volume ratio, and oxide supports reduce sintering, thereby enhancing catalyst stability and lifetime.\cite{dai2018,astruc2020}
We assess these interactions using Cu$_2$, Ag$_2$, and Au$_2$ on MgO,\cite{shi2024} a model system that combines the prototypical rock-salt oxide support with catalytically relevant coinage metals, whose localized binding provides a stringent test for fMLIPs.\cite{barcaro2005,luo2016}

Adsorbate--surface interactions are assessed through adsorption energies spanning a chemically diverse range of adsorbates and surfaces,\cite{sharada2019,trepte2022,kleryoung} as well as adsorption site preferences for the notorious case of CO on Cu(111) and Pt(111).\cite{janthon2017,fanta2025}
Adsorption energies are central to heterogeneous catalysis because reactants must first adsorb before reacting at an active site.
Their importance is encapsulated by the Sabatier principle,\cite{sabatier1920} which relates adsorption energies to catalytic activity: adsorbates that bind too strongly poison the surface, while weak binding leads to desorption before activation.\cite{hammer2000,somorjai2010}
Beyond adsorption energetics, the CO site preference task provides a stringent test of the ability of fMLIPs to resolve subtle energy differences between competing adsorption motifs and reproduce the underlying DFT potential energy surface (PES).
A model is considered accurate here if it reproduces the site preference predicted by its underlying DFT functional, not the experimental site preference.\cite{feibelman2001,campbell2026}

Finally, reaction energetics are examined through barriers to dissociative chemisorption of H$_2$, N$_2$, and CH$_4$ on transition metal surfaces,\cite{tchakoua2023} and through barriers of elementary steps within representative catalytic reaction networks: CO$_2$ hydrogenation to methanol on Cu(111)\cite{niu} and the water--gas shift reaction (WGSR) on a range of Cu surfaces, spanning ideal terraces and increasingly undercoordinated active sites.\cite{hwang}
This progression from pristine surfaces to highly undercoordinated sites moves from conventional catalytic models toward more realistic ones, constituting a particularly demanding test of transferability.
These tests probe the ability of fMLIPs to describe the relative energetics of catalytic reaction pathways, including transition states and changes in bonding topology, and identify catalytic environments that remain challenging for current foundation models.

Table~\ref{tab:hetcat26_composition} summarizes the composition of \textit{HetCat26}; detailed descriptions of all benchmark datasets are provided in the Supporting Information.
Following common practice in the benchmarking of electronic-structure methods, the fMLIPs were evaluated by single point calculations on the reference DFT structures, which isolates the accuracy of the PES from differences in the optimized structures.
In practice, however, fMLIPs are used to locate minima and transition states and to run molecular dynamics (MD) simulations.
In the final section, we therefore examine how energies and structures change when reference DFT structures are reoptimized with the fMLIPs, and use two of the best-performing models in finite-temperature simulations of CO on Pt(111) and Cu(111).

\begin{table}[ht!]
    \centering
    \caption{Overview of the datasets in \textit{HetCat26}.
    For each property, the table reports the number of data points and the reference from which the benchmark task is obtained.
    For the ADS41 and CADS34 datasets, marked with an asterisk, subsets of 36 and 32 data points, respectively, were used throughout the analysis, except in the final section, which addresses the excluded systems (O-bearing adsorbates on W-, Ni-, Co-, and Mo-containing surfaces).}
    \label{tab:hetcat26_composition}
    \begin{tabular}{
    >{\centering\arraybackslash}m{0.56\textwidth}
    >{\centering\arraybackslash}m{0.26\textwidth}
    >{\centering\arraybackslash}m{0.10\textwidth}
    }
    \toprule
    \textbf{Property} & \textbf{\# Datapoints} & \textbf{Ref.} \\
    \midrule
    Surface energies & 321 & \cite{tran2016} \\
    \addlinespace
    \makecell{Adatom--vacancy formation energies} & 16 & \cite{xu2023} \\
    \addlinespace
    \makecell{Segregation and aggregation energies} & 35 & \cite{berger2026atlas} \\
    \addlinespace
    \makecell{Metal--metal oxide interactions} & 24 & \cite{shi2024} \\
    \addlinespace
    \makecell{Adsorption energies (CADS34)*} & 34 & \cite{kleryoung} \\
    \addlinespace
    \makecell{Adsorption energies (ADS41)*} & 41 & \cite{sharada2019,trepte2022} \\
    \addlinespace
    \makecell{Adsorption site preferences} & 7 &  \cite{janthon2017,fanta2025} \\
    \addlinespace
    \makecell{Reaction barriers (dissoc. chemisorption - SBH17)} & 15 & \cite{tchakoua2023} \\
    \addlinespace
    \makecell{Reaction barriers (CO$_2$ hydrogenation on Cu(111))} & 42 & \cite{niu} \\
    \addlinespace
    \makecell{Reaction barriers (WGSR on Cu)} & 144 & \cite{hwang} \\
    \bottomrule
    \end{tabular}
\end{table}

\subsection{Pre-trained models investigated in this work}

Developing and validating fMLIPs against DFT results is a rapidly evolving field, so a clear criterion for model selection is needed.
Our guiding principle is consistency in the level of theory between the training data and the reference data, so that discrepancies can be attributed to model error rather than to differences in the underlying electronic structure method.
Most available large-scale datasets have been computed with the Perdew--Burke--Ernzerhof (PBE) functional.\cite{perdew1996}
For this reason, the reference calculations in \textit{HetCat26} employ PBE, although this functional is known to describe some catalytic properties poorly.\cite{gerrits2020,araujo2022,berger2021,berger2023}
Accordingly, we consider only fMLIPs trained on PBE-based, general-purpose materials datasets, including OMat24,\cite{barroso_luque2024} MPTrj,\cite{deng2023} Alexandria (and its subsampled, filtered version sAlex),\cite{schmidt2021,wang2023,schmidt2023} and MatPES-PBE.\cite{kaplan2025}
The few large-scale catalysis-specific datasets, such as OC20\cite{chanussot2021} and AQCat25,\cite{allam2026} would in principle be natural candidates for these benchmark tests.
However, their level of theory differs from that of our reference data, which would confound functional differences with model error.
For the same reason, we do not consider models trained on popular non-PBE datasets such as MatPES-r$^2$SCAN\cite{kaplan2025} and MAD.\cite{mazitov2025}
For the CADS34 and the CO$_2$ hydrogenation on Cu(111) datasets, D3 dispersion with Becke--Johnson damping\cite{grimme2010} is added to the fMLIP predictions to match the reference calculations.

We selected 18 models that span a broad range of datasets and architectures and thus provide a representative snapshot of the current landscape.
Given the rapid progress in this field, this selection is necessarily non-exhaustive.
Thirteen of these models have also been evaluated on Matbench Discovery, which allows us to compare their performance on general materials properties with that on heterogeneous catalysis and thus to assess their transferability.
Table~\ref{tab:mlip_models} lists the fMLIPs considered.

\begin{table}[!htp]
    \centering
    \caption{Summary of the fMLIPs investigated in this work.
    For each model, the table reports the dataset(s) and procedure used during model development, the release date, and the corresponding reference.
    The $+$ symbol indicates training on combined datasets, while the $\rightarrow$ symbol indicates fine-tuning.
    For models trained across multiple levels of theory or with separate prediction heads (SevenNet-MF-ompa, SevenNet-Omni-i12, MACE-MH-1-OMAT, and UMA-s-1p2p1), marked with an asterisk, we report the datasets relevant here.}
    \label{tab:mlip_models}
    \begin{tabular}{
    >{\centering\arraybackslash}m{0.28\textwidth} 
    >{\centering\arraybackslash}m{0.46\textwidth} 
    >{\centering\arraybackslash}m{0.12\textwidth} 
    >{\centering\arraybackslash}m{0.04\textwidth}
    }
    \toprule
    \textbf{fMLIP} & \textbf{Training} &
    \textbf{Release} &
    \textbf{Ref.} \\
    \midrule
    MatterSim-v1-5M & MatterSim & May 2024 & \cite{yang2024} \\ \addlinespace
    MACE-MPA-0 & MPTrj + sAlex & Dec. 2024 & \cite{batatia2025} \\ \addlinespace
    MACE-MP-0b3 & MPTrj & Dec 2024 & \cite{batatia2025} \\ \addlinespace
    MACE-OMAT-0 & OMat24 & Jan. 2025 & \cite{batatia2025} \\ \addlinespace
    eSEN-30M-OAM & OMat24 $\rightarrow$ MPTrj + sAlex & Mar. 2025 & \cite{fu2025} \\ \addlinespace
    MACE-MATPES-PBE-0 & OMat24 $\rightarrow$ MatPES-PBE & Mar. 2025 & \cite{batatia2025} \\ \addlinespace
    SevenNet-MF-ompa & * MPtrj + sAlex & Mar. 2025 & \cite{kim2025} \\ \addlinespace
    ORB-v3 & MPTrj + sAlex & Apr. 2025 & \cite{rhodes2025} \\ \addlinespace
    NequIP-OAM-L & OMat24 $\rightarrow$ MPTrj + sAlex & Aug. 2025 & \cite{kavanagh2026} \\ \addlinespace
    MACE-MH-1-OMAT & * OMat24 & Oct. 2025 & \cite{batatia2025mh1} \\ \addlinespace
    NequIP-OAM-XL & OMat24 $\rightarrow$ MPTrj + sAlex & Nov. 2025 & \cite{kavanagh2026} \\ \addlinespace
    PET-OAM-XL & OMat24 $\rightarrow$ MPTrj + sAlex & Jan. 2026 & \cite{bigi2026} \\
    \addlinespace
    SevenNet-Omni-i12 & * MPtrj + sAlex & Jan. 2026 & \cite{kim2026} \\ \addlinespace 
    TACE-OAM-L & OMat24 $\rightarrow$ MPTrj + sAlex & Apr. 2026 & \cite{xu2025} \\ \addlinespace
    EquFlashV2 & OMat24 $\rightarrow$ MPTrj + sAlex & Jun. 2026 & \cite{lee2025} \\ \addlinespace
    GRACE-3L-OAM-L & OMat24 $\rightarrow$ MPTrj + sAlex & Jul. 2026 & \cite{lysogorskiy2025} \\ \addlinespace
    TECE-OAM-RRA-1.0 & OMat24 $\rightarrow$ MPTrj + sAlex & Jul. 2026 & \cite{xu2026} \\ \addlinespace
    UMA-s-1p2p1 & * OMat24 & Aug. 2026 &\cite{wood2025} \\ \addlinespace
    \bottomrule
    \end{tabular}
\end{table}

\section{Results and Discussion}

\subsection{Overall Performance of Pre-trained MLIPs on \textit{HetCat26}}

\begin{figure}[htbp]
    \includegraphics[width=14.1cm,height=\textheight,keepaspectratio]{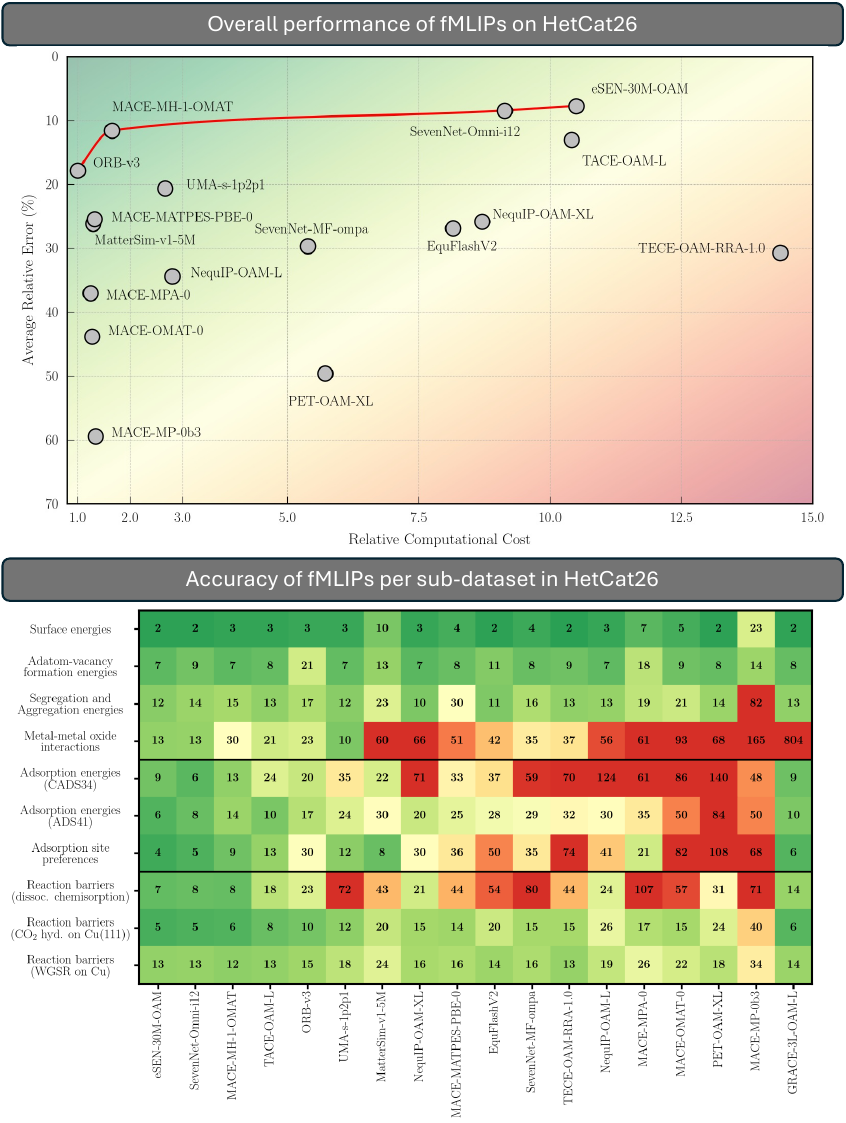}
    \caption{
    Overall and per-dataset performance of the eighteen investigated fMLIPs.
    The Pareto plot at the top summarizes accuracy, reported as the average ``relative error'' across all datasets as defined in the main text, against computational cost, reported as a multiple of that of the cheapest fMLIP investigated, ORB-v3.
    The Pareto front (red) identifies the models with the best accuracy at a given cost.
    The heatmap at the bottom breaks down the relative error per dataset for each fMLIP; green corresponds to low and red to high relative errors (0\% and 60\% respectively).
    }
    \label{fig:overall_performance}
\end{figure}

For each dataset, model accuracy is quantified as a relative error in \%: the mean absolute error (MAE) in the property of interest with respect to the reference DFT results, normalized by the average magnitude of the DFT values.
The computational cost of a model is defined as the total time required to complete the inferences associated with the benchmark on an NVIDIA GeForce RTX 5080 GPU.
These quantities, together, provide a measure of the practical quality and efficiency of each model.
The overall performance of all investigated fMLIPs across \textit{HetCat26} is summarized in the Pareto plot in the top panel of Figure~\ref{fig:overall_performance}.

Despite substantial overlap in their training data, many models differ considerably in performance.
The eSEN-30M-OAM, SevenNet-Omni-i12, and MACE-MH-1-OMAT models exhibit the lowest average relative errors across the datasets, at 7.8\%, 8.5\%, and 11.6\%, respectively.
This makes them particularly promising candidates for heterogeneous catalysis applications out of the box, without further fine-tuning.
For eSEN-30M-OAM, this is in line with other benchmarking efforts in the field.\cite{moon2025,kempen2026}
Notably, the three models that combine multi-head architectures with cross-domain training on datasets that include catalysis-specific data (SevenNet-Omni-i12, UMA-s-1p2p1, and MACE-MH-1-OMAT) perform well, suggesting that such approaches are promising for the development of universal models.

Strong performance for individual properties does not necessarily translate into strong overall performance, and vice versa, as highlighted in the bottom panel of Figure~\ref{fig:overall_performance}.
For example, although ORB-v3 performs well overall, obtaining the fifth lowest average relative error (17.9\%), it is the least accurate of all models for adatom--vacancy formation energies (21.5\%).
Conversely, TECE-OAM-RRA-1.0 performs best for surface energies (1.8\%), yet ranks within the lowest half overall, with an average relative error of 30.7\%.
In addition, the GRACE-3L-OAM-L model shows very strong performance across all quantities, except metal--metal oxide interactions.
These observations highlight the importance of evaluating fMLIPs across diverse catalytic properties rather than relying on a single benchmark category, and of assessing performance for the specific application of interest before use.

Beyond accuracy, computational efficiency (inference speed and memory demand) is also an important consideration for practical applications.
Several models based on the MACE architecture,\cite{batatia2022} together with ORB-v3 and MatterSim-v1-5M, are among the fastest models investigated here.
At the high-accuracy end of the Pareto front, eSEN-30M-OAM, SevenNet-Omni-i12 and MACE-MH-1-OMAT stand out.
While eSEN-30M-OAM is the most accurate of the three, MACE-MH-1-OMAT is only slightly behind and roughly five times as fast, making it a particularly attractive compromise between accuracy and computational cost.

\begin{figure}[ht!]
    \includegraphics[width=16cm,height=\textheight,keepaspectratio]{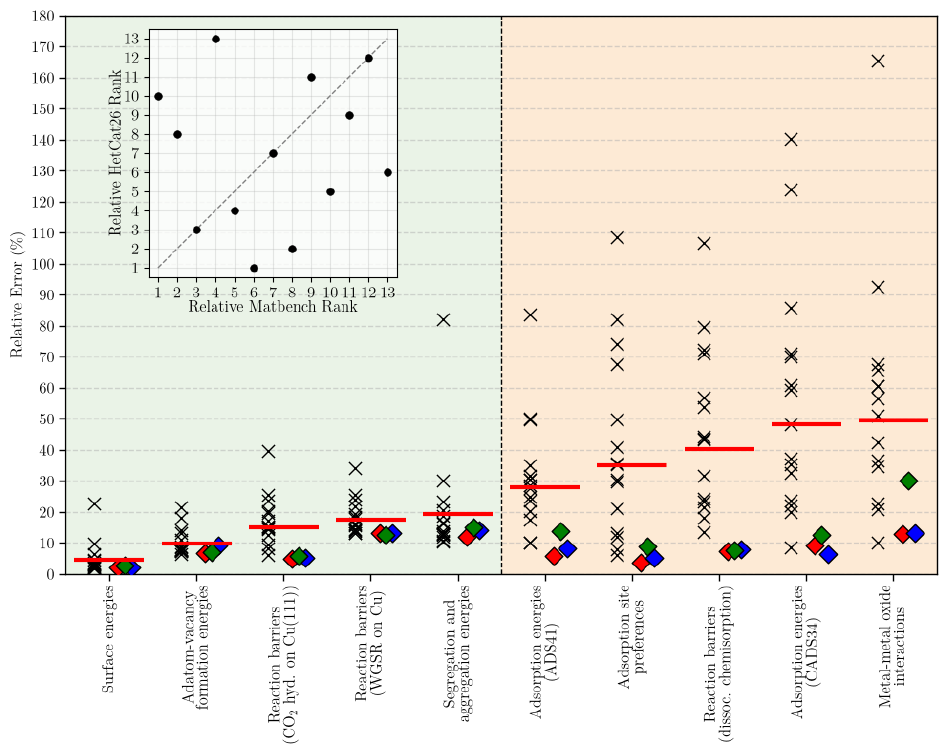}

    \caption{
    Relative errors of fMLIPs for individual datasets in \textit{HetCat26}, together with a comparison of their aggregate performance against that on Matbench Discovery.
    Most fMLIPs are represented by crosses; diamonds indicate models that perform well across all tests:
    eSEN-30M-OAM (red), MACE-MH-1-OMAT (green) and SevenNet-Omni-i12 (blue).
    Red lines indicate the average relative error across all fMLIPs for a given dataset (the abnormal result for GRACE-3L-OAM-L for metal--metal oxide interactions is not included for this calculation).
    For the thirteen fMLIPs tested on both Matbench Discovery and \textit{HetCat26}, the inset in the top left shows weak correlation in overall performance on the two benchmarks.
    The relative Matbench rank is determined from the models' Combined Performance Score (CPS v1),\cite{riebesell2025} using the default weighting of the constituent metrics.
    The \textit{HetCat26} rank is determined by the models' average relative error across all datasets.}
    \label{fig:metric_summary}
\end{figure}

The overall performance of fMLIPs on \textit{HetCat26} differs substantially from the rankings obtained on existing materials-focused benchmarks.
For example, PET-OAM-XL, which ranks among the leading models on Matbench Discovery\cite{riebesell2025} (at the time of model selection), performs comparatively poorly on \textit{HetCat26}, owing to its less accurate description of adsorbate--surface interactions.
As shown in the inset of Figure~\ref{fig:metric_summary}, performance on Matbench Discovery, as measured by the Combined Performance Score (CPS v1), is only weakly predictive of performance on \textit{HetCat26}.
Strong performance for general materials therefore does not necessarily translate into accurate descriptions of the diverse chemical environments encountered in heterogeneous catalysis, highlighting the need for dedicated benchmarking in this field.

\subsection{The Good, the Bad, and the Ugly: Which Aspects of Heterogeneous Catalysis Challenge Pre-trained Models?}

The overall rankings discussed above conceal substantial variation across individual benchmark categories.
Some catalytic properties are reproduced with remarkable accuracy by current fMLIPs, whereas others remain challenging and continue to limit transferability for the majority of models.
Figure~\ref{fig:metric_summary} summarizes performance for the individual tests in \textit{HetCat26}.
In the following, we focus on the overarching trends that emerge from these results; fMLIP--DFT parity plots for all metrics and all models are provided in the Supporting Information.

\begin{figure}[ht!]
    \includegraphics[width=16cm,height=\textheight,keepaspectratio]{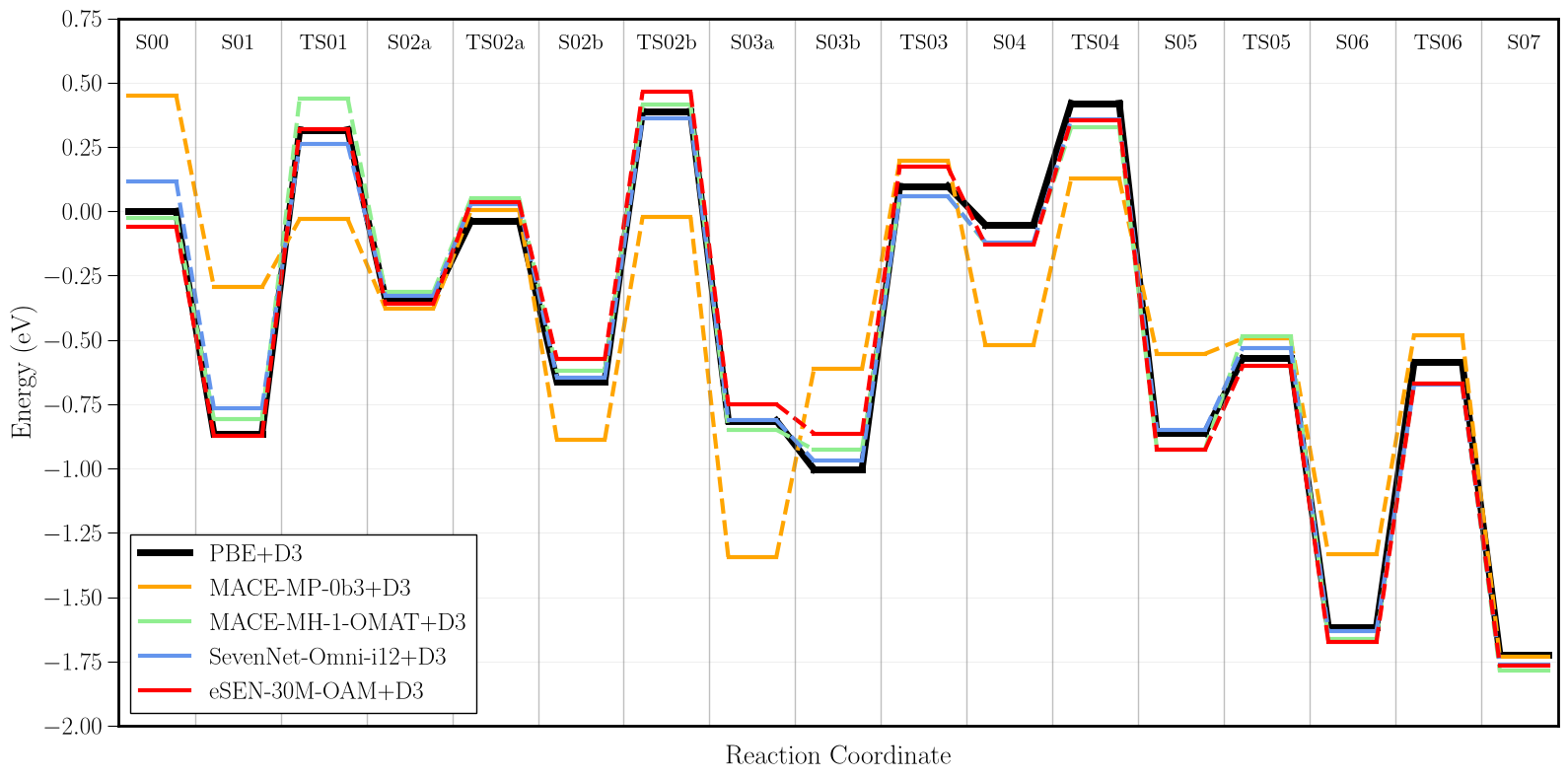}
    \caption{Comparison of the PBE+D3 and four fMLIP+D3 reaction pathways for the hydrogenation of CO$_2$ to methanol on Cu(111).
    The labels at the top denote the states along the reaction pathway; details are available in the Supporting Information.
    The reference PBE+D3 pathway is set to 0~eV at the initial state.
    The vertical offset of each fMLIP pathway is determined by minimizing the average separation from the corresponding reference states across the full sequence of elementary steps, rather than by aligning the initial states, so that the comparison is not disproportionately influenced by any single step.
    The reaction energy profiles aligned to the initial state are provided in the Supporting Information (Figure~S11).
    The eSEN-30M-OAM+D3, SevenNet-Omni-i12+D3 and MACE-MH-1-OMAT+D3 pathways agree closely with the reference.
    MACE-MP-0b3 performs more poorly, likely as a result of the low coverage of configurations far from equilibrium in MPTrj.\cite{kaplan2025}}
    \label{fig:pathway}
\end{figure}

Current fMLIPs accurately capture five benchmark tasks: surface energies, adatom--vacancy formation energies, segregation and aggregation energies, and reaction barriers for both CO$_2$ hydrogenation on Cu(111) and the WGSR on Cu.
Surface energies, for a diverse set of 321 surfaces, are reproduced with the lowest average relative error across all models (4.6\%), a marked improvement over early fMLIPs (CHGNet and M3GNet).\cite{focassio2025}
The energies associated with adatom--vacancy formation (i.e., surface atom migration), are also predicted accurately (9.9\%), despite involving undercoordinated environments such as vacancies, adatoms, step edges, and kinks.
For segregation and aggregation energies, all models except MACE-MP-0b3 perform well (19.3\%).
Reaction energetics for the hydrogenation of CO$_2$ on Cu(111) and the WGSR at various Cu active sites are, perhaps surprisingly, among the better-described quantities in \textit{HetCat26}, with average relative errors across all fifteen models of 15.2\% and 17.6\%, respectively.
For the best-performing model in each case, these correspond to MAEs in reaction barriers of 45~meV for CO$_2$ hydrogenation and 116~meV for the WGSR, or relative errors of 4.8 and 12.4\%, respectively.
Figure~\ref{fig:pathway} illustrates this for the COOH--CO pathway (see Supporting Information for details) in the hydrogenation of CO$_2$ on Cu(111).
eSEN-30M-OAM+D3, SevenNet-Omni-i12+D3 and MACE-MH-1-OMAT+D3 follow the reference PBE+D3 energy profile closely across the full sequence of elementary steps.
This level of accuracy, out-of-the-box, approaches that traditionally associated with bespoke MLIPs.

\begin{figure}[htbp]
    \includegraphics[width=12.4cm,height=\textheight,keepaspectratio]{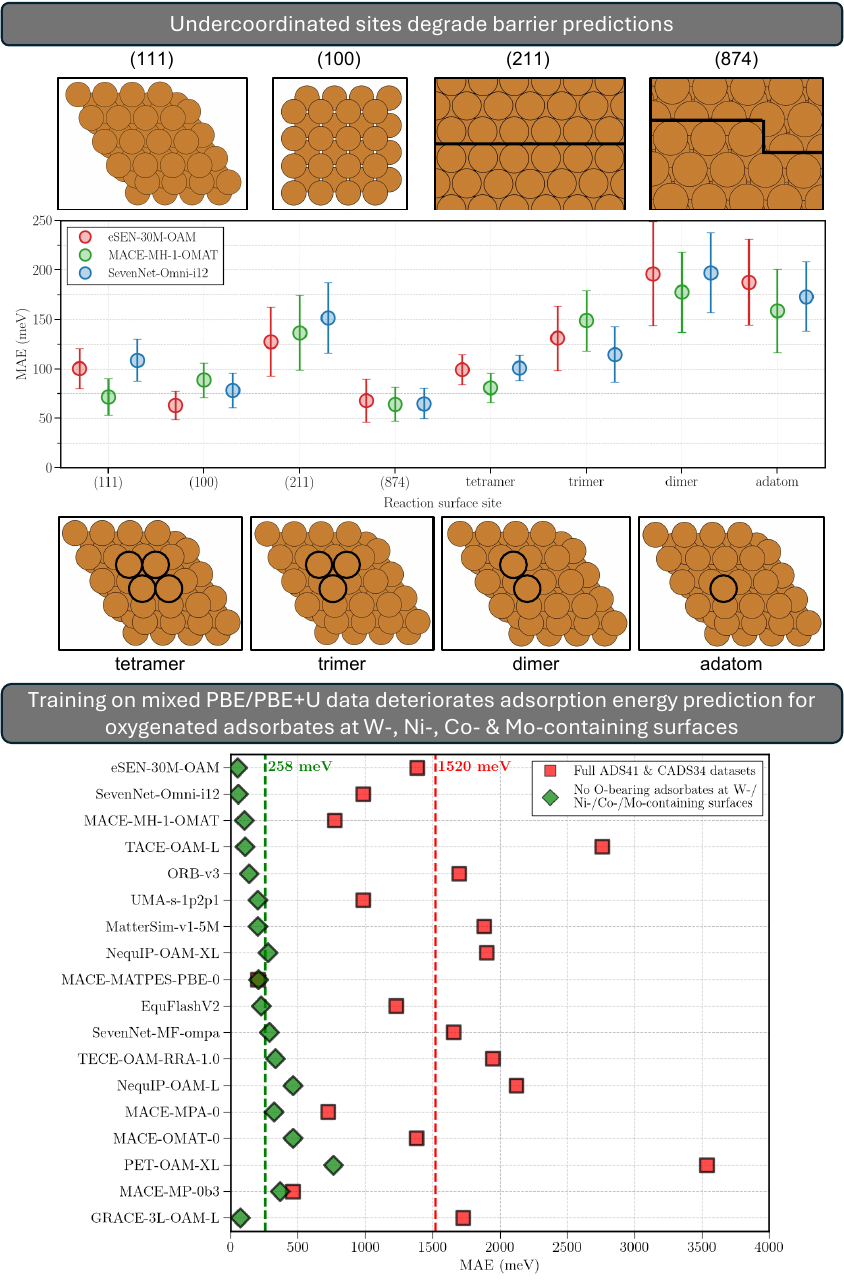}
    \caption{
    Limitations of fMLIPs for modeling in heterogeneous catalysis.
    The top panel shows that the accuracy of eSEN-30M-OAM, SevenNet-Omni-i12 and MACE-MH-1-OMAT in predicting reaction barriers within the WGSR network on different Cu sites generally worsens as undercoordination increases.
    The bottom panel shows MAEs across the ADS41 and CADS34 adsorption energy datasets when systems with O-bearing adsorbates on W-, Ni-, Co-, and Mo-containing surfaces are included (red) and excluded (green); all but one model improve markedly upon their exclusion.
    }
    \label{fig:fmlip_limitations}
\end{figure}

The encouraging performance for the WGSR dataset nevertheless masks a clear dependence on the nature of the active site, as illustrated in the top panel of Figure~\ref{fig:fmlip_limitations}.
For the best-performing model, eSEN-30M-OAM, errors in reaction barriers are lowest for extended Cu surfaces, with MAEs ranging from 63 to 128~meV across the Cu(100), Cu(111), Cu(211), and Cu(874) facets.
As the coordination of the active site decreases, the MAE increases from 99~meV for Cu tetramers to 131~meV for Cu trimers and 196~meV for Cu dimers supported on Cu(111), comparable to the 188~meV error for a single Cu adatom.
Overall, although not a perfect trend, eSEN-30M-OAM generally becomes less accurate for more undercoordinated environments.
These observations hold for other top-performing models SevenNet-Omni-i12 and MACE-MH-1-OMAT.
Given the importance of such sites in realistic catalysts,\cite{yang2025_2,elnabawy2025,xu2025_2,chen2025_atomistic,zhang2020ensembles,poths2024thermodynamic} this is a priority for future dataset development.

Larger errors are generally observed for quantities that depend on adsorbate--surface interactions.
From the ADS41 and CADS34 datasets, we exclude O-bearing adsorbates on W-, Ni-, Co-, and Mo-containing surfaces from the analysis; these systems are discussed below.
The average relative errors (across all models) for these two datasets, for CO adsorption on Cu(111) and Pt(111), and for dissociative chemisorption (SBH17) lie between 27.8\% and 48.2\%.
The best-performing models nevertheless achieve relative errors below 10\% for all four datasets.
The underlying physics can therefore be captured by current fMLIPs, but its accurate description remains strongly model dependent.
PET-OAM-XL illustrates this particularly clearly: it is one of the most accurate model for surface energies (2.0\%), yet exhibits MAEs of 771~meV for ADS41, 753~meV for CADS34, and as much as 1305~meV for CO adsorption on Pt(111) and Cu(111), corresponding to relative errors of 83.6, 140.3 and 108.5\% respectively.
Its reduced performance for adsorption is consistent with results for the MamunHighT2019 dataset\cite{mamun2019} in CatBench.\cite{moon2025}
Moreover, of the eighteen models evaluated here, as shown in Figure~S8 in the Supporting Information, only nine correctly recover the adsorption site preference ordering obtained from the reference DFT calculations for CO on Cu(111) and Pt(111).
These observations suggest that the most severe limitation of many fMLIPs is not the description of reactive steps, but the accurate treatment of how molecules adsorb from the gas phase onto the surface.

Another challenging benchmark test is the description of metal--metal oxide interactions.
Excluding GRACE-3L-OAM-L, which is an outlier for this dataset, the average relative error across all models is around 50\%, and only three models obtain a relative error below 20\%.
This dataset combines two features that make it particularly demanding: the metal clusters are highly undercoordinated, which tests the transferability of the models, and the interaction energies are comparatively small, which inflates the relative error.

An additional caveat emerges for systems in ADS41 and CADS34 where oxygen-containing species approach W-, Ni-, Co-, and Mo-containing surfaces.
Most fMLIPs exhibit strongly reduced accuracy for these systems, consistent with recent observations.\cite{warford2026}
This behavior originates from the use of training datasets that combine PBE and PBE+U reference calculations; in the Materials Project, the +U correction is applied to the \textit{d}-states of V, W, Fe, Ni, Co, Cr, Mo, and Mn in the presence of oxygen or fluorine.
Because PBE and PBE+U correspond to distinct potential energy surfaces, mixing the two levels of theory introduces inconsistencies into the training data.
Local atomic environments encountered during inference may then resemble those associated with Hubbard-corrected reference data, despite the benchmark systems being described by reference data without the correction.
The impact can be substantial, as shown by the large improvement in model performance upon excluding the affected systems, namely O-bearing adsorbates on W-, Ni-, Co-, and Mo-containing surfaces, as shown in the bottom panel of Figure~\ref{fig:fmlip_limitations}. 
These are the only affected combinations, as the other problematic combinations are not present in either ADS41 or CADS34.
MACE-MATPES-PBE-0, fine-tuned exclusively on PBE data, is the only model that does not exhibit this deterioration.
Consistency of the underlying electronic structure method is thus a critical aspect of foundation model dataset development.

Collectively, the results reveal a hierarchy of transferability across properties relevant to heterogeneous catalysis.
Current fMLIPs accurately describe surface energetics and, somewhat unexpectedly, the relative energetics of catalytic reaction pathways, including transition state structures far from equilibrium.
By contrast, adsorption and interaction of species with surfaces (both standard adsorbates with surfaces and metal clusters with metal--oxide surfaces) remain more challenging.

\subsection{Locating stationary points with fMLIPs}

So far, our analysis has relied on DFT reference structures and the energy differences derived from them.
Although this is common practice in benchmarking electronic structure methods, practical workflows require the fMLIPs to locate stationary points or to run MD simulations themselves.
Here, we assess their performance in locating stationary points.
To this end, we optimize the structures in the ADS41 dataset with the fMLIPs (excluding oxygen-bearing adsorbates on Ni-, Co-containing surfaces) and analyze the resulting changes in energies and structures.
In addition, we assess the ability of the fMLIPs to locate transition states, and the resulting changes in energies and structures, by comparison with the DFT transition state structures of SBH17.
Computational details and full results are provided in the Supporting Information.

\begin{figure}[htbp]
    \includegraphics[width=14.1cm,height=\textheight,keepaspectratio]{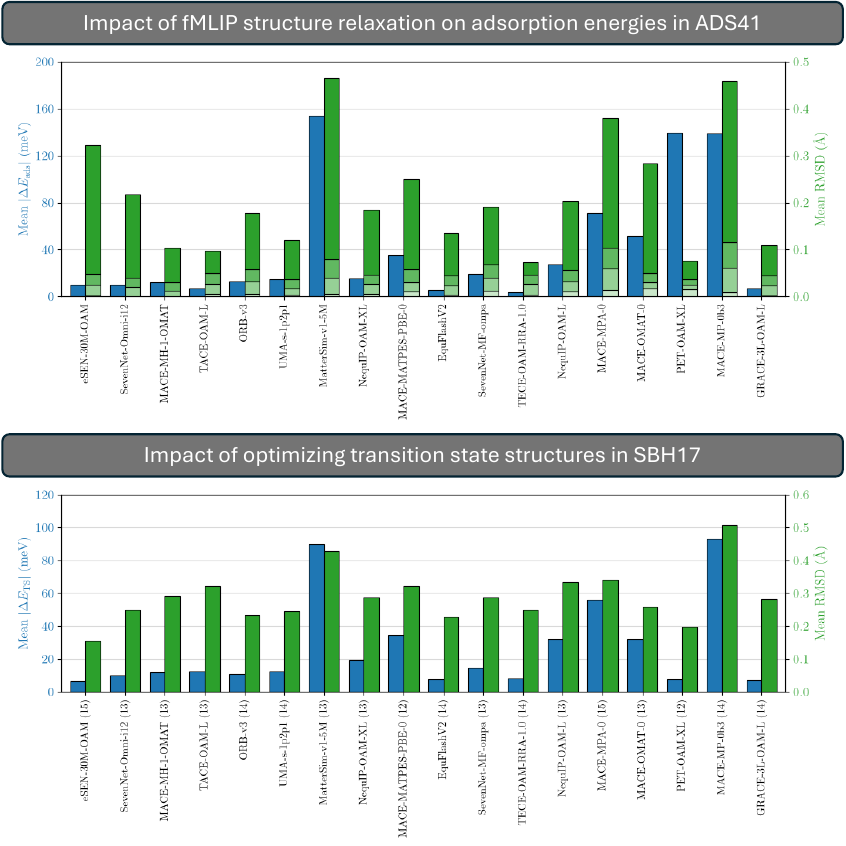}
    \caption{
    Impact of structure optimizations with fMLIPs.
    In the top panel, the effect of fMLIP structure relaxation on adsorption energies in the ADS41 dataset is reported.
    The blue bars give the mean absolute change in adsorption energy, between single point evaluations on the reference structures and evaluations after relaxation with each model ($\Delta E_\mathrm{ads} = E_\mathrm{ads}^\mathrm{relaxed} - E_\mathrm{ads}^\text{single point}$).
    The green bars give the mean RMSD between the relaxed and reference structures (lighter to darker green: gas phase molecules, clean slabs, combined system slabs, combined system adsorbates).
    In the bottom panel, the refinement of transition states in the SBH17 dataset by fMLIPs is considered. 
    The blue bars give the average absolute change in transition state energy ($\Delta E_\mathrm{TS} = E_\mathrm{model}(\mathbf{R}_\mathrm{TS}^\mathrm{model}) - E_\mathrm{model}(\mathbf{R}_\mathrm{TS}^\mathrm{DFT})$) across each model’s successful transition state refinements (the number of which is given in brackets and can reach a maximum of 15), and the green bars show the corresponding mean RMSD of the molecule from the DFT structure.
    }
    \label{fig:fmlip_relaxation}
\end{figure}

As shown in the top panel of Figure~\ref{fig:fmlip_relaxation}, the average absolute change in adsorption energy upon relaxation ranges from 4~meV (TECE-OAM-RRA-1.0) to 154~meV (MatterSim-v1-5M).
Since the mean absolute reference value is 922 meV, the changes due to structural relaxation are almost negligible for the best performing models.
Structural deviations from the DFT structures are small for the gas-phase molecules, the clean slabs, and the surface atoms of the adsorbate--slab systems.
The largest deviations occur for the adsorbate atoms in these systems.
Their root-mean-square deviation (RMSD), averaged over all models, is 0.16~\AA{} and ranges from only 0.03~\AA{} (TECE-OAM-RRA-1.0) to 0.39~\AA{} (MatterSim-v1-5M).
The detailed results in the Supporting Information show that these structural changes are dominated by weakly bound species, which can be displaced considerably relative to the surface at little energetic cost.

Focusing now on transition states, the bottom panel of Figure~\ref{fig:fmlip_relaxation} shows that the 18 models locate a transition state for most systems, between 12 (MACE-MATPES-PBE-0, PET-OAM-XL) and all 15 (eSEN-30M-OAM, MACE-MPA-0) of the 15 systems.
For many models, the energy of the refined transition state differs from the DFT reference by less than 20~meV on average, and the RMSD of the reacting molecule is below 0.30~\AA{}.
The best performing model, eSEN-30M-OAM, shows an average deviation of 6.6 meV in transition state energies and an average RMSD of 0.15~\AA{}.
These results suggest that fMLIPs can reliably refine transition states for the systems considered here.

Beyond locating stationary points, a key strength of fMLIPs is that they make finite-temperature simulations affordable: MD simulations for finite-temperature effects and enhanced-sampling techniques for rare events capture anharmonic and entropic effects that are inaccessible from stationary points on the PES alone.
To demonstrate this, we perform umbrella sampling (US) simulations of CO on Pt(111) and Cu(111) with two of the best-performing models (eSEN-30M-OAM and MACE-MH-1-OMAT).
All simulations remained stable with the fMLIPs out of the box.
Details of the simulations and the resulting free energy surfaces are provided in the Supporting Information.

\section{Conclusion}

Pre-trained MLIPs are rapidly becoming central tools for atomistic simulations across materials science.
Their transferability to heterogeneous catalysis has nevertheless remained an open question, despite growing interest in simulating catalytic applications with such models.
To address this gap, we introduce \textit{HetCat26}, a benchmark framework designed specifically to assess the performance and transferability of foundation MLIPs across key aspects of heterogeneous catalysis, including surface energetics, metal--metal oxide interactions, adsorbate--surface interactions, and catalytic reaction networks.

Benchmarking eighteen foundation MLIPs reveals substantial differences in performance despite considerable overlap in their underlying training datasets.
Three models, eSEN-30M-OAM, SevenNet-Omni-i12 and MACE-MH-1-OMAT, consistently achieve high accuracy out-of-the-box across the diverse benchmark categories.
At the same time, performance on existing materials-focused benchmarks proves weakly predictive of performance in heterogeneous catalysis, highlighting the need for dedicated benchmark frameworks.

Beyond single point evaluations on DFT reference structures, the best models also perform well in practical workflows.
Structure optimizations change adsorption energies by only a few meV and preserve the reference structures, and transition states are refined close to the DFT reference.
Umbrella sampling simulations with two of the best-performing models remain stable, demonstrating that fMLIPs make finite-temperature free energy calculations for catalytic systems accessible.

Our results reveal distinct levels of transferability across different aspects of heterogeneous catalysis.
Current foundation MLIPs describe surface energies and adatom--vacancy formation well, both of which underpin surface structure, and also capture energy barriers along two representative catalytic reaction pathways.
Their performance deteriorates, however, for properties that depend sensitively on adsorbate--surface interactions.
The main limitation of current foundation MLIPs for heterogeneous catalysis is therefore not the description of reaction energetics itself, but the accurate treatment of the interaction between gas phase species and surfaces.

Beyond providing a benchmark for model evaluation, \textit{HetCat26} identifies classes of catalytic environments that remain insufficiently represented in current training datasets and therefore constitute promising targets for future dataset development.
It further highlights the importance of training data consistency, corroborating recent findings that mixing PBE and PBE+U reference calculations can substantially deteriorate model performance.\cite{warford2026}
Future foundation MLIPs will likely benefit not only from expanded datasets covering challenging catalytic environments, but also from greater consistency in the underlying electronic structure methodology.
We anticipate that \textit{HetCat26} will serve as a community benchmark for guiding both dataset construction and model development, enabling increasingly transferable foundation MLIPs for heterogeneous catalysis.
All reference structures and energies are archived alongside analysis scripts, and we intend to make \textit{HetCat26} available through live leaderboards so that new models can be assessed against a consistent reference as the field advances.

\begin{acknowledgement}
FB acknowledges support from the Alexander von Humboldt Foundation through a Feodor Lynen Research Fellowship, from the Isaac Newton Trust through an Early Career Fellowship, and from Churchill College, Cambridge, through a Postdoctoral By-Fellowship. 
This work has been funded by the European Union (ERC, n-AQUA, 101071937).
Views and opinions expressed are, however, those of the authors only and do not necessarily reflect those of the European Union or the European Research Council Executive Agency.
Neither the European Union nor the granting authority can be held responsible for them. 
This work was performed using resources provided by the Cambridge Service for Data Driven Discovery (CSD3) operated by the University of Cambridge Research Computing Service (www.csd3.cam.ac.uk), provided by Dell EMC and Intel using Tier-2 funding from the Engineering and Physical Sciences Research Council (capital grant EP/T022159/1), and DiRAC funding from the Science and Technology Facilities Council (https://www.dirac.ac.uk), with additional access through a University of Cambridge EPSRC Core Equipment Award (EP/\-X034712/1). We additionally acknowledge computational support and resources from the UK National High-Performance Computing Service, Advanced Research Computing High End Resource (ARCHER2).
Access for ARCHER2 was obtained via the Materials Chemistry Consortium (MCC), funded by EPSRC grant references EP/X035859 and EP/F067496. 
Further computational support and resources were provided by YOUNG, the Tier-2 High Performance Computing Hub in Materials and Molecular Modeling (MMM), which is partially funded by EPSRC grant reference EP/T022213. 
We also acknowledge the EuroHPC Joint Undertaking for awarding project ID EHPC-REG-2024R02-130 access to Leonardo at CINECA (Italy), and project ID EHPC-REG-2025R02-112 access to JUPITER at Jülich (Germany). 
MM gratefully acknowledges support from a John Wilfrid Linnett Visiting Professorship with the Department of Chemistry, and a Visiting Fellowship with Sidney Sussex College, Cambridge University.
Work at UW-Madison was supported by US Department of Energy, Basic Energy Sciences (DOE-BES), Division of Chemical Sciences, Catalysis Science Program, grant number DE-FG02-05ER15731, and used resources of the National Energy Research Scientific Computing Center, a DOE Office of Science User Facility supported by the Office of Science of the U.S. Department of Energy under Contract No. DE-AC02-05CH11231 using NERSC award BES-ERCAP0035227.
\end{acknowledgement}

\begin{suppinfo}
The Supporting Information is available free of charge at https:// ...
It includes the rationale behind the construction of \textit{HetCat26} and details about the sub-datasets contained within.
In addition, the Supporting Information details the computational setup for the fMLIP calculations and fMLIP--DFT parity plots are available for each individual sub-dataset and model.
\end{suppinfo}

\clearpage

\bibliography{refs}

@article{baiker2000,
  author = {Baiker, A.},
  title = {Utilization of carbon dioxide in heterogeneous catalytic synthesis},
  journal = {Appl. Organomet. Chem.},
  year = {2000},
  volume = {14},
  number = {12},
  pages = {751},
  publisher = {John Wiley \& Sons, Ltd.}
}

@article{marcilly2003,
  author = {C. Marcilly},
  title = {Present status and future trends in catalysis for refining and petrochemicals},
  journal = {J. Catal.},
  volume = {216},
  number = {1},
  pages = {47},
  year = {2003},
  publisher = {Elsevier}
}

@article{armor2011,
  author = {J. N. Armor},
  title = {A history of industrial catalysis},
  journal = {Catal. Today},
  volume = {163},
  number = {1},
  pages = {3},
  year = {2011},
  publisher = {Elsevier}
}

@article{liu2014,
  title = {Ammonia synthesis catalyst 100 years: Practice, enlightenment and challenge},
  author = {H. Liu},
  journal = {Chin. J. Catal.},
  volume = {35},
  number = {10},
  pages = {1619},
  year = {2014},
  publisher = {Elsevier}
}

@article{bowker2022,
  author = {Bowker, M. and DeBeer, S. and Dummer, N. F. and Hutchings, G. J. and Scheffler, M. and Schüth, F. and Taylor, S. H. and Tüysüz, H.},
  title = {Advancing Critical Chemical Processes for a Sustainable Future: Challenges for Industry and the Max Planck–Cardiff Centre on the Fundamentals of Heterogeneous Catalysis (FUNCAT)},
  journal = {Angew. Chem. Int. Ed.},
  volume = {61},
  number = {50},
  pages = {e202209016},
  year = {2022},
  publisher = {Wiley-VCH}
}

@article{norskov2009,
  author = {N{\o}rskov, J. K. and Bligaard, T. and Rossmeisl, J. and Christensen, C. H.},
  title = {Towards the computational design of solid catalysts},
  journal = {Nat. Chem.},
  year = {2009},
  volume = {1},
  pages = {37},
  publisher = {Nature Publishing Group}
}

@article{schlogl2015,
  author = {Schl{\"o}gl, R.},
  title = {Heterogeneous Catalysis},
  journal = {Angew. Chem. Int. Ed.},
  year = {2015},
  volume = {54},
  number = {11},
  pages = {3465},
  publisher = {Wiley-VCH}
}

@article{swetlana2015,
    author = {Schauermann, S. and Freund, H.-J.},
    title = {Model Approach in Heterogeneous Catalysis: Kinetics and Thermodynamics of Surface Reactions},
    journal = {Acc. Chem. Res.},
    volume = {48},
    number = {10},
    pages = {2775},
    year = {2015},
    publisher = {American Chemical Society}
}

@article{honkala2005,
  author = {Honkala, K. and Hellman, A. and Remediakis, I. N. and Logadottir, A. and Carlsson, A. and Dahl, S. and Christensen, C. H. and N{\o}rskov, J. K.},
  title = {Ammonia Synthesis from First-Principles Calculations},
  journal = {Science},
  year = {2005},
  volume = {307},
  number = {5709},
  pages = {555},
  publisher = {American Association for the Advancement of Science}
}

@article{kandoi2006,
  author = {Kandoi, S. and Greeley, J. and Sanchez-Castillo, M. A. and Evans, S. T. and Gokhale, A. A. and Dumesic, J. A. and Mavrikakis, Manos},
  title = {Prediction of experimental methanol decomposition rates on platinum from first principles},
  journal = {Top. Catal.},
  year = {2006},
  volume = {37},
  pages = {17},
  publisher = {Springer}
}

@article{chen2021,
author = {Chen, B. W. J. and Xu, L. and Mavrikakis, M.},
title = {Computational Methods in Heterogeneous Catalysis},
journal = {Chem. Rev.},
volume = {121},
number = {2},
pages = {1007},
year = {2021},
publisher = {American Chemical Society}
}

@article{chen2025,
  author = {Chen, B. W. J. and Mavrikakis, M.},
  title = {Modeling the impact of structure and coverage on the reactivity of realistic heterogeneous catalysts},
  journal = {Nat. Chem. Eng.},
  volume = {2},
  pages = {181},
  year = {2025},
  publisher = {Nature Publishing Group}
}

@article{norskov2002,
  author = {N{\o}rskov, J. K. and Bligaard, T. and Logadottir, A. and Bahn, S. and Hansen, L. B. and Bollinger, M. and Bengaard, H. and Hammer, B. and Sljivancanin, Z. and Mavrikakis, M. and Xu, Y. and Dahl, S. and Jacobsen, C. J. H.},
  title = {Universality in Heterogeneous Catalysis},
  journal = {J. Catal.},
  year = {2002},
  volume = {209},
  number = {2},
  pages = {275},
  publisher = {Elsevier}
}

@article{michaelides2003,
  author = {Michaelides, A. and Liu, Z.-P. and Zhang, C. J. and Alavi, A. and King, D. A. and Hu, P.},
  title = {Identification of General Linear Relationships between Activation Energies and Enthalpy Changes for Dissociation Reactions at Surfaces},
  journal = {J. Am. Chem. Soc.},
  year = {2003},
  volume = {125},
  number = {13},
  pages = {3704},
  publisher = {American Chemical Society}
}

@article{bligaard2004,
  author = {T. Bligaard and J. K. Nørskov and S. Dahl and J. Matthiesen and C. H. Christensen and J. Sehested},
  title = {The Brønsted–Evans–Polanyi relation and the volcano curve in heterogeneous catalysis},
  journal = {J. Catal.},
  volume = {224},
  number = {1},
  pages = {206},
  year = {2004},
  publisher = {Elservier}
}

@article{liu2011,
    author = {Liu, Bin and Greeley, Jeffrey},
    title = {Decomposition Pathways of Glycerol via C-H, O-H, and C-C Bond Scission on Pt(111): A Density Functional Theory Study},
    journal = {J. Phys. Chem. C},
    volume = {115},
    number = {40},
    pages = {19702},
    year = {2011},
    publisher = {American Chemical Society}
}

@article{bozal-ginesta2025,
  author = {Bozal-Ginesta, C. and Pablo-Garc{\'i}a, S. and Choi, C. and Taranc{\'o}n, A. and Aspuru-Guzik, A.},
  title = {Developing machine learning for heterogeneous catalysis with experimental and computational data},
  journal = {Nat. Rev. Chem.},
  volume = {9},
  pages = {601},
  year = {2025},
  publisher = {Nature Publishing Group}
}

@article{deringer2019,
author = {Deringer, V. L. and Caro, M. A. and Csányi, G.},
title = {Machine Learning Interatomic Potentials as Emerging Tools for Materials Science},
journal = {Adv. Mater.},
volume = {31},
number = {46},
pages = {1902765},
year = {2019},
publisher = {Wiley-VCH}
}

@article{ko2023,
  author = {Ko, T. W. and Ong, S. P.},
  title = {Recent advances and outstanding challenges for machine learning interatomic potentials},
  journal = {Nat. Comput. Sci.},
  volume = {3},
  pages = {998},
  year = {2023},
  publisher = {Nature Publishing Group}
}

@article{wang2024,
  author = {Wang, G. and Wang, C. and Zhang, X. and Li, Z. and Zhou, J. and Sun, Z.},
  title = {Machine learning interatomic potential: Bridge the gap between small-scale models and realistic device-scale simulations},
  journal = {iScience},
  volume = {27},
  number = {5},
  pages = {109673},
  year = {2024},
  publisher = {Elsevier}
}

@article{jacobs2025,
  author = {Ryan Jacobs and Dane Morgan and Siamak Attarian and Jun Meng and Chen Shen and Zhenghao Wu and Clare Yijia Xie and Julia H. Yang and Nongnuch Artrith and Ben Blaiszik and Gerbrand Ceder and Kamal Choudhary and Gabor Csanyi and Ekin Dogus Cubuk and Bowen Deng and Ralf Drautz and Xiang Fu and Jonathan Godwin and Vasant Honavar and Olexandr Isayev and Anders Johansson and Boris Kozinsky and Stefano Martiniani and Shyue Ping Ong and Igor Poltavsky and KJ Schmidt and So Takamoto and Aidan P. Thompson and Julia Westermayr and Brandon M. Wood},
  title = {A practical guide to machine learning interatomic potentials -- {Status} and future},
  journal = {Curr. Opin. Solid State Mater. Sci.},
  volume = {35},
  pages = {101214},
  year = {2025},
  publisher = {Elsevier}
}

@article{chen2023,
  author = {Chen, Dongxiao and Shang, Cheng and Liu, Zhi-Pan},
  title = {Machine-learning atomic simulation for heterogeneous catalysis},
  journal = {npj Comput. Mater.},
  volume = {9},
  pages = {2},
  year = {2023},
  publisher = {Nature Publishing Group}
}

@article{olajide2025,
  author = {Olajide, G. and Baral, K. and Ezendu, S. and Soyemi, A. and Szilv{\'a}si, T.},
  title = {Application of machine learning interatomic potentials in heterogeneous catalysis},
  journal = {J. Catal.},
  year = {2025},
  volume = {448},
  pages = {116202},
  publisher = {Elsevier}
}

@article{omranpour2025,
  author = {Omranpour, A. and Elsner, J. and Lausch, K. N. and Behler, J.},
  title = {Machine Learning Potentials for Heterogeneous Catalysis},
  journal = {ACS Catal.},
  year = {2025},
  volume = {15},
  number = {3},
  pages = {1616},
  publisher = {American Chemical Society}
}

@article{xie2026,
author = {Xie, W. and Han, Y. and Wu, C. and Hu, P.},
title = {Smarter Data: Rethinking
Data Generation for Machine
Learning Potentials in Heterogeneous Catalysis},
journal = {JACS Au},
volume = {6},
number = {6},
pages = {3081},
year = {2026},
publisher = {American Chemical Society}
}

@article{yuan2026,
  author = {Yuan, E. C.-Y. and Liu, Y. and Chen, J. and Zhong, P. and Raja, S. and Kreiman, T. and Vargas, S. and Xu, W. and Head-Gordon, M. and Yang, C. and Blau, S. M. and Cheng, B. and Krishnapriyan, A. and Head-Gordon, T.},
  title = {Foundation models for atomistic simulation of chemistry and materials},
  journal = {Nat. Rev. Chem.},
  year = {2026},
  volume = {10},
  pages = {212},
  publisher = {Nature Publishing Group}
}

@article{batatia2025,
author = {Batatia, Ilyes and Benner, Philipp and Chiang, Yuan and Elena, Alin M. and Kovács, Dávid P. and Riebesell, Janosh and Advincula, Xavier R. and Asta, Mark and Avaylon, Matthew and Baldwin, William J. and Berger, Fabian and Bernstein, Noam and Bhowmik, Arghya and Bigi, Filippo and Blau, Samuel M. and Cărare, Vlad and Ceriotti, Michele and Chong, Sanggyu and Darby, James P. and De, Sandip and Della Pia, Flaviano and Deringer, Volker L. and Elijošius, Rokas and El-Machachi, Zakariya and Fako, Edvin and Falcioni, Fabio and Ferrari, Andrea C. and Gardner, John L. A. and Gawkowski, Mikołaj J. and Genreith-Schriever, Annalena and George, Janine and Goodall, Rhys E. A. and Grandel, Jonas and Grey, Clare P. and Grigorev, Petr and Han, Shuang and Handley, Will and Heenen, Hendrik H. and Hermansson, Kersti and Ho, Cheuk Hin and Hofmann, Stephan and Holm, Christian and Jaafar, Jad and Jakob, Konstantin S. and Jung, Hyunwook and Kapil, Venkat and Kaplan, Aaron D. and Karimitari, Nima and Kermode, James R. and Kourtis, Panagiotis and Kroupa, Namu and Kullgren, Jolla and Kuner, Matthew C. and Kuryla, Domantas and Liepuoniute, Guoda and Lin, Chen and Margraf, Johannes T. and Magdău, Ioan-Bogdan and Michaelides, Angelos and Moore, J. Harry and Naik, Aakash A. and Niblett, Samuel P. and Norwood, Sam Walton and O’Neill, Niamh and Ortner, Christoph and Persson, Kristin A. and Reuter, Karsten and Rosen, Andrew S. and Rosset, Louise A. M. and Schaaf, Lars L. and Schran, Christoph and Shi, Benjamin X. and Sivonxay, Eric and Stenczel, Tamás K. and Sutton, Christopher and Svahn, Viktor and Swinburne, Thomas D. and Tilly, Jules and van der Oord, Cas and Vargas, Santiago and Varga-Umbrich, Eszter and Vegge, Tejs and Vondrák, Martin and Wang, Yangshuai and Witt, William C. and Wolf, Thomas and Zills, Fabian and Csányi, Gábor},
title = {A foundation model for atomistic materials chemistry},
journal = {J. Chem. Phys.},
volume = {163},
number = {18},
pages = {184110},
year = {2025}
}

@article{yang2025,
  author = {Yang, C. and Wu, C. and Xie, W. and Xie, D. and Hu, P.},
  title = {General reactive element-based machine learning potentials for heterogeneous catalysis},
  journal = {Nat. Catal.},
  volume = {8},
  pages = {891},
  year = {2025},
  publisher = {Nature Publishing Group}
}

@article{perego2024,
  author = {Perego, S. and Bonati, L.},
  title = {Data efficient machine learning potentials for modeling catalytic reactivity via active learning and enhanced sampling},
  journal = {npj Comput. Mater.},
  year = {2024},
  volume = {10},
  number = {1},
  pages = {291},
  publisher = {Nature Publishing Group}
}

@article{ma2026,
  author = {Ma, J. and Fu, X. and Xie, W. and Hu, P.},
  title = {From Pretrained to Precision: Fine-Tuning Universal Interatomic Potentials for Accurate Catalytic Reaction Simulations},
  journal = {J. Chem. Theory Comput.},
  year = {2026},
  volume = {22},
  number = {4},
  pages = {1920},
  publisher = {American Chemical Society}
}

@article{cheula2026,
  author = {Cheula, R. and Andersen, M. and Kitchin, J. R.},
  title = {Fine-tuning universal machine learning potentials for transition state search in surface catalysis},
  journal = {npj Comput. Mater.},
  year = {2026},
  note = {Early Release, DOI: https://doi.org/10.1038/s41524-026-02228-1}
}

@article{riebesell2025,
  author = {Riebesell, J. and Goodall, R. E. A. and Benner, P. and Chiang, Y. and Deng, B. and Ceder, G. and Asta, M. and Lee, A. A. and Jain, A. and Persson, K. A.},
  title = {A framework to evaluate machine learning crystal stability predictions},
  journal = {Nat. Mach. Intell.},
  year = {2025},
  volume = {7},
  pages = {836},
  publisher = {Nature Publishing Group}
}

@article{peng2026,
  author = {Peng, A. and Cai, C. and Guo, M. and Zhang, D. and Zhang, C. and Jiang, W. and Wang, Y. and Loew, A. and Wu, C. and E, W. and Zhang, L. and Wang, H.},
  title = {{LAMBench}: a benchmark for large atomistic models},
  journal = {npj Comput. Mater.},
  year = {2026},
  volume = {12},
  pages = {62},
  publisher = {Nature Publishing Group}
}

@article{wu2026,
  author = {Wu, C. and Yang, C. and Fang, Z. and Xie, W. and Xie, D. and Hu, P.},
  title = {Hitchhikers Guide To Training More General Machine Learning Potentials in Heterogeneous Catalysis},
  journal = {J. Chem. Theory Comput.},
  volume = {22},
  number = {4},
  pages = {1970},
  year = {2026},
  publisher = {American Chemical Society}
}

@article{moon2025,
  author = {Moon, J. and Jeon, U. and Choung, S. and Han, J. W.},
  title = {CatBench Framework for Benchmarking Machine Learning Interatomic Potentials in Adsorption Energy Predictions for Heterogeneous Catalysis},
  journal = {Cell Rep. Phys. Sci.},
  year = {2025},
  volume = {6},
  number = {12},
  pages = {102968},
  publisher = {Elsevier}
}

@article{kempen2026,
  author = {Kempen, L. H. E. and Cheula, R. and Andersen, M.},
  title = {How accurate are foundational machine learning interatomic potentials for heterogeneous catalysis?},
  journal = {J. Chem. Phys.},
  year = {2026},
  volume = {164},
  number = {19},
  pages = {194119},
  publisher = {AIP Publishing}
}

@article{loveday2026,
  author = {Loveday, O. and Ka{\'z}mierczak, K. and L{\'o}pez, N.},
  title = {Challenges and Opportunities of Pretrained Machine Learning Interatomic Potentials in Heterogeneous Catalysis},
  journal = {ACS Catal.},
  year = {2026},
  volume = {16},
  number = {5},
  pages = {4113},
  publisher = {American Chemical Society}
}

@book{balluffi2005,
  author = {Balluffi, R. W. and Allen, S. M. and Carter, W. C.},
  title = {Kinetics of Materials},
  publisher = {John Wiley \& Sons},
  year = {2005},
  isbn = {978-0-471-24689-3}
}

@article{tran2016,
  author = {Tran, R. and Xu, Z. and Radhakrishnan, B. and Winston, D. and Sun, W. and Persson, K. A. and Ong, S. P.},
  title = {Surface energies of elemental crystals},
  journal = {Sci. Data.},
  year = {2016},
  volume = {3},
  pages = {160080},
  publisher = {Nature Publishing Group}
}

@article{zhuang2016,
  author = {Zhuang, H. and Tkalych, A. J. and Carter, E. A.},
  title = {Surface Energy as a Descriptor of Catalytic Activity},
  journal = {J. Phys. Chem. C},
  year = {2016},
  volume = {120},
  number = {41},
  pages = {23698},
  publisher = {American Chemical Society}
}

@article{xu2023,
  author = {Xu, L. and Papanikolaou, K. G. and Lechner, B. A. J. and Je, L. and Somorjai, G. A. and Salmeron, M. and Mavrikakis, M.},
  title = {Formation of active sites on transition metals through reaction-driven migration of surface atoms},
  journal = {Science},
  year = {2023},
  volume = {380},
  number = {6640},
  pages = {70},
  publisher = {American Association for the Advancement of Science}
}

@article{shi2024,
  author = {Shi, B. X. and Wales, D. J. and Michaelides, A. and Myung, C. W.},
  title = {Going for Gold(-Standard): Attaining Coupled Cluster Accuracy in Oxide-Supported Nanoclusters},
  journal = {J. Chem. Theory Comput.},
  volume = {20},
  number = {12},
  pages = {5306},
  year = {2024},
  publisher = {American Chemical Society}
}

@article{morales_vidal2026,
  author = {Morales-Vidal, Jordi and Lian, Zan and Pinheiro Ara{\'u}jo, Thaylan and Mitchell, Sharon and P{\'e}rez-Ram{\'i}rez, Javier and L{\'o}pez, N{\'u}ria},
  title = {Phases and Architectures in Metal/Metal Oxide Systems Driven by Strong Metal--Support Interactions},
  journal = {ACS Phys. Chem. Au},
  year = {2026},
  volume = {6},
  number = {2},
  pages = {340-347},
}

@article{dai2018,
  author = {Dai, Y. and Lu, P. and Cao, Z. and Campbell, C. T. and Xia, Y.},
  title = {The physical chemistry and materials science behind sinter-resistant catalysts},
  journal = {Chem. Soc. Rev.},
  year = {2018},
  volume = {47},
  number = {12},
  pages = {4314},
  publisher = {Royal Society of Chemistry}
}

@article{astruc2020,
  author = {Astruc, D.},
  title = {Introduction: Nanoparticles in Catalysis},
  journal = {Chem. Rev.},
  year = {2020},
  volume = {120},
  number = {2},
  pages = {461},
  publisher = {American Chemical Society}
}

@article{sharada2019,
  author = {Sharada, S. M. and Karlsson, R. K. B. and Maimaiti, Y. and Voss, J. and Bligaard, T.},
  title = {Adsorption on transition metal surfaces: Transferability and accuracy of DFT using the ADS41 dataset},
  journal = {Phys. Rev. B},
  volume = {100},
  pages = {035439},
  year = {2019},
  publisher = {American Physical Society}
}

@article{trepte2022,
  author = {Trepte, K. and Voss, J.},
  title = {Data-driven and constrained optimization of semi-local exchange and nonlocal correlation functionals for materials and surface chemistry},
  journal = {J. Comput. Chem.},
  volume = {43},
  number = {16},
  pages = {1104},
  year = {2022},
  publisher = {Wiley}
}

@article{kleryoung,
  note = {In-house dataset. See section S1.2.1 of the Supporting Information (SI) for details.}
}

@article{janthon2017,
  author = {Janthon, P. and Vi{\~n}es, F. and Sirijaraensre, J. and Limtrakul, J. and Illas, F.},
  title = {Adding Pieces to the {CO/Pt(111)} Puzzle: The Role of Dispersion},
  journal = {J. Phys. Chem. C},
  year = {2017},
  volume = {121},
  number = {7},
  pages = {3970},
  publisher = {American Chemical Society}
}

@article{fanta2025,
  author = {Fanta, R. and Bajdich, M.},
  title = {Resolution of Selectivity Steps of {CO} Reduction Reaction on Copper by Quantum {Monte} {Carlo}},
  journal = {J. Phys. Chem. Lett.},
  year = {2025},
  volume = {16},
  number = {6},
  pages = {1494},
  publisher = {American Chemical Society}
}

@book{sabatier1920,
  author = {Sabatier, P.},
  title = {La Catalyse en Chimie Organique},
  publisher = {B{\'e}ranger},
  year = {1920}
}

@article{hammer2000,
  author = {Hammer, B. and N{\o}rskov, J. K.},
  title = {Theoretical surface science and catalysis---calculations and concepts},
  journal = {Adv. Catal.},
  year = {2000},
  volume = {45},
  pages = {71},
  publisher = {Elsevier}
}

@book{somorjai2010,
  author = {Somorjai, G. A. and Li, Y.},
  title = {Introduction to Surface Chemistry and Catalysis},
  publisher = {John Wiley \& Sons},
  year = {2010},
  isbn = {978-0-470-50823-7}
}

@article{feibelman2001,
  author = {Feibelman, P. J. and Hammer, B. and N{\o}rskov, J. K. and Wagner, F. and Scheffler, M. and Stumpf, R. and Watwe, R. and Dumesic, J.},
  title = {The {CO/Pt(111)} Puzzle},
  journal = {J. Phys. Chem. B},
  year = {2001},
  volume = {105},
  number = {18},
  pages = {4018},
  publisher = {American Chemical Society}
}

@article{campbell2026,
author = {C. T. Campbell},
title = {Calorimetric methods for measuring the energies of adsorbed catalytic reaction intermediates, of surface metal atoms in catalyst materials, and of surface reaction steps in other technologies},
journal = {Adv. Sci. Instrum.},
volume = {1},
number = {1},
pages = {100010},
year = {2026},
publisher = {Elsevier}
}

@article{tchakoua2023,
  author = {Tchakoua, T. and Gerrits, N. and Smeets, E. W. F. and Kroes, G.-J.},
  title = {{SBH17}: Benchmark Database of Barrier Heights for Dissociative Chemisorption on Transition Metal Surfaces},
  journal = {J. Chem. Theory Comput.},
  volume = {19},
  number = {1},
  pages = {245},
  year = {2023},
  publisher = {American Chemical Society}
}

@article{niu,
  note = {In-house dataset. See section S1.3.2 of the Supporting Information (SI) for details.}
}

@article{hwang,
  note = {In-house dataset. See section S1.3.3 of the Supporting Information (SI) for details.}
}

@article{barroso_luque2024,
  author = {L. Barroso-Luque and M. Shuaibi and X. Fu and B. M. Wood and M. Dzamba and M. Gao and A. Rizvi and M. Uyttendaele and C. L. Zitnick and Z. W. Ulissi},
  title = {Open Materials 2024 ({OMat24}) Inorganic Materials Dataset and Models},
  journal = {Nat. Comput. Sci.},
  volume = {6},
  pages = {642},
  year = {2026}
}

@article{deng2023,
  author = {Deng, B. and Zhong, P. and Jun, K. and Riebesell, J. and Han, K. and Bartel, C. J. and Ceder, G.},
  title = {{CHGNet} as a pretrained universal neural network potential for charge-informed atomistic modelling},
  journal = {Nat. Mach. Intell.},
  volume = {5},
  pages = {1031},
  year = {2023},
  publisher = {Nature Publishing Group}
}

@article{schmidt2021,
  author = {J. Schmidt  and L. Pettersson  and C. Verdozzi  and S. Botti  and M. A. L. Marques},
  title = {Crystal graph attention networks for the prediction of stable materials},
  journal = {Sci. Adv.},
  volume = {7},
  number = {49},
  pages = {eabi7948},
  year = {2021},
  publisher = {American Association for the Advancement of Science}
}

@article{wang2023,
  author = {H.-C. Wang and J. Schmidt and M. A. L. Marques and L. Wirtz and A. H. Romero},
  title = {Symmetry-based computational search for novel binary and ternary {2D} materials},
  journal = {2D Mater.},
  volume = {10},
  pages = {035007},
  year = {2023},
  publisher = {IOP Publishing}
}

@article{schmidt2023,
  author = {Schmidt, J. and Hoffmann, N. and Wang, H.-C. and Borlido, P. and Carri{\c{c}}o, P. J. M. A. and Cerqueira, T. F. T. and Botti, S. and Marques, M. A. L.},
  title = {Machine-Learning-Assisted Determination of the Global Zero-Temperature Phase Diagram of Materials},
  journal = {Adv. Mater.},
  volume = {35},
  number = {22},
  pages = {2210788},
  year = {2023},
  publisher = {Wiley}
}

@article{kaplan2025,
  author = {Kaplan, A. D. and Liu, R. and Qi, J. and Ko, T. W. and Deng, B. and Riebesell, J. and Ceder, G. and Persson, K. A. and Ong, S. P.},
  title = {A Foundational Potential Energy Surface Dataset for Materials},
  note = {arXiv:2503.04070},
  year = {2025}
}

@article{chanussot2021,
  author = {Chanussot, Lowik and Das, Abhishek and Goyal, Siddharth and Lavril, Thibaut and Shuaibi, Muhammed and Riviere, Morgane and Tran, Kevin and Heras-Domingo, Javier and Ho, Caleb and Hu, Weihua and Palizhati, Aini and Sriram, Anuroop and Wood, Brandon and Yoon, Junwoong and Parikh, Devi and Zitnick, C. Lawrence and Ulissi, Zachary},
  title = {Open Catalyst 2020 ({OC20}) Dataset and Community Challenges},
  journal = {ACS Catal.},
  volume = {11},
  number = {10},
  pages = {6059},
  year = {2021},
  publisher = {American Chemical Society}
}

@article{allam2026,
author = {Allam, Omar and Wander, Brook and Kim, SungYeon and Plesch, Rudi and Sours, Tyler and Chu, Jia-Min and Ludwig, Thomas and Kim, Jiyoon and Wang, Rodrigo and Agarwal, Shivang and Rask, Alan and Fleury, Alexandre and Wang, Chuhong and Wildman, Andrew and Mustard, Thomas and Ryczko, Kevin and Abruzzo, Paul and Nish, A. J. and Singh, Aayush R.},
title = {{AQCat25}: unlocking spin-aware, high-fidelity machine learning potentials for heterogeneous catalysis},
journal = {npj Comput. Mater.},
volume = {12},
pages = {226},
year = {2026},
publisher = {Nature Publishing Group}
}

@article{perdew1996,
  author = {Perdew, J. P. and Burke, K. and Ernzerhof, M.},
  title = {Generalized Gradient Approximation Made Simple},
  journal = {Phys. Rev. Lett.},
  volume = {77},
  number = {18},
  pages = {3865},
  year = {1996},
  publisher = {American Physical Society}
}

@article{gerrits2020,
author = {Gerrits, N. and Smeets, E. W. F. and Vuckovic, S. and Powell, A. D. and Doblhoff-Dier, K. and Kroes, G.-J.},
title = {Density Functional Theory for Molecule–Metal
Surface Reactions: When Does the Generalized Gradient Approximation Get It Right, and What to Do If It Does Not},
journal = {J. Phys. Chem. Lett.},
volume = {11},
number = {24},
pages = {10552},
year = {2020},
publisher = {American Chemical Society}
}

@article{araujo2022,
author = {Araujo, R. B. and Rodrigues, G. L. S. and dos Santos, E. C. and Pettersson, L. G. M.},
title = {Adsorption energies on transition metal surfaces: towards an accurate and balanced description},
journal = {Nat. Commun.},
volume = {13},
pages = {6853},
year = {2022},
publisher = {Nature Publishing Group}
}

@article{grimme2010,
  author = {Grimme, Stefan and Antony, Jens and Ehrlich, Stephan and Krieg, Helge},
  title = {A consistent and accurate ab initio parametrization of density functional dispersion correction ({DFT-D}) for the 94 elements {H-Pu}},
  journal = {J. Chem. Phys.},
  volume = {132},
  number = {15},
  pages = {154104},
  year = {2010},
  publisher = {American Institute of Physics}
}

@article{mazitov2025,
  author = {Mazitov, A. and Bigi, F. and Kellner, M. and Pegolo, P. and Tisi, D. and Fraux, G. and Pozdnyakov, S. and Loche, P. and Ceriotti, M.},
  title = {{PET-MAD} as a lightweight universal interatomic potential for advanced materials modeling},
  journal = {Nat. Commun.},
  volume = {16},
  pages = {10653},
  year = {2025},
  publisher = {Nature Publishing Group}
}

@article{yang2024,
  author = {Han Yang and Chenxi Hu and Yichi Zhou and Xixian Liu and Yu Shi and Jielan Li and Guanzhi Li and Zekun Chen and Shuizhou Chen and Claudio Zeni and Matthew Horton and Robert Pinsler and Andrew Fowler and Daniel Zügner and Tian Xie and Jake Smith and Lixin Sun and Qian Wang and Lingyu Kong and Chang Liu and Hongxia Hao and Ziheng Lu},
  title = {MatterSim: A Deep Learning Atomistic Model Across Elements, Temperatures and Pressures},
  note = {arXiv:2405.04967},
  year = {2024}
}

@article{lysogorskiy2025,
  author = {Lysogorskiy, Y. and Bochkarev, A. and Drautz, R.},
  title = {Graph atomic cluster expansion for foundational machine learning interatomic potentials},
  journal = {npj Comput. Mater.},
  volume = {12},
  pages = {114},
  year = {2026}
}

@article{fu2025,
  author = {Xiang Fu and Brandon M. Wood and Luis Barroso-Luque and Daniel S. Levine and Meng Gao and Misko Dzamba and C. Lawrence Zitnick},
  title = {Learning Smooth and Expressive Interatomic Potentials for Physical Property Prediction},
  note = {arXiv:2502.12147},
  year = {2025}
}

@article{kim2025,
  author = {Jaesun Kim and Jisu Kim and Jaehoon Kim and Jiho Lee and  Yutack Park and  Youngho Kang and  Seungwu Han},
  title = {Data-Efficient Multifidelity Training for High-Fidelity Machine Learning Interatomic Potentials},
  journal = {J. Am. Chem. Soc.},
  volume = {147},
  number = {1},
  pages = {1042},
  year = {2025},
  publisher = {American Chemical Society}
}

@article{rhodes2025,
  author = {Benjamin Rhodes and Sander Vandenhaute and Vaidotas Šimkus and James Gin and Jonathan Godwin and Tim Duignan and Mark Neumann},
  title = {Orb-v3: atomistic simulation at scale},
  note = {arXiv:2504.06231},
  year = {2025}
}

@article{wood2025,
  author = {Brandon M. Wood and Misko Dzamba and Xiang Fu and Meng Gao and Muhammed Shuaibi and Luis Barroso-Luque and Kareem Abdelmaqsoud and Vahe Gharakhanyan and John R. Kitchin and Daniel S. Levine and Kyle Michel and Anuroop Sriram and Taco Cohen and Abhishek Das and Ammar Rizvi and Sushree Jagriti Sahoo and Zachary W. Ulissi and C. Lawrence Zitnick},
  title = {UMA: A Family of Universal Models for Atoms},
  note = {arXiv:2506.23971},
  year = {2025}
}

@article{kavanagh2026,
    author = {Seán R. Kavanagh and Chuin Wei Tan and Menghang Wang and Marc L. Descoteaux and Gabriel de Miranda Nascimento and Ulrik Unneberg and Laura Zichi and Francesco Libbi and Norma Rivano and Austin Glover and Vivek Bharadwaj and Anders Johansson and William C. Witt and Albert Musaelian and Boris Kozinsky},
    title = {Fast and Accurate Foundation Models for Equivariant Machine-Learned Interatomic Potentials},
    note = {arXiv:2607.28461},
    year = {2026}
}

@article{batatia2025mh1,
  author = {Batatia, Ilyes and Lin, Chen and Hart, Joseph and Kasoar, Elliott and Elena, Alin M. and Norwood, Sam Walton and Wolf, Thomas and Cs{\'a}nyi, G{\'a}bor},
  title = {Cross Learning between Electronic Structure Theories for Unifying Molecular, Surface, and Inorganic Crystal Foundation Force Fields},
  note = {arXiv:2510.25380},
  year = {2025}
}

@article{xu2025,
  author = {Zemin Xu and Wenbo Xie and P. Hu},
  title = {Spectral/Spatial Tensor Atomic Cluster Expansion with Universal Embeddings in Cartesian Space},
  note = {arXiv:2509.14961},
  year = {2025}
}

@article{bigi2026,
  author = {Bigi, Filippo and Pegolo, Paolo and Mazitov, Arslan and Ceriotti, Michele},
  title = {Pushing the limits of unconstrained machine-learned interatomic potentials},
  note = {arXiv:2601.16195},
  year = {2026}
}

@inproceedings{batatia2022,
  title={{MACE}: Higher Order Equivariant Message Passing Neural Networks for Fast and Accurate Force Fields},
  author={Ilyes Batatia and David Peter Kovacs and Gregor N. C. Simm and Christoph Ortner and Gabor Csanyi},
  booktitle={Advances in Neural Information Processing Systems},
  editor={Alice H. Oh and Alekh Agarwal and Danielle Belgrave and Kyunghyun Cho},
  year={2022},
  url={https://openreview.net/forum?id=YPpSngE-ZU}
}

@article{focassio2025,
  author = {Focassio, B. and Freitas, L. P. M. and Schleder, G. R.},
  title = {Performance Assessment of Universal Machine Learning Interatomic Potentials: Challenges and Directions for Materials' Surfaces},
  journal = {ACS Appl. Mater. Interfaces},
  volume = {17},
  number = {9},
  pages = {13111},
  year = {2025},
  publisher = {American Chemical Society}
}

@article{yang2025_2,
  author = {Yao Yang and Julian Feijóo and Marc Figueras-Valls and Chubai Chen and Chuqiao Shi and Maria V. Fonseca Guzman and Yves Murhabazi Maombi and Shikai Liu and Pulkit Jain and Valentín Briega-Martos and Zhengxing Peng and Yu Shan and Geonhui Lee and Michael Rebarchik and Lang Xu and Christopher J. Pollock and Jianbo Jin and Nathan E. Soland and Cheng Wang and Miquel B. Salmeron and Zhu Chen and Yimo Han and Manos Mavrikakis and Peidong Yang},
  title = {Operando probing dynamic migration of copper carbonyl during electrocatalytic CO$_2$ reduction},
  journal = {Nat. Catal.},
  volume = {8},
  pages = {579},
  year = {2025},
  publisher = {Nature Publishing Group}
}

@article{elnabawy2025,
  author = {Elnabawy, A. O. and Mavrikakis, M.},
  title = {Nanocluster Active Sites Formed on Heterogeneous Thermal Catalysts and Electrocatalysts by Operando Reactive Environments},
  journal = {ACS Catal.},
  volume = {15},
  number = {11},
  pages = {9919},
  year = {2025},
  publisher = {American Chemical Society}
}

@article{xu2025_2,
  author = {Xu, L. and Rebarchik, M. and Bhandari, S. and Mavrikakis, M.},
  title = {Adsorbate-induced adatom formation on Au-Cu bimetallic alloys and its possible consequences for CO$_2$ electroreduction},
  journal = {Surf. Sci.},
  volume = {751},
  pages = {122613},
  year = {2025},
  publisher = {Elsevier}
}

@article{greiner2018free,
  title={Free-atom-like d states in single-atom alloy catalysts},
  author={Greiner, Mark T and Jones, TE and Beeg, Sebastian and Zwiener, Leon and Scherzer, Michael and Girgsdies, Frank and Piccinin, S and Armbr{\"u}ster, Marc and Knop-Gericke, Axel and Schl{\"o}gl, Robert},
  journal={Nat. Chem.},
  volume={10},
  number={10},
  pages={1008},
  year={2018},
  publisher={Nature Publishing Group UK London}
}

@article{berger2025dopant,
  title={When Are Dopant d-States Free-Atom-Like? Periodic Trends and Confinement Effects in Single-Atom Alloys},
  author={Berger, Fabian and Michaelides, Angelos},
  journal={J. Am. Chem. Soc.},
  volume={147},
  number={41},
  pages={37079},
  year={2025},
  publisher={ACS Publications}
}

@article{ruban1999surface,
  title={Surface segregation energies in transition-metal alloys},
  author={Ruban, AV and Skriver, Hans Lomholt and N{\o}rskov, Jens Kehlet},
  journal={Phys. Rev. B},
  volume={59},
  number={24},
  pages={15990},
  year={1999},
  publisher={APS}
}

@article{wang2020surface,
  title={Surface facet dependence of competing alloying mechanisms},
  author={Wang, Yicheng and Papanikolaou, Konstantinos G and Hannagan, Ryan T and Patel, Dipna A and Balema, Tedros A and Cramer, Laura A and Kress, Paul L and Stamatakis, Michail and Sykes, E Charles H},
  journal={J. Chem. Phys.},
  volume={153},
  pages={244702},
  year={2020},
  publisher={AIP Publishing}
}

@article{karageorgiou2026mechanisms,
  title={Mechanisms for the formation of active sites in single-atom alloys},
  author={Karageorgiou, Ioannis and Michaelides, Angelos and Berger, Fabian},
  journal={Nanoscale},
  volume={18},
  pages={9709},
  year={2026},
  publisher={Royal Society of Chemistry}
}

@article{hannagan2020single,
  title={Single-atom alloy catalysis},
  author={Hannagan, Ryan T and Giannakakis, Georgios and Flytzani-Stephanopoulos, Maria and Sykes, E Charles H},
  journal={Chem. Rev.},
  volume={120},
  number={21},
  pages={12044},
  year={2020},
  publisher={ACS Publications}
}

@article{berger2024bringing,
  title={Bringing molecules together: Synergistic coadsorption at dopant sites of single atom alloys},
  author={Berger, Fabian and Schumann, Julia and R{\'e}ocreux, Romain and Stamatakis, Michail and Michaelides, Angelos},
  journal={J. Am. Chem. Soc.},
  volume={146},
  number={41},
  pages={28119},
  year={2024},
  publisher={ACS Publications}
}

@article{berger2026atlas,
  title={An Atlas and Design Rules for Single-and Dual-Atom Alloys},
  author={Berger, Fabian and Wang, Yicheng and Sykes, E Charles H and Michaelides, Angelos},
  note={arXiv:2609.19087},
  year={2026}
}

@article{mamun2019,
author = {Osman Mamun and Kirsten T. Winther and Jacob R. Boes and Thomas Bligaard},
title = {High-throughput calculations of catalytic properties of bimetallic alloy surfaces},
journal = {Sci. Data},
volume = {6},
pages = {76},
year = {2019}
}

@article{warford2026,
  author = {Warford, T. and Thiemann, F. L. and Cs{\'a}nyi, G.},
  title = {Better without {U}: impact of selective {Hubbard} {U} correction on foundational {MLIPs}},
  journal = {Mach. Learn.: Sci. Technol.},
  volume = {7},
  pages = {035033},
  year = {2026},
  publisher = {IOP Publishing}
}

@article{antczak2007,
  author = {Antczak, Grazyna and Ehrlich, Gert},
  title = {Jump processes in surface diffusion},
  journal = {Surf. Sci. Rep.},
  volume = {62},
  number = {2},
  pages = {39},
  year = {2007},
  publisher = {Elsevier}
}

@article{larsen2017,
  author = {Larsen, Ask Hjorth and Mortensen, Jens J{\o}rgen and Blomqvist, Jakob and Castelli, Ivano E. and Christensen, Rune and Du{\l}ak, Marcin and Friis, Jesper and Groves, Michael N. and Hammer, Bj{\o}rk and Hargus, Cory and Hermes, Eric D. and Jennings, Paul C. and Jensen, Peter Bjerre and Kermode, James and Kitchin, John R. and Kolsbjerg, Esben Leonhard and Kubal, Joseph and Kaasbjerg, Kirsten and Lysgaard, Steen and Maronsson, J{\'o}n Bergmann and Maxson, Tristan and Olsen, Thomas and Pastewka, Lars and Peterson, Andrew and Rostgaard, Carsten and Schi{\o}tz, Jakob and Sch{\"u}tt, Ole and Strange, Mikkel and Thygesen, Kristian S. and Vegge, Tejs and Vilhelmsen, Lasse and Walter, Michael and Zeng, Zhenhua and Jacobsen, Karsten W.},
  title = {The atomic simulation environment---a {Python} library for working with atoms},
  journal = {J. Phys.: Condens. Matter},
  volume = {29},
  number = {27},
  pages = {273002},
  year = {2017},
  publisher = {IOP Publishing}
}

@article{kresse1996efficient,
  title={Efficient iterative schemes for ab initio total-energy calculations using a plane-wave basis set},
  author={Kresse, Georg and Furthm{\"u}ller, J{\"u}rgen},
  journal={Phys. Rev. B},
  volume={54},
  pages={11169},
  year={1996},
  publisher={APS}
}

@article{kresse1999ultrasoft,
  title={From ultrasoft pseudopotentials to the projector augmented-wave method},
  author={Kresse, Georg and Joubert, Daniel},
  journal={Phys. Rev. B},
  volume={59},
  pages={1758},
  year={1999},
  publisher={APS}
}

@article{kresse1996efficiency,
  title={Efficiency of ab-initio total energy calculations for metals and semiconductors using a plane-wave basis set},
  author={Kresse, Georg and Furthm{\"u}ller, J{\"u}rgen},
  journal={Comput. Mater. Sci.},
  volume={6},
  number={1},
  pages={15},
  year={1996},
  publisher={Elsevier}
}

@article{kresse1993ab,
  title={Ab initio molecular dynamics for liquid metals},
  author={Kresse, Georg and Hafner, J{\"u}rgen},
  journal={Phys. Rev. B},
  volume={47},
  pages={558},
  year={1993},
  publisher={APS}
}

@article{kresse1994ab,
  title={Ab initio molecular-dynamics simulation of the liquid-metal--amorphous-semiconductor transition in germanium},
  author={Kresse, Georg and Hafner, J{\"u}rgen},
  journal={Phys. Rev. B},
  volume={49},
  pages={14251},
  year={1994},
  publisher={APS}
}

@article{kresse1994norm,
  title={Norm-conserving and ultrasoft pseudopotentials for first-row and transition elements},
  author={Kresse, Georg and Hafner, Jurgen},
  journal={J. Phys.: Condens. Matter},
  volume={6},
  number={40},
  pages={8245},
  year={1994},
  publisher={IOP Publishing}
}

@article{chen2025_atomistic,
author = {Chen, Dongxiao and Sautet, Philippe},
title = {Atomistic Landscape of Pt Nanoparticles via Machine Learning: How Size Effect and Hydrogen Adsorption Govern Structural Ensembles and Catalytic Activity},
journal = {Angew. Chem. Int. Ed.},
volume = {65},
number = {2},
pages = {e19209},
year = {2026}
}

@article{zhang2020ensembles,
  title={Ensembles of metastable states govern heterogeneous catalysis on dynamic interfaces},
  author={Zhang, Zisheng and Zandkarimi, Borna and Alexandrova, Anastassia N},
  journal={Acc. Chem. Res.},
  volume={53},
  number={2},
  pages={447},
  year={2020},
  publisher={ACS Publications}
}

@article{poths2024thermodynamic,
  title={Thermodynamic equilibrium versus kinetic trapping: thermalization of cluster catalyst ensembles can extend beyond reaction time scales},
  author={Poths, Patricia and Vargas, Santiago and Sautet, Philippe and Alexandrova, Anastassia N},
  journal={ACS Catal.},
  volume={14},
  number={7},
  pages={5403},
  year={2024},
  publisher={ACS Publications}
}

@article{barcaro2005,
  author = {Barcaro, Giovanni and Fortunelli, Alessandro},
  title = {The Interaction of Coinage Metal Clusters with the {MgO}(100) Surface},
  journal = {J. Chem. Theory Comput.},
  year = {2005},
  volume = {1},
  number = {5},
  pages = {972}
}

@article{luo2016,
  author = {Luo, Zhixun and Castleman, Jr., A. W. and Khanna, Shiv N.},
  title = {Reactivity of Metal Clusters},
  journal = {Chem. Rev.},
  year = {2016},
  volume = {116},
  number = {23},
  pages = {14456},
}

@article{kim2026,
  title = {Optimizing cross-domain transfer for universal machine learning interatomic potentials},
  author = {Kim, Jaesun and You, Jinmu and Park, Yutack and Lim, Yunsung and Kang, Yujin and Kim, Jisu and Jeon, Haekwan and Ju, Suyeon and Hong, Deokgi and Lee, Seung Yul and Choi, Saerom and Kim, Yongdeok and Lee, Jae W. and Han, Seungwu},
  journal = {Nature Communications},
  volume = {17},
  pages = {3432},
  year = {2026},
}

@article{xu2026,
  title = {Edge Cluster Expansion with Radial Rotary Attention for Interatomic Potentials},
  author = {Xu, Zemin and Xie, Wenbo and Hu, P.},
  note = {arXiv:2607.10664},
  year = {2026},
}

@inproceedings{lee2025,
  title = {{FlashTP}: Fused, Sparsity-Aware Tensor Product for Machine Learning Interatomic Potentials},
  author = {Lee, Seung Yul and Kim, Hojoon and Park, Yutack and Jeong, Dawoon and Han, Seungwu and Park, Yeonhong and Lee, Jae W.},
  booktitle = {Proceedings of the 42nd International Conference on Machine Learning},
  series = {Proceedings of Machine Learning Research},
  volume = {267},
  pages = {33143},
  year = {2025},
}

@article{bussi2007,
  author = {Bussi, Giovanni and Donadio, Davide and Parrinello, Michele},
  title = {Canonical sampling through velocity rescaling},
  journal = {J. Chem. Phys.},
  volume = {126},
  pages = {014101},
  year = {2007},
}

@article{plumed2019,
  author = {{The PLUMED consortium}},
  title = {Promoting transparency and reproducibility in enhanced molecular simulations},
  journal = {Nat. Methods},
  volume = {16},
  pages = {670},
  year = {2019},
}

@article{stecher2014,
  author = {Stecher, Thomas and Bernstein, Noam and Cs{\'a}nyi, G{\'a}bor},
  title = {Free Energy Surface Reconstruction from Umbrella Samples Using Gaussian Process Regression},
  journal = {J. Chem. Theory Comput.},
  volume = {10},
  number = {9},
  pages = {4079},
  year    = {2014},
}

@article{wei2025,
  author = {Wei, Zhiyuan and He, Jin-En and Li, Siwu and Chen, Zhe-Ning and Zhang, Lu and Chen, Jun},
  title = {Resolving the ``{CO} puzzle'': Disentangling electronic structure and dynamic effects via an operando dynamics framework},
  journal = {J. Chem. Phys.},
  volume = {163},
  pages = {234705},
  year = {2025},
}

@article{berger2021,
  author = {Berger, Fabian and Rybicki, Marcin and Sauer, Joachim},
  title = {Adsorption and cracking of propane by zeolites of different pore size},
  journal = {Journal of Catalysis},
  year = {2021},
  volume = {395},
  pages = {117},
}

@article{berger2023,
  author = {Berger, Fabian and Rybicki, Marcin and Sauer, Joachim},
  title = {Molecular Dynamics with Chemical Accuracy-Alkane Adsorption in Acidic Zeolites},
  journal = {ACS Catalysis},
  year = {2023},
  volume = {13},
  pages = {2011},
}

\end{document}

\clearpage

\tableofcontents

\clearpage

\section{Benchmark datasets details}

\subsection{Catalyst surfaces}

The first category of benchmark tests within \textit{HetCat26} pertains to catalyst surfaces.
Foundation machine learning interatomic potentials (fMLIPs) are evaluated in their prediction of surface energies of transition metals, adatom--vacancy formation energies, segregation and aggregation energies and metal--metal oxide support interactions.

\subsubsection{Surface energies}

Surface energies are critical thermodynamic quantities that govern the equilibrium shape of crystalline materials \cite{balluffi2005}, and contribute to the reactivity observed at surfaces \cite{zhuang2016}. 
This has motivated the creation of a database of surface energies for elemental crystals \cite{tran2016} using density functional theory (DFT) calculations employing the Perdew--Burke--Ernzerhof (PBE) functional\cite{perdew1996} (full computational settings are available in the publication introducing the database). 
In that work, well-converged surface energies were obtained by performing the bulk and slab calculations within the Oriented Unit Cell (OUC) framework, thus ensuring consistent reciprocal-space integration grids between the two systems.
In this work, for the bulk structure, the conventional unit cell (CUC) was instead employed, since fMLIP inference does not require the use of such an integration grid.
For a surface $(hkl)$, the surface energy, $\gamma_{hkl}$, is obtained by subtracting the bulk energy per atom, $\epsilon_{\mathrm{bulk}}$, times the number of slab atoms, $n_{\mathrm{slab}}$, from the total slab energy, $E_{hkl}^{\mathrm{slab}}$, normalizing by the slab area, $A_{\mathrm{slab}}$:

\begin{equation}
\gamma_{hkl} = \frac{E_{hkl}^{\mathrm{slab}} - \epsilon_{\mathrm{bulk}} n_{\mathrm{slab}}}{2 A_{\mathrm{slab}}}
\label{eq:surface_energy}
\end{equation}

\noindent{Given the focus of our work on heterogeneous catalysis, a subset of surfaces were chosen from the full dataset: surfaces associated with the most stable polymorph of each transition metal element were selected and procured via the Crystalium app\cite{tran2016}. 
Table \ref{tab:surfaces} summarizes the surfaces considered here.}

As noted by Focassio \textit{et al.} in their investigation of early fMLIPs for surface energy prediction\cite{focassio2025}, errors in surface energies are expected to arise primarily from the description of surface structures rather than bulk structures; fMLIPs are trained on bulk data, including MPtrj\cite{deng2023} (such that very high accuracy is expected for the bulk structures).\\\\

\begin{table}[ht!]
\centering
\caption{Transition metal surfaces included within \textit{HetCat26}. 
An asterisk is used to indicate that a surface is reconstructed. 
In total, 321 surfaces of varying complexity are included.}
\label{tab:surfaces}
\begin{tabularx}{\textwidth}{l}
\toprule
\multicolumn{1}{c}{\textbf{Metal surfaces}} \\
\midrule
\footnotesize{Ag(mp-124) $(100)$, $(110)$, $(111)$, $(210)$, $(211)$, $(221)$, $(310)$, $(311)$, $(320)$, $(321)$, $(322)$, $(331)$, $(332)$, *$(110)$} \\
\footnotesize{Au(mp-81) $(100)$, $(110)$, $(111)$, $(210)$, $(211)$, $(221)$, $(310)$, $(311)$, $(320)$, $(321)$, $(322)$, $(331)$, $(332)$, *$(110)$} \\
\footnotesize{Co(mp-54) $(0001)$, $(2\bar{1}\bar{1}2)$, $(10\bar{1}0)$, $(10\bar{1}1)$, $(10\bar{1}2)$,
$(11\bar{2}0)$,
$(11\bar{2}1)$, $(20\bar{2}1)$, $(21\bar{3}0)$, $(21\bar{3}1)$, $(21\bar{3}2)$,  $(22\bar{4}1)$} \\
\footnotesize{Cr(mp-90) $(100)$, $(110)$, $(111)$, $(210)$, $(211)$, $(221)$, $(310)$, $(311)$, $(320)$, $(321)$, $(322)$, $(331)$, $(332)$} \\
\footnotesize{Cu(mp-30) $(100)$, $(110)$, $(111)$, $(210)$, $(211)$, $(221)$, $(310)$, $(311)$, $(320)$, $(321)$, $(322)$, $(331)$, $(332)$} \\
\footnotesize{Fe(mp-13) $(100)$, $(110)$, $(111)$, $(210)$, $(211)$, $(221)$, $(310)$, $(311)$, $(320)$, $(321)$, $(322)$, $(331)$, $(332)$} \\
\footnotesize{Hf(mp-103) $(0001)$, $(2\bar{1}\bar{1}2)$, $(10\bar{1}0)$, $(10\bar{1}1)$, $(10\bar{1}2)$,
$(11\bar{2}0)$,
$(11\bar{2}1)$, $(20\bar{2}1)$, $(21\bar{3}0)$, $(21\bar{3}1)$, $(21\bar{3}2)$,  $(22\bar{4}1)$} \\
\footnotesize{Ir(mp-101) $(100)$, $(110)$, $(111)$, $(210)$, $(211)$, $(221)$, $(310)$, $(311)$, $(320)$, $(321)$, $(322)$, $(331)$, $(332)$, *$(110)$} \\
\footnotesize{Mn(mp-35) $(001)$, $(110)$, $(111)$} \\
\footnotesize{Mo(mp-129) $(100)$, $(110)$, $(111)$, $(210)$, $(211)$, $(221)$, $(310)$, $(311)$, $(320)$, $(321)$, $(322)$, $(331)$, $(332)$} \\
\footnotesize{Nb(mp-75) $(100)$, $(110)$, $(111)$, $(210)$, $(211)$, $(221)$, $(310)$, $(311)$, $(320)$, $(321)$, $(322)$, $(331)$, $(332)$} \\
\footnotesize{Ni(mp-23) $(100)$, $(110)$, $(111)$, $(210)$, $(211)$, $(221)$, $(310)$, $(311)$, $(320)$, $(321)$, $(322)$, $(331)$, $(332)$} \\
\footnotesize{Os(mp-49) $(0001)$, $(2\bar{1}\bar{1}2)$, $(10\bar{1}0)$, $(10\bar{1}1)$, $(10\bar{1}2)$,
$(11\bar{2}0)$,
$(11\bar{2}1)$, $(20\bar{2}1)$, $(21\bar{3}0)$, $(21\bar{3}1)$, $(21\bar{3}2)$,  $(22\bar{4}1)$} \\
\footnotesize{Pd(mp-2) $(100)$, $(110)$, $(111)$, $(210)$, $(211)$, $(221)$, $(310)$, $(311)$, $(320)$, $(321)$, $(322)$, $(331)$, $(332)$} \\
\footnotesize{Pt(mp-126) $(100)$, $(110)$, $(111)$, $(210)$, $(211)$, $(221)$, $(310)$, $(311)$, $(320)$, $(321)$, $(322)$, $(331)$, $(332)$, *$(110)$} \\
\footnotesize{Re(mp-8) $(0001)$, $(2\bar{1}\bar{1}2)$, $(10\bar{1}0)$, $(10\bar{1}1)$, $(10\bar{1}2)$,
$(11\bar{2}0)$,
$(11\bar{2}1)$, $(20\bar{2}1)$, $(21\bar{3}0)$, $(21\bar{3}1)$, $(21\bar{3}2)$,  $(22\bar{4}1)$} \\
\footnotesize{Rh(mp-74) $(100)$, $(110)$, $(111)$, $(210)$, $(211)$, $(221)$, $(310)$, $(311)$, $(320)$, $(321)$, $(322)$, $(331)$, $(332)$} \\
\footnotesize{Ru(mp-33) $(0001)$, $(2\bar{1}\bar{1}2)$, $(10\bar{1}0)$, $(10\bar{1}1)$, $(10\bar{1}2)$,
$(11\bar{2}0)$,
$(11\bar{2}1)$, $(20\bar{2}1)$, $(21\bar{3}0)$, $(21\bar{3}1)$, $(21\bar{3}2)$,  $(22\bar{4}1)$} \\
\footnotesize{Sc(mp-67) $(0001)$, $(2\bar{1}\bar{1}2)$, $(10\bar{1}0)$, $(10\bar{1}1)$, $(10\bar{1}2)$,
$(11\bar{2}0)$,
$(11\bar{2}1)$, $(20\bar{2}1)$, $(21\bar{3}0)$, $(21\bar{3}1)$, $(21\bar{3}2)$,  $(22\bar{4}1)$} \\
\footnotesize{Ta(mp-50) $(100)$, $(110)$, $(111)$, $(210)$, $(211)$, $(221)$, $(310)$, $(311)$, $(320)$, $(321)$, $(322)$, $(331)$, $(332)$} \\
\footnotesize{Tc(mp-113) $(0001)$, $(2\bar{1}\bar{1}2)$, $(10\bar{1}0)$, $(10\bar{1}1)$, $(10\bar{1}2)$,
$(11\bar{2}0)$,
$(11\bar{2}1)$, $(20\bar{2}1)$, $(21\bar{3}0)$, $(21\bar{3}1)$, $(21\bar{3}2)$,  $(22\bar{4}1)$} \\
\footnotesize{Ti(mp-72) $(0001)$, $(2\bar{1}\bar{1}2)$, $(10\bar{1}0)$, $(10\bar{1}1)$, $(10\bar{1}2)$,
$(11\bar{2}0)$,
$(11\bar{2}1)$, $(20\bar{2}1)$, $(21\bar{3}0)$, $(21\bar{3}1)$, $(21\bar{3}2)$,  $(22\bar{4}1)$} \\
\footnotesize{V(mp-146) $(100)$, $(110)$, $(111)$, $(210)$, $(211)$, $(221)$, $(310)$, $(320)$, $(321)$, $(322)$, $(331)$, $(332)$} \\
\footnotesize{W(mp-91) $(100)$, $(110)$, $(111)$, $(210)$, $(211)$, $(221)$, $(310)$, $(311)$, $(320)$, $(321)$, $(322)$, $(331)$, $(332)$} \\
\footnotesize{Y(mp-112) $(0001)$, $(2\bar{1}\bar{1}2)$, $(10\bar{1}0)$, $(10\bar{1}1)$, $(10\bar{1}2)$,
$(11\bar{2}0)$,
$(11\bar{2}1)$, $(20\bar{2}1)$, $(21\bar{3}0)$, $(21\bar{3}1)$, $(21\bar{3}2)$,  $(22\bar{4}1)$} \\
\footnotesize{Zr(mp-131) $(0001)$, $(2\bar{1}\bar{1}2)$, $(10\bar{1}0)$, $(10\bar{1}1)$, $(10\bar{1}2)$,
$(11\bar{2}0)$,
$(11\bar{2}1)$, $(20\bar{2}1)$, $(21\bar{3}0)$, $(21\bar{3}1)$, $(21\bar{3}2)$,  $(22\bar{4}1)$} \\
\bottomrule
\end{tabularx}
\end{table}

\begin{figure}[H]
    \centering
    \includegraphics[width=0.61\textwidth]{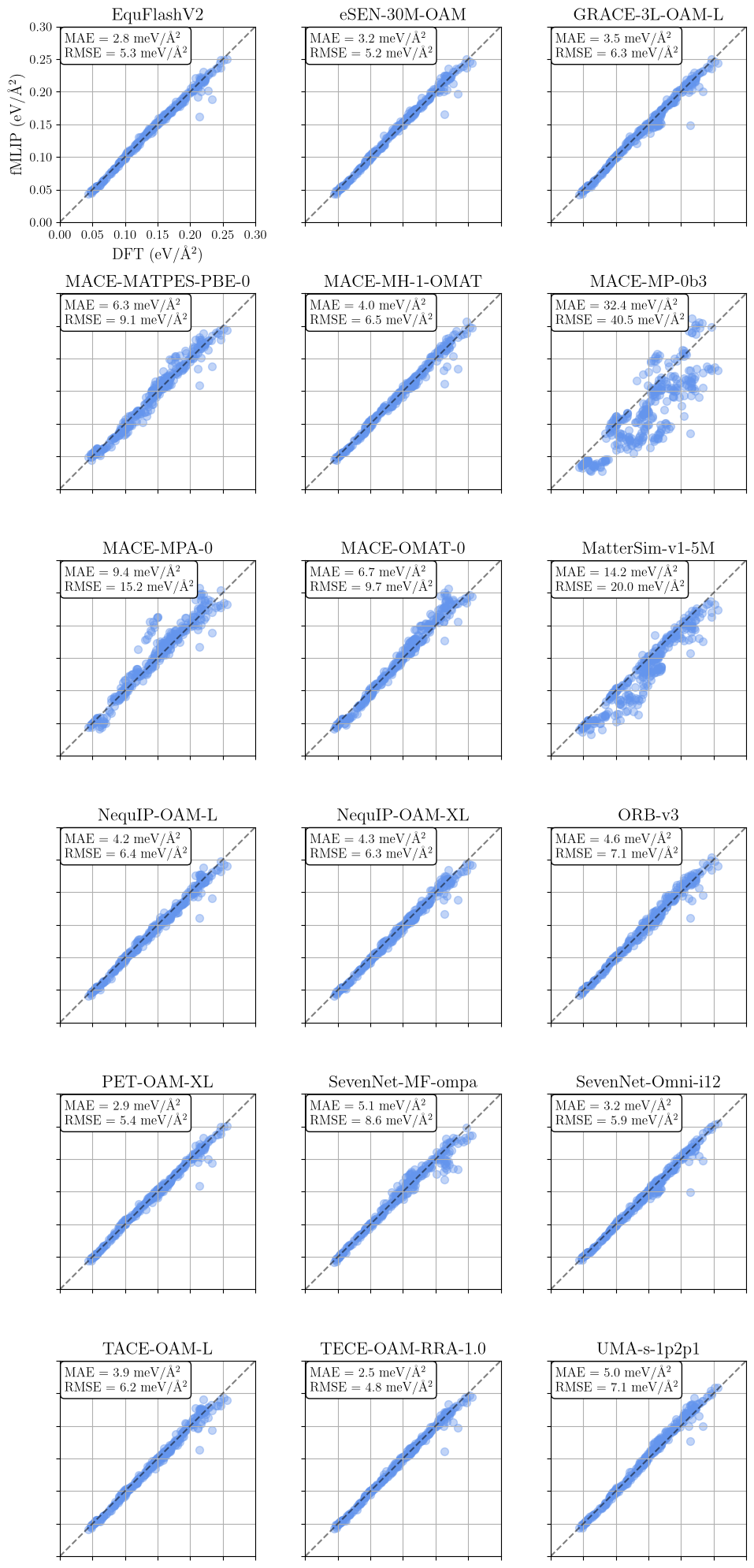}

    \caption{Parity plots comparing the fMLIPs' predicted surface energies to the reference PBE values. 
    The dashed line represents perfect agreement between predicted and reference values. 
    For each model, the MAE and RMSE are reported in the top left corner.}
    
\end{figure}

\subsubsection{Adatom--vacancy formation energies}

The migration of surface atoms is another important factor in determining surface structure \cite{antczak2007}. 
The energetics of ejection of atoms from step-edges or kinks, creating a vacancy and leading to adatom formation on an adjacent terrace has previously been studied in detail\cite{xu2023}.
The authors of the cited study used a two-slab model: one slab is the ``source'' from which a surface atom is removed (giving a vacancy), while the second slab is the ``sink'' where the removed atom is placed as an adatom. 
The total energy difference between these configurations (e.g. for ejection from a step-edge $E_{211+\rm vac}$ and $E_{111+\rm adatom}$) and two clean slabs ($E_{111}$ and $E_{211}$) provides the thermodynamic cost of adatom--vacancy formation.
The energies for adatom ejection from a step-edge $E_{\rm form}^{\rm vac} 
\overset{(211)}{\underset{(111)}{\downarrow}}$ and from a kink $E_{\rm form}^{\rm vac} 
\overset{(874)}{\underset{(111)}{\downarrow}}$ are given by Eq.~\eqref{eq:step-edge} and Eq.~\eqref{eq:kink} respectively. 
Such energies were considered for eight fcc transition metals (Ag, Cu, Au, Pd, Ni, Rh, Pt, and Ir) using the PBE functional\cite{perdew1996} (full computational settings are available in the cited study).

\begin{equation}
E_{\rm form}^{\rm vac} 
\overset{(211)}{\underset{(111)}{\downarrow}}
=
E_{111+\rm adatom} + E_{211+\rm vac} - E_{111} - E_{211}
\label{eq:step-edge}
\end{equation}

\begin{equation}
E_{\rm form}^{\rm vac} 
\overset{(874)}{\underset{(111)}{\downarrow}}
=
E_{111+\rm adatom} + E_{874+\rm vac} - E_{111} - E_{874}
\label{eq:kink}
\end{equation}

\clearpage

\begin{figure}[H]
    \centering
    \includegraphics[width=0.61\textwidth]{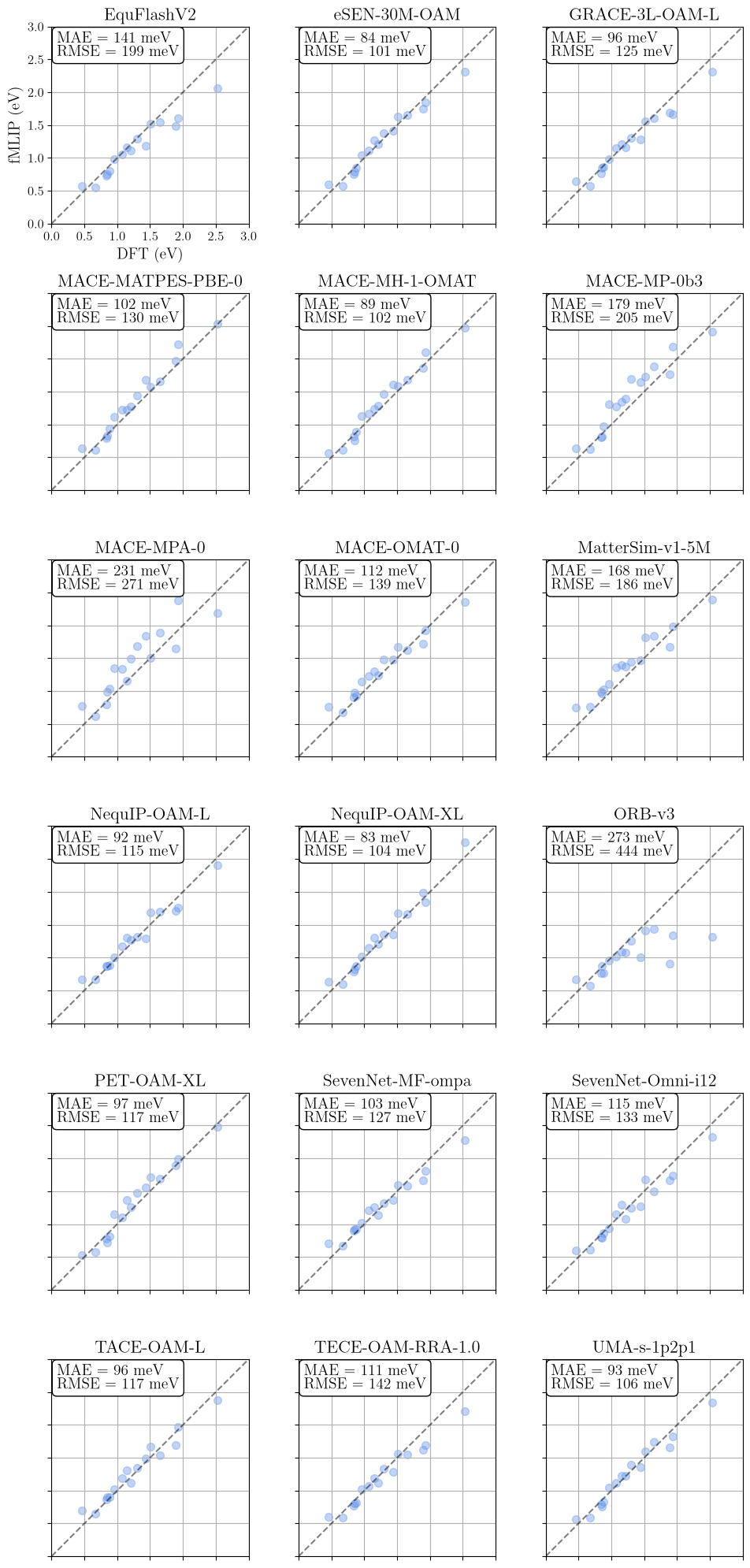}

    \caption{Parity plots comparing the fMLIPs' predicted adatom--vacancy formation energies to the reference PBE values. 
    The dashed line represents perfect agreement between predicted and reference values. 
    For each model, the MAE and RMSE are reported in the top left corner.}

\end{figure}

\subsubsection{Segregation and aggregation energies}

Dilute alloys are of particular interest because the electronic structure of their active sites differs markedly from that of the bulk materials\cite{greiner2018free,berger2025dopant} that dominate current training datasets, giving rise to distinct stability,\cite{ruban1999surface,berger2026atlas} formation energetics,\cite{wang2020surface,karageorgiou2026mechanisms} and reactivity.\cite{hannagan2020single,berger2024bringing}
The segregation of isolated dopants between the bulk and the surface of a host metal, forming single-atom alloys (SAAs), and their aggregation into dopant dimers, forming dual-atom alloys (DAAs), have previously been studied in detail.\cite{berger2026atlas}
The segregation energy, $E^{\mathrm{seg}}$, is defined as the total energy difference between a slab with the dopant in a bulk-like layer ($E_{\mathrm{bulk}}^{\mathrm{SAA}}$) and a slab with the dopant in the surface layer ($E_{\mathrm{surface}}^{\mathrm{SAA}}$):

\begin{equation}
E^{\mathrm{seg}} = E_{\mathrm{bulk}}^{\mathrm{SAA}} - E_{\mathrm{surface}}^{\mathrm{SAA}}
\label{eq4}
\end{equation}

\noindent{Segregation energies were considered for the ten 4$d$ transition-metal dopants (Y, Zr, Nb, Mo, Tc, Ru, Rh, Pd, Ag, and Cd) in a Cu host.}

The aggregation energy of dopants $i$ and $j$ is given by the total energy of the DAA, in which $i$ and $j$ occupy nearest-neighbor surface sites ($E_{ij,\mathrm{surface}}^{\mathrm{DAA}}$), relative to the two corresponding surface SAAs ($E_{i,\mathrm{surface}}^{\mathrm{SAA}}$ and $E_{j,\mathrm{surface}}^{\mathrm{SAA}}$) and referenced to the pristine host surface ($E^{\mathrm{host}}$):

\begin{equation}
E_{ij}^{\mathrm{agg}} = E_{ij,\mathrm{surface}}^{\mathrm{DAA}} + E^{\mathrm{host}} - E_{i,\mathrm{surface}}^{\mathrm{SAA}} - E_{j,\mathrm{surface}}^{\mathrm{SAA}}
\label{eq5}
\end{equation}

\noindent{The aggregation energies comprise the homonuclear dimers of all ten 4$d$ metals (Y--Y, Zr--Zr, Nb--Nb, Mo--Mo, Tc--Tc, Ru--Ru, Rh--Rh, Pd--Pd, Ag--Ag, and Cd--Cd) and 15 heteronuclear dimers (Y--Rh, Y--Cd, Zr--Ru, Zr--Rh, Zr--Ag, Nb--Tc, Nb--Rh, Nb--Pd, Mo--Rh, Tc--Ru, Tc--Rh, Ru--Rh, Rh--Pd, Rh--Ag, and Rh--Cd).}
All reference calculations use the PBE functional;\cite{perdew1996} full computational settings are given in Ref.~\cite{berger2026atlas}.

\clearpage

\begin{figure}[H]
    \centering
    \includegraphics[width=0.64\textwidth]{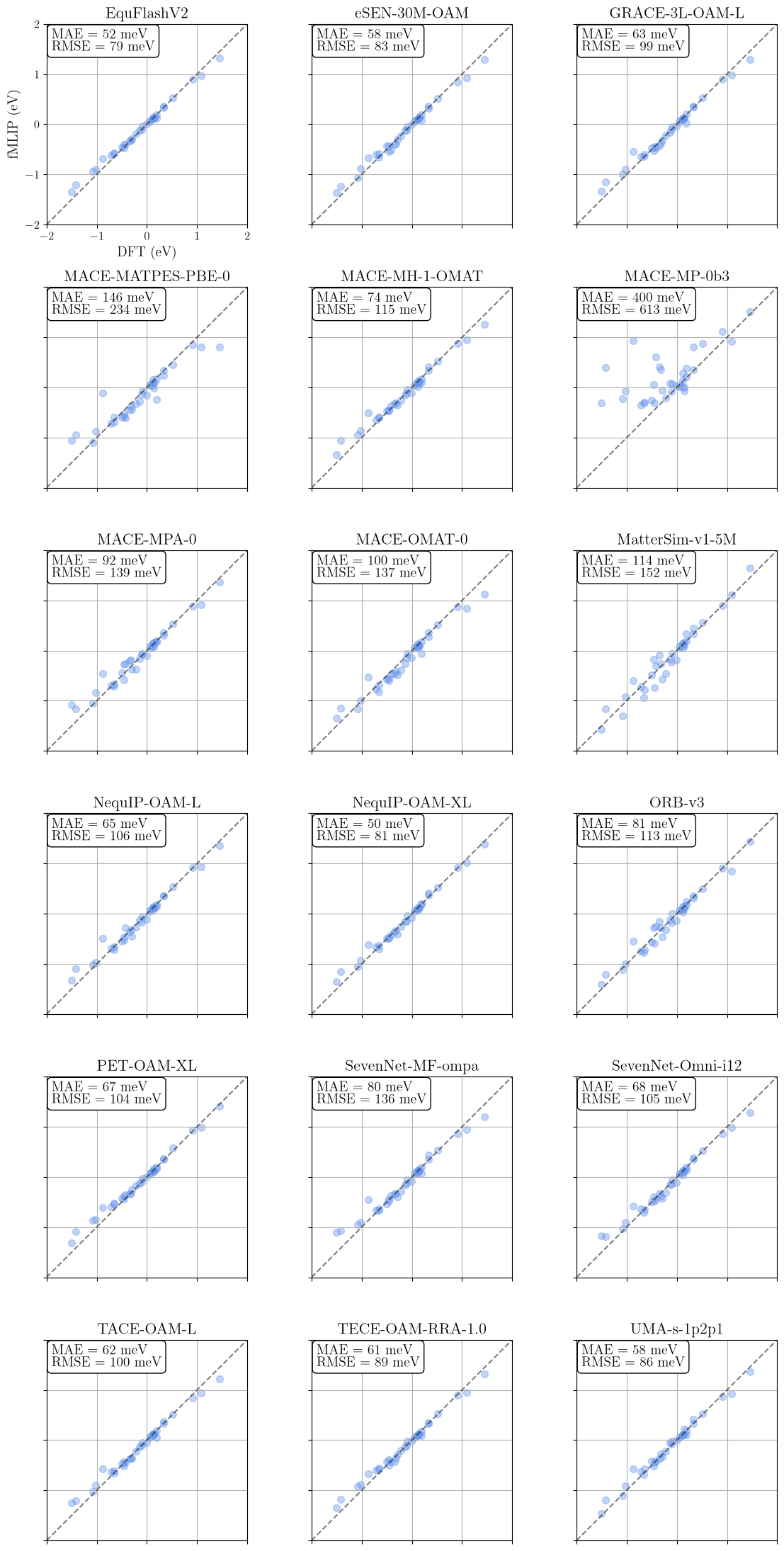}

    \caption{Parity plots comparing the fMLIPs' predicted segregation and aggregation energies to the reference PBE values. 
    The dashed line represents perfect agreement between predicted and reference values. 
    For each model, the MAE and RMSE are reported in the top left corner.}

\end{figure}

\subsubsection{Metal--support interactions}

Metal nanoparticles are widely used in catalytic processes owing to their high surface-area-to-volume ratio.\cite{astruc2020}
Sintering, in which the particles coalesce and lose catalytic surface area, is a critical deactivation mechanism, and supports, commonly metal oxides, are used to minimize it.\cite{dai2018}

We assess how well fMLIPs capture metal--support interactions using coinage metal dimers on MgO as a well-defined model system.
MgO is an archetypal, well-characterized oxide support,\cite{barcaro2005} and small Cu, Ag, and Au clusters are of direct catalytic interest.\cite{luo2016}
Dimers contain both a metal--metal bond and a metal--support interaction in their simplest form.\cite{shi2024}
Because the interaction is localized on one or two metal atoms, the interaction energy is sensitive to the binding site and orientation, which makes it a stringent test for fMLIPs.
The interaction energies, $E_{\mathrm{int}}$, of $\mathrm{Cu_2}$, $\mathrm{Ag_2}$, and $\mathrm{Au_2}$ on MgO were recently computed with the PBE functional\cite{perdew1996} in Ref.~\cite{shi2024} as

\begin{equation}
E_{\mathrm{int}} = E_{\mathrm{M_2/MgO}} - E_{\mathrm{M_2}} - E_{\mathrm{MgO}},
\label{eq:metal-support}
\end{equation}

\noindent{where $E_{\mathrm{M_2/MgO}}$ is the total energy of the dimer adsorbed on the support, $E_{\mathrm{M_2}}$ is the total energy of the gas-phase dimer, and $E_{\mathrm{MgO}}$ is the total energy of the isolated support.}
This dataset comprises 24 structures, corresponding to dimers at different positions and orientations on the surface.
Full computational settings are given in Ref.~\cite{shi2024}.

In practice, supported nanoparticles in catalysis are typically larger than dimers.
To move toward more realistic systems and to increase the diversity of metal--support interactions, we also examine extended interfaces consisting of rutile TiO$_2$ (r-TiO$_2$), CeO$_2$, and In$_2$O$_3$ layers on Pd and Pt slabs (six systems in total).\cite{morales_vidal2026}
These systems are not included in the main benchmark because the reference calculations apply a Hubbard U correction to the Ce 4$f$ and Ti 3$d$ states, whereas the fMLIP training data do not apply U to these elements.
The interaction energies are calculated as

\begin{equation}
E_{\mathrm{int}} = \frac{1}{n_{\mathrm{Ms}}}\left[ E_{\mathrm{ox/M}} - \frac{n_{\mathrm{M}}^{\mathrm{ox/M}}}{n_{\mathrm{M}}^{\mathrm{M}}}\,E_{\mathrm{M}} - E_{\mathrm{ox}} \right],
\label{eq:smsi}
\end{equation}

\noindent{where $E_{\mathrm{ox/M}}$ is the total energy of the oxide layer on the metal slab, $E_{\mathrm{M}}$ is the total energy of the clean metal slab, and $E_{\mathrm{ox}}$ is the total energy of the isolated oxide layer.}
Because the interface and the clean slab structures contain different numbers of metal atoms, $n_{\mathrm{M}}^{\mathrm{ox/M}}$ and $n_{\mathrm{M}}^{\mathrm{M}}$, respectively, the slab energy is scaled accordingly.
The interaction energy is normalized by $n_{\mathrm{Ms}}$, the number of Ce, In, or Ti atoms at the interface.
Parity plots comparing fMLIP and DFT results for this dataset are shown in Figure~S5.

\clearpage

\begin{figure}[H]
    \centering
    \includegraphics[width=0.62\textwidth]{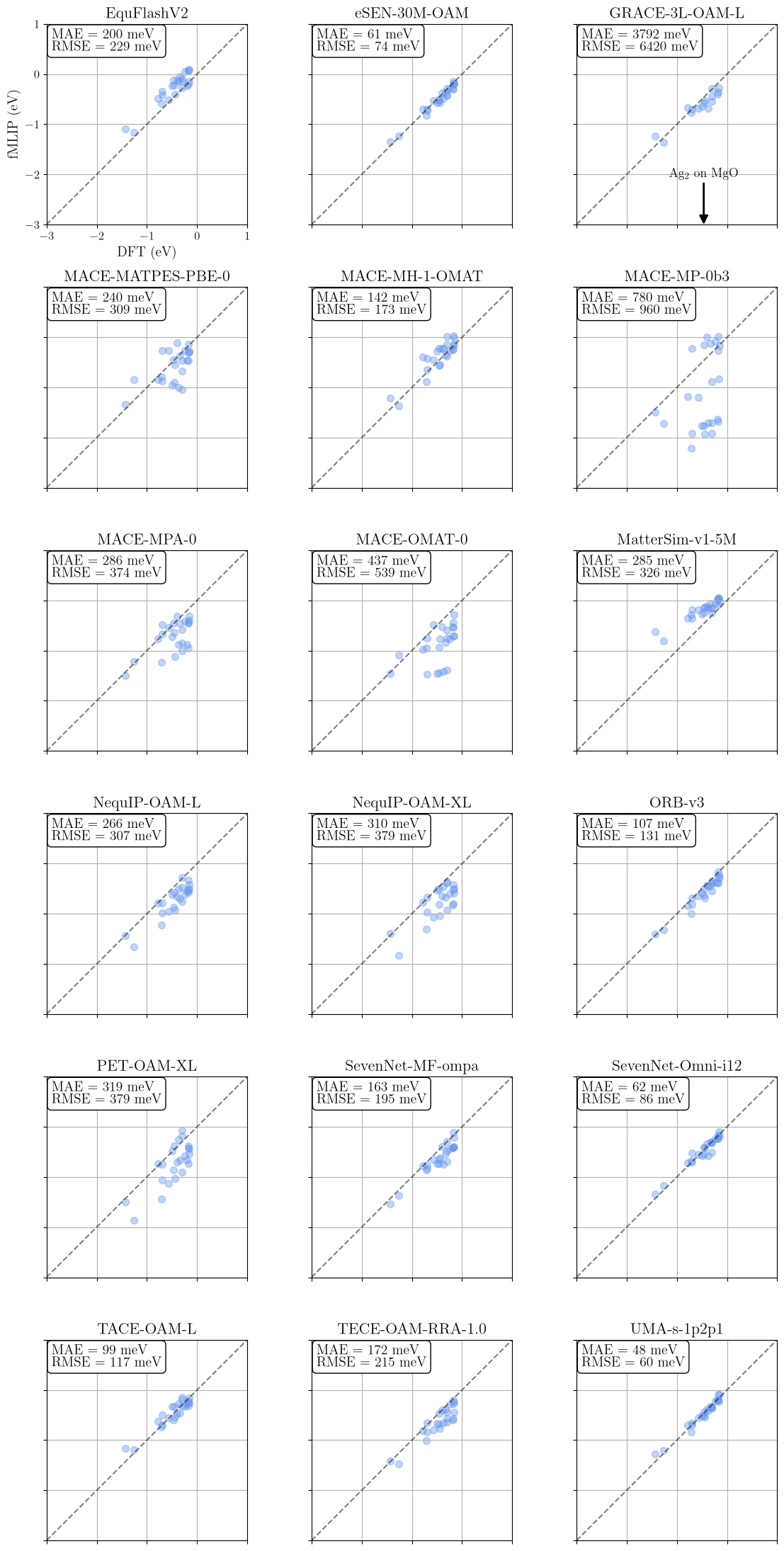}

    \caption{Parity plots comparing the fMLIPs' predicted metal--metal oxide interactions to the reference PBE values for the dataset involving coinage metal dimers on MgO. 
    The dashed line represents perfect agreement between predicted and reference values. 
    For each model, the MAE and RMSE are reported in the top left corner.}

\end{figure}

\begin{figure}[H]
    \centering
    \includegraphics[width=0.60\textwidth]{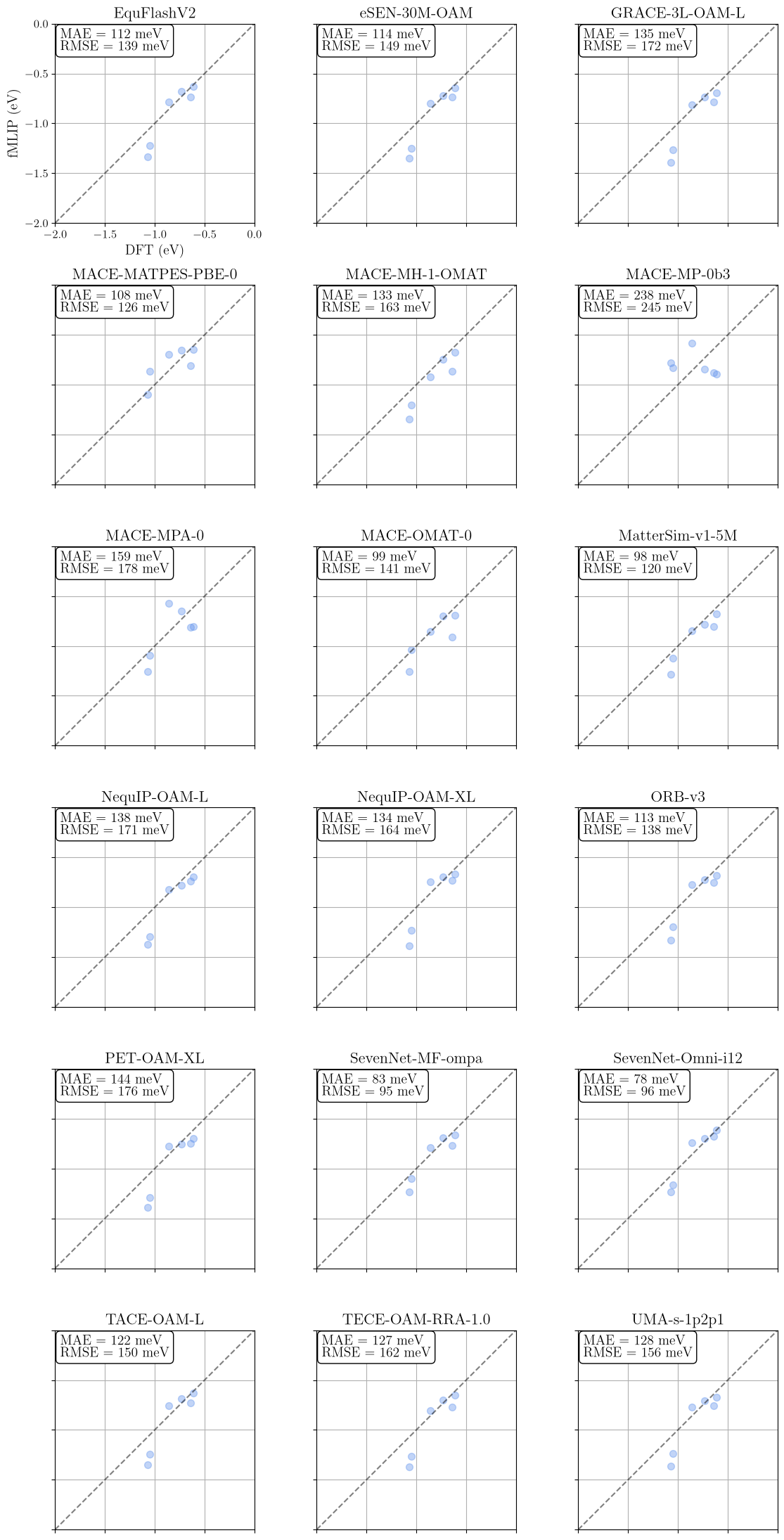}

    \caption{Parity plots comparing the fMLIPs' predicted metal--metal oxide interactions to the reference PBE values for the dataset involving CeO$_2$, In$_2$O$_3$ and r-TiO$_2$ layers on Pt/Pd. 
    The dashed line represents perfect agreement between predicted and reference values.
    The comparison is not like-for-like: deviations may reflect the missing U correction rather than model error, and agreement may partly result from error cancellation between the isolated layer and the interface.
    For each model, the MAE and RMSE are reported in the top left corner.
    The mean absolute reference value is 827~meV.}

\end{figure}

\subsection{Adsorbate--surface interactions}

The second category of \textit{HetCat26} moves beyond catalytic surfaces, focusing on the interactions between these and adsorbed reactants.
Two datasets focused on interaction energies (ADS41\cite{sharada2019} and CADS34) together with an analysis of adsorption site preferences in the famous case of CO on Cu(111) and Pt(111) surfaces\cite{feibelman2001} are used to test fMLIPs.

\subsubsection{Adsorption energies}

The Sabatier principle embodies the importance of adsorption energies.\cite{sabatier1920} It links adsorption energies to catalytic activity, with optimal catalysts showing intermediate adsorbate binding: if binding is too strong, the surface becomes poisoned, while if binding is too weak, desorption occurs instead of activation.
Two complementary datasets are included in \textit{HetCat26} to probe the description of the interaction of adsorbates with surfaces by fMLIPs. 
The first is the ADS41 dataset which comprises 41 adsorbate--surface combinations (standard adsorbates on transition metal surfaces) \cite{sharada2019} as summarized in Table~\ref{tab:ads41}. 
PBE geometries and energies were obtained from Trepte \textit{et al.}. \cite{trepte2022}
Adsorption energies, $E_{\mathrm{ads}}$, are computed as:

\begin{equation}
E_{\mathrm{ads}} = E_{\mathrm{combined}} - E_{\mathrm{slab}} - E_{\mathrm{adsorbate}}
\label{eq:adsorption}
\end{equation}

\noindent where $E_{\mathrm{slab}}$ and $E_{\mathrm{adsorbate}}$ are the total energies of the isolated slab and adsorbate respectively, and $E_{\mathrm{combined}}$ is the total energy of the combined system.

To consider a diverse set of surfaces beyond exclusively transition metals, we also include the CADS34 dataset, consisting of 34 systems as outlined in Table~\ref{tab:cads34}.
For this dataset, the reference calculations are interaction energies rather than adsorption energies, the difference being that interaction energies are calculated at a fixed geometry ($E_{\mathrm{slab}}$ and $E_{\mathrm{adsorbate}}$ are obtained from the same structure as $E_{\mathrm{combined}}$).
Calculations employed the PBE functional\cite{perdew1996}, with Grimme's D3 dispersion correction\cite{grimme2010} with Becke--Johnson damping (D3(BJ)).
%Full details are provided in the paper introducing CADS34 (coming on arXiv soon).

\clearpage

\begin{table}[H]
\centering
\caption{Adsorption systems within the ADS41 dataset. 
O-bearing adsorbates on Co- or Ni-containing surfaces, denoted by an asterisk, are excluded from all analyses, except in the final section of the main text, which focuses specifically on these systems.}
\label{tab:ads41}
\small
\begin{tabularx}{\linewidth}{Y}
\toprule
\textbf{Adsorption systems} \\
\midrule
C$_6$H$_6$ + Ag$(111)$ $\rightarrow$ C$_6$H$_6$@Ag$(111)$ \\
C$_6$H$_6$ + Au$(111)$ $\rightarrow$ C$_6$H$_6$@Au$(111)$ \\
C$_6$H$_6$ + Cu$(111)$ $\rightarrow$ C$_6$H$_6$@Cu$(111)$ \\
C$_6$H$_6$ + Pt$(111)$ $\rightarrow$ C$_6$H$_6$@Pt$(111)$ \\
C$_2$H$_4$ + Pt$(111)$ $\rightarrow$ [CCH$_3$+H]@Pt$(111)$ \\
C$_2$H$_6$ + Pt$(111)$ $\rightarrow$ C$_2$H$_6$@Pt$(111)$ \\
C$_3$H$_8$ + Pt$(111)$ $\rightarrow$ C$_3$H$_8$@Pt$(111)$ \\
C$_4$H$_{10}$ + Pt$(111)$ $\rightarrow$ C$_4$H$_{10}$@Pt$(111)$ \\
CH$_2$I$_2$ + Pt$(111)$ $\rightarrow$ [CH+H+I+I]@Pt$(111)$ \\
CH$_3$I + Pt$(111)$ $\rightarrow$ CH$_3$I@Pt$(111)$ \\
CH$_3$I + Pt$(111)$ $\rightarrow$ [CH$_3$+I]@Pt$(111)$ \\
CH$_4$ + Pt$(111)$ $\rightarrow$ CH$_4$@Pt$(111)$ \\
* CO + Co$(001)$ $\rightarrow$ CO@Co$(001)$ \\
CO + Cu$(111)$ $\rightarrow$ CO@Cu$(111)$ \\
CO + Ir$(111)$ $\rightarrow$ CO@Ir$(111)$ \\
* CO + Ni$(111)$ $\rightarrow$ CO@Ni$(111)$ \\
CO + Pd$(100)$ $\rightarrow$ CO@Pd$(100)$ \\
CO + Pd$(111)$ $\rightarrow$ CO@Pd$(111)$ \\
CO + Pt$(111)$ $\rightarrow$ CO@Pt$(111)$ \\
CO + Rh$(111)$ $\rightarrow$ CO@Rh$(111)$ \\
CO + Ru$(001)$ $\rightarrow$ CO@Ru$(001)$ \\
C$_6$H$_{10}$ + Pt$(111)$ $\rightarrow$ C$_6$H$_{10}$@Pt$(111)$ \\
H$_2$O + $\frac{1}{3}$[O@Pt(111)] $\rightarrow$ $\frac{2}{3}$[(H$_2$O··OH)@Pt(111)] \\
H$_2$O + Pt$(111)$ $\rightarrow$ H$_2$O@Pt$(111)$ \\
H$_2$ + Ni$(100)$ $\rightarrow$ [H+H]@Ni$(100)$ \\
H$_2$ + Ni$(111)$ $\rightarrow$ [H+H]@Ni$(111)$ \\
H$_2$ + Pd$(111)$ $\rightarrow$ [H+H]@Pd$(111)$ \\
H$_2$ + Pt$(111)$ $\rightarrow$ [H+H]@Pt$(111)$ \\
H$_2$ + Rh$(111)$ $\rightarrow$ [H+H]@Rh$(111)$ \\
I$_2$ + Pt$(111)$ $\rightarrow$ [I+I]@Pt$(111)$ \\
CH$_3$OH + Pt$(111)$ $\rightarrow$ CH$_3$OH@Pt$(111)$ \\
C$_{10}$H$_8$ + Pt$(111)$ $\rightarrow$ C$_{10}$H$_8$@Pt$(111)$ \\
NH$_3$ + Cu$(100)$ $\rightarrow$ NH$_3$@Cu$(100)$ \\
* NO + Ni$(100)$ $\rightarrow$ [N+O]@Ni$(100)$ \\
NO + Pd$(100)$ $\rightarrow$ NO@Pd$(100)$ \\
NO + Pd$(111)$ $\rightarrow$ NO@Pd$(111)$ \\
NO + Pt$(111)$ $\rightarrow$ NO@Pt$(111)$ \\
* O$_2$ + Ni$(100)$ $\rightarrow$ [O+O]@Ni$(100)$ \\
* O$_2$ + Ni$(111)$ $\rightarrow$ [O+O]@Ni$(111)$ \\
O$_2$ + Pt$(111)$ $\rightarrow$ [O+O]@Pt$(111)$ \\
O$_2$ + Rh$(100)$ $\rightarrow$ [O+O]@Rh$(100)$ \\
\bottomrule
\end{tabularx}
\end{table}

\clearpage

\begin{table}[H]
\centering
\caption{Adsorption systems within the CADS34 dataset. 
O-bearing adsorbates on Mo- or W-containing surfaces, denoted by an asterisk, are excluded from all analyses, except in the final section of the main text, which focuses specifically on these systems.}
\label{tab:cads34}
\small
\begin{tabularx}{\linewidth}{Y}
\toprule
\textbf{Adsorption systems} \\
\midrule
C$_6$H$_6$ + Au $\rightarrow$ C$_6$H$_6$@Au \\
H$_2$O + hBN $\rightarrow$ H$_2$O@hBN \\
C$_2$H$_4$ + Chabazite $\rightarrow$ C$_2$H$_4$@Chabazite \\
C$_3$H$_8$ + Chabazite $\rightarrow$ C$_3$H$_8$@Chabazite \\
H$_2$O + Chabazite $\rightarrow$ H$_2$O@Chabazite \\
H$_2$ + Carbon nanotube (ext) $\rightarrow$ H$_2$@Carbon nanotube (ext) \\
H$_2$ + Carbon nanotube (int) $\rightarrow$ H$_2$@Carbon nanotube (int) \\
H$_2$O + Carbon nanotube (ext) $\rightarrow$ H$_2$O@Carbon nanotube (ext) \\
H$_2$O + Carbon nanotube (int) $\rightarrow$ H$_2$O@Carbon nanotube (int) \\
CO + Cu $\rightarrow$ CO@Cu \\
CO$_2$ + Graphene $\rightarrow$ CO$_2$@Graphene \\
H$_2$O + Graphene $\rightarrow$ H$_2$O@Graphene \\
CH$_3$OH + Kaolinite(Al) $\rightarrow$ CH$_3$OH@Kaolinite(Al) \\
H$_2$O + Kaolinite(Al) $\rightarrow$ H$_2$O@Kaolinite(Al) \\
CH$_3$OH + Kaolinite(Si) $\rightarrow$ CH$_3$OH@Kaolinite(Si) \\
H$_2$O + Kaolinite(Si) $\rightarrow$ H$_2$O@Kaolinite(Si) \\
C$_2$H$_6$ + MgO $\rightarrow$ C$_2$H$_6$@MgO \\
C$_6$H$_6$ + MgO $\rightarrow$ C$_6$H$_6$@MgO \\
CH$_4$ + MgO $\rightarrow$ CH$_4$@MgO \\
CO + MgO $\rightarrow$ CO@MgO \\
H$_2$O + MgO $\rightarrow$ H$_2$O@MgO \\
N$_2$O + MgO $\rightarrow$ N$_2$O@MgO \\
NH$_3$ + MgO $\rightarrow$ NH$_3$@MgO \\
* H$_2$O + MoS$_2$ $\rightarrow$ H$_2$O@MoS$_2$ \\
H$_2$O + NaCl $\rightarrow$ H$_2$O@NaCl \\
H$_2$ + (Na+Graphene) $\rightarrow$ [H+H]@(Na+Graphene) \\
CH$_4$ + Pt $\rightarrow$ CH$_4$@Pt \\
H$_2$O + TiO$_2$(anatase) $\rightarrow$ H$_2$O@TiO$_2$(anatase) \\
NH$_3$ + TiO$_2$(anatase) $\rightarrow$ NH$_3$@TiO$_2$(anatase) \\
CH$_3$OH + TiO$_2$(rutile) $\rightarrow$ CH$_3$OH@TiO$_2$(rutile) \\
CH$_4$ + TiO$_2$(rutile) $\rightarrow$ CH$_4$@TiO$_2$(rutile) \\
CO$_2$ + TiO$_2$(rutile) $\rightarrow$ CO$_2$@TiO$_2$(rutile) \\
H$_2$O + TiO$_2$(rutile) $\rightarrow$ H$_2$O@TiO$_2$(rutile) \\
* H$_2$O + WSe$_2$ $\rightarrow$ H$_2$O@WSe$_2$ \\
\bottomrule
\end{tabularx}
\end{table}

\clearpage

\begin{figure}[H]
    \centering
    \includegraphics[width=0.62\textwidth]{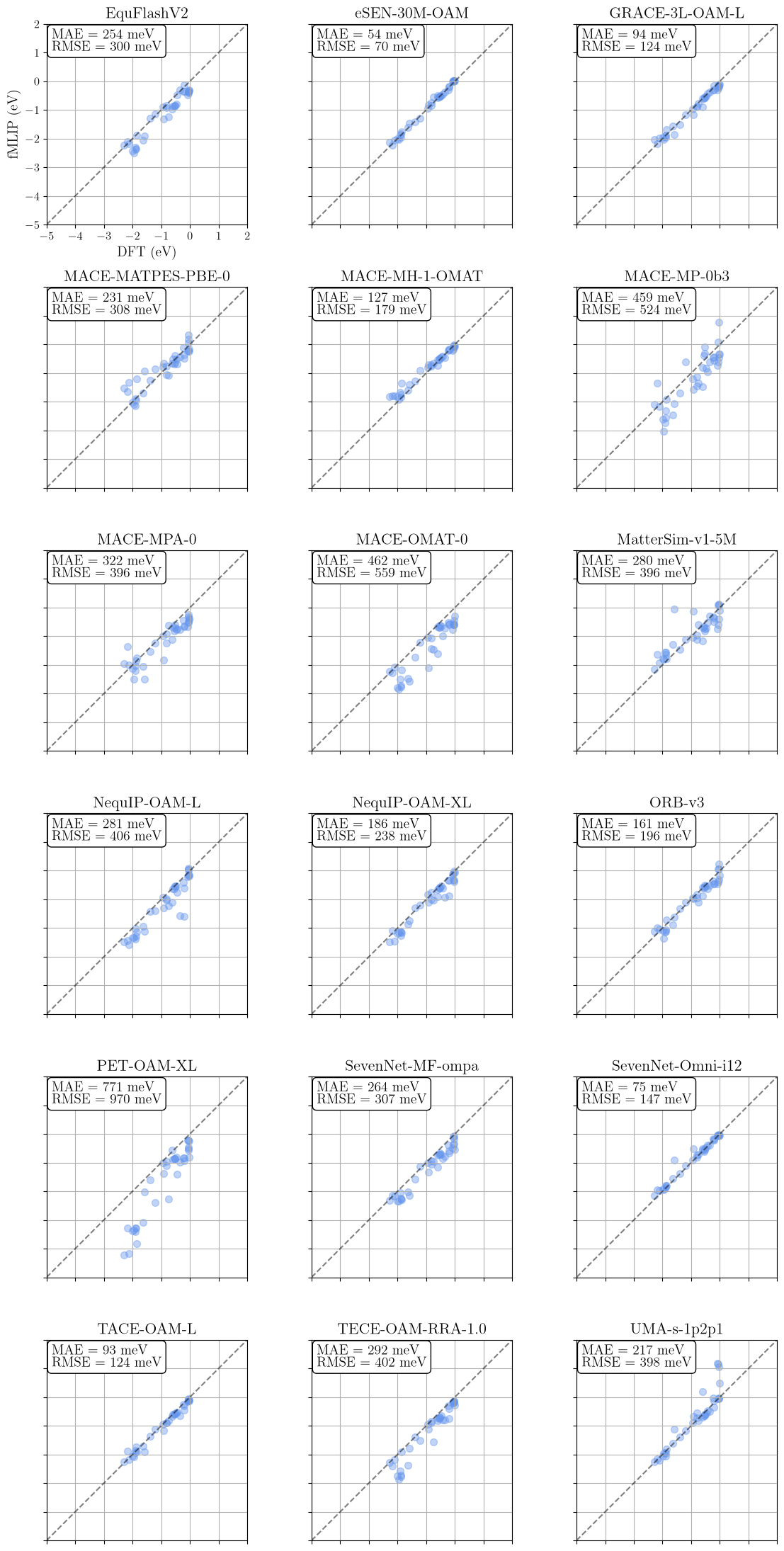}

    \caption{Parity plots comparing the fMLIPs' predicted adsorption energies to the reference PBE values. 
    The dashed line represents perfect agreement between predicted and reference values.
    Systems with O-bearing adsorbates on Ni- and Co-containing surfaces are excluded from the ADS41 dataset.
    For each model, the MAE and RMSE are reported in the top left corner.}

\end{figure}

\begin{figure}[H]
    \centering
    \includegraphics[width=0.63\textwidth]{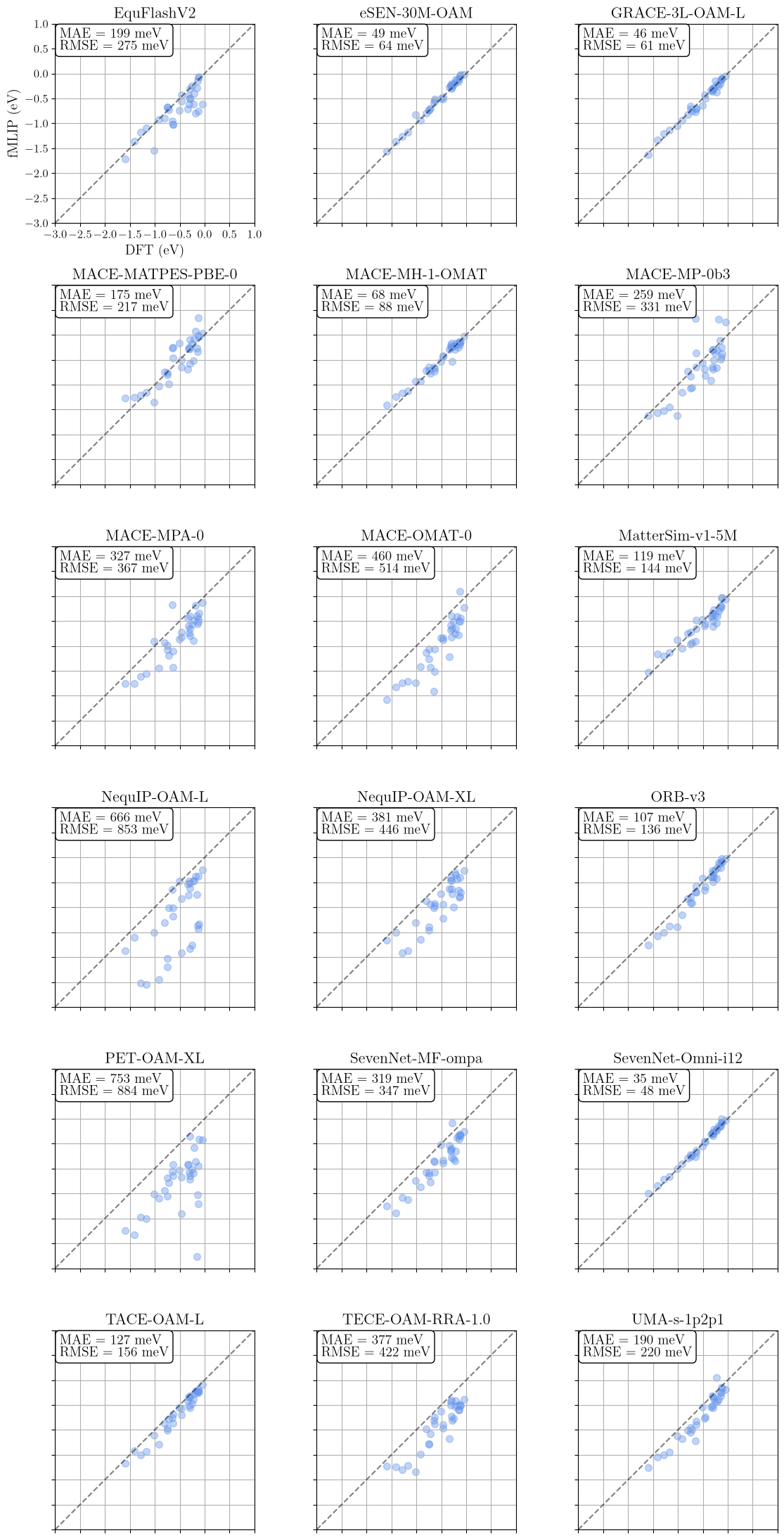}

    \caption{Parity plots comparing the fMLIPs' predicted adsorption energies to the reference PBE+D3 values. 
    The dashed line represents perfect agreement between predicted and reference values.
    Systems with O-bearing adsorbates on W- and Mo-containing surfaces are excluded from the CADS34 dataset.
    For each model, the MAE and RMSE are reported in the top left corner.}

\end{figure}

\subsubsection{Adsorption site preferences}

There are multiple possible sites at which adsorption can occur on a (111) surface: these include the ``top'', ``bridge'', ``fcc/hcp'' sites, which correspond to one-, two-, and three-fold coordination with surface atoms, respectively.
In previous work, adsorption energies of CO at the top, fcc-hollow, and hcp-hollow sites on Pt(111) \cite{janthon2017}, and top, fcc-hollow, hcp-hollow, and bridge sites on Cu(111)\cite{fanta2025}, have been computed with the PBE functional\cite{perdew1996} (full
computational settings are available in the cited studies).
Whether fMLIPs reproduce the relative preference for adsorption at different sites with respect to the PBE results is a stringent test of their sensitivity. 
Here, a model is considered accurate if it reproduces the relative ordering of PBE calculations, which differ from experimental observations. This mismatch between DFT and experiment has become known as the ``CO Puzzle''\cite{feibelman2001}. 

\clearpage

\begin{figure}[H]
    \centering
    \includegraphics[width=0.59\textwidth]{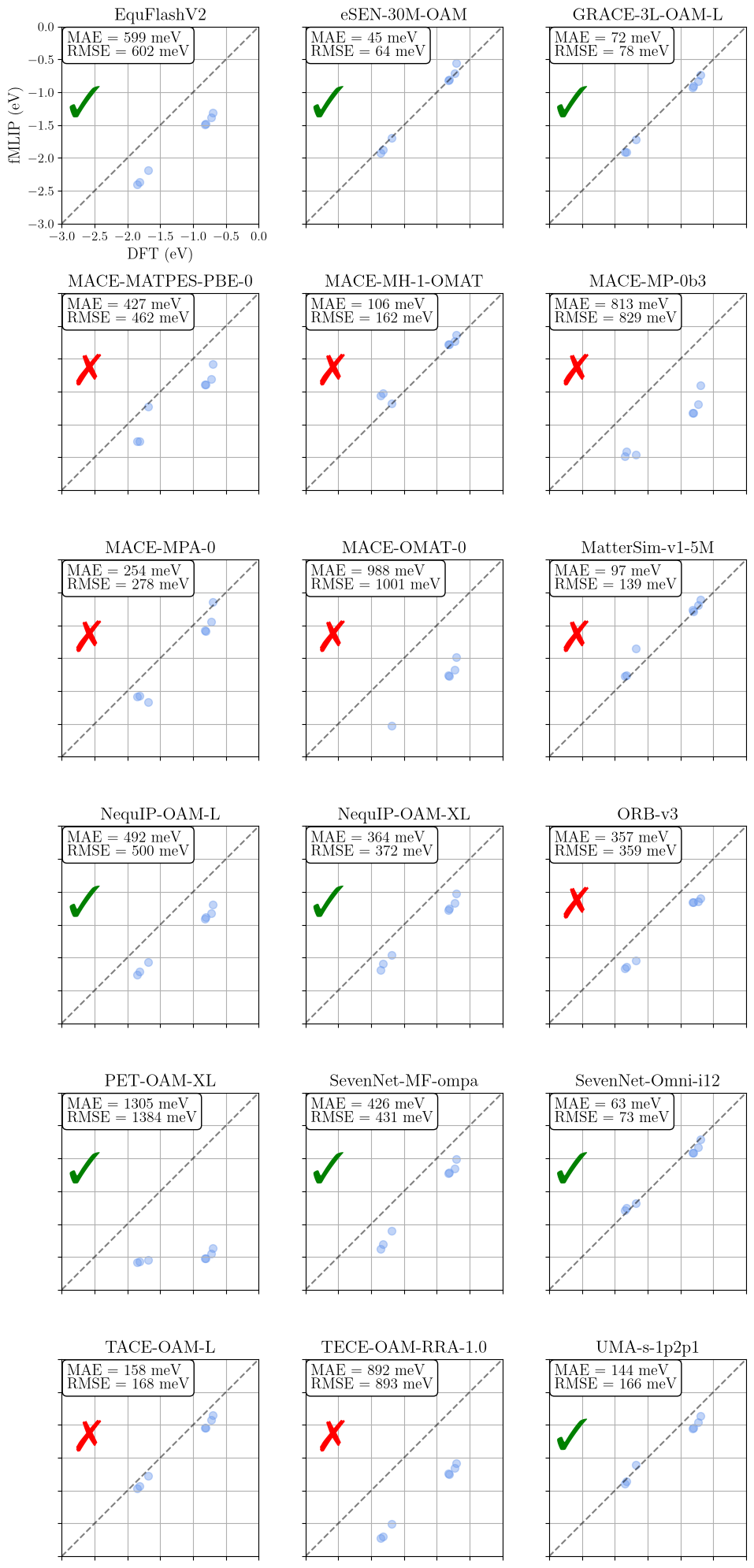}

    \caption{Parity plots comparing the fMLIPs' predicted adsorption energies to the reference PBE values. 
    The dashed line represents perfect agreement between predicted and reference values.
    For each model, the MAE and RMSE are reported in the top left corner. 
    A green tick indicates that the DFT site preference ordering is reproduced by the fMLIP; a red cross shows that it is not.}

\end{figure}

\subsection{Reactions at surfaces}
 
The third section of \textit{HetCat26} tests fMLIPs on reproducing reference energy barriers for dissociative chemisorption as well as within two representative catalytic reaction networks.

\subsubsection{Reaction barriers (dissociative chemisorption)}

Dissociative chemisorption describes the step within which a bond in a molecule is broken as it approaches a surface. 
The SBH17 dataset consists of 17 dissociative chemisorption reactions of common adsorbates on transition metal surfaces \cite{tchakoua2023}. The systems considered are summarized in Table \ref{tab:sbh17} and dissociative chemisorption barriers, $E_{\mathrm{barrier}}^{\mathrm{diss}}$, are computed using:

\begin{equation}
E_{\mathrm{barrier}}^{\mathrm{diss}} = E_{\mathrm{TS}} - E_{\mathrm{IS}}
\label{eq:sbh17}
\end{equation}

\noindent{where $E_{\mathrm{TS}}$ is the total energy of the transition state structure and $E_{\mathrm{IS}}$ is that of the initial state structure. 
Reference PBE barrier heights and the corresponding structures were only available for 15 of the 17 systems. 
Detailed computational settings are available in the cited publication.\\}

\begin{table}[H]
\centering
\caption{Systems within the SBH17 dataset incorporated in \textit{HetCat26} to test how accurately fMLIPs reproduce reference PBE dissociative chemisorption energy barriers.}
\label{tab:sbh17}
\footnotesize
\begin{tabularx}{\linewidth}{Y}
\toprule
\textbf{Dissociative chemisorption systems} \\
\midrule
H$_2$ + Cu(100) \\
H$_2$ + Cu(110) \\
H$_2$ + Pt(111) \\
H$_2$ + Ru(0001) \\
H$_2$ + Ni(111) \\
H$_2$ + Ag(111) \\
N$_2$ + Ru(0001) \\
N$_2$ + Ru(10$\bar{1}$0) \\
CH$_4$ + Ni(111) \\
CH$_4$ + Ni(100) \\
CH$_4$ + Ni(211) \\
CH$_4$ + Pt(111) \\
CH$_4$ + Pt(211) \\
CH$_4$ + Ir(111) \\
CH$_4$ + Ru(0001) \\
\bottomrule
\end{tabularx}
\end{table}

\clearpage

\begin{figure}[H]
    \centering
    \includegraphics[width=0.60\textwidth]{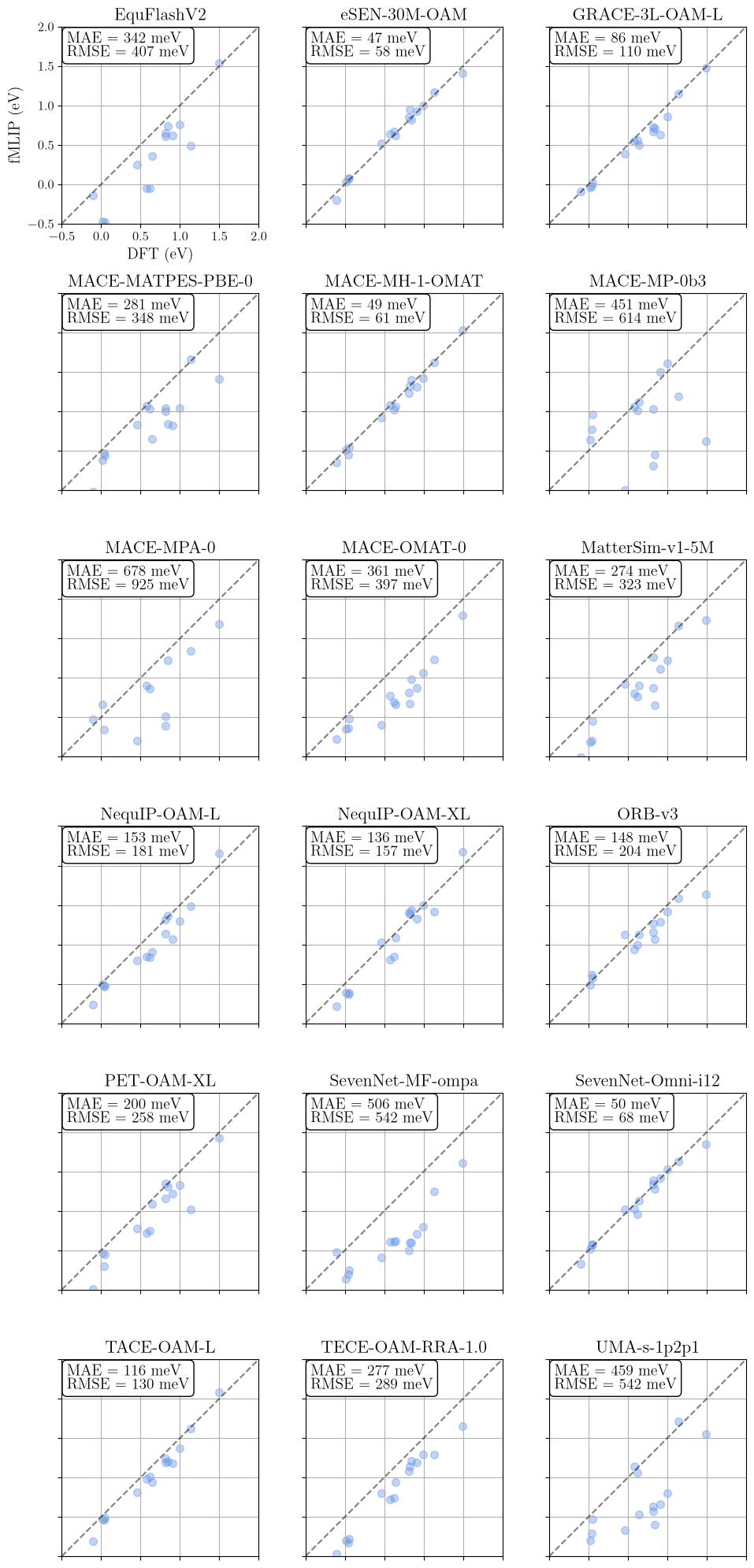}

    \caption{Parity plots comparing the fMLIPs' predicted reaction barriers to dissoc. chemisorption to the reference PBE values. 
    The dashed line represents perfect agreement between predicted and reference values. 
    For each model, the MAE and RMSE are reported in the top left corner.}

\end{figure}

\subsubsection{Reaction barriers (CO$_2$ hydrogenation on Cu(111))}

CO$_2$ hydrogenation to methanol is a critically important reaction.
Its benefits are twofold: it combines CO$_2$ utilization with methanol production, enabling the usage of atmospheric CO$_2$, while giving access to a potential energy carrier.
The full reaction is: 

\[
\mathrm{CO_2 + 3H_2 \rightarrow CH_3OH + H_2O}
\]

\noindent{In reality, the actual reaction network involves a multitude of steps and interlinked pathways. 
Four possible pathways for the conversion of CO$_2$ to methanol, labelled by some key intermediates they go via, were calculated with DFT.
The four pathways (COOH–CO, COOH–COHOH, HCOO–H$_2$COO, HCOO–HCOOH), some of which share elementary reaction steps, are set out in Table \ref{tab:reaction_paths}.
Considering both forward and backward steps, energy barriers for 42 elementary steps across the network of reactions are included and used to test the predictions of fMLIPs.}

The reference DFT calculations against which the fMLIPs are tested employed the PBE functional\cite{perdew1996} as implemented in the Vienna Ab Initio Simulation Package (VASP)\cite{kresse1993ab,kresse1994ab,kresse1994norm,kresse1996efficiency,kresse1996efficient,kresse1999ultrasoft}, with plane-wave basis sets for the valence electrons and the projector augmented-wave (PAW) method with standard potentials for the core electrons.
With the PAW method, a plane-wave kinetic energy cutoff of 400 eV was utilized.
Grimme's D3 dispersion correction with Becke–Johnson damping (D3(BJ)) was used to treat long-range dispersion interactions\cite{grimme2010}.
Brillouin zone integrations for the surface calculations were performed using a (7$\times$7$\times$1) Monkhorst–Pack \textbf{k}-point mesh. 
Geometry optimizations were considered converged when the residual forces on all unconstrained atoms were less than 0.01 eV\AA$^{-1}$, with an energy convergence criterion of 1 $\times$ 10$^{-6}$ eV.

\clearpage

%\begin{landscape}
\begin{sidewaystable}[p]
\centering
\caption{List of reaction pathways (labelled by some of the key intermediates they go via) and their composing elementary reaction steps within the hydrogenation of CO$_2$ to methanol on a Cu(111) surface.
In Figure 4 of the main text, the labels at the top (above the computed pathways) denote the corresponding states along the COOH–CO pathway, with S00, S01, and TS01 corresponding sequentially to the first, second, and third entries in the table respectively, and so on.}
\label{tab:reaction_paths}
\small
\setlength{\tabcolsep}{6pt}

\begin{tabularx}{\linewidth}{YYYY}
\toprule
\textbf{COOH--CO} & \textbf{COOH--COHOH} & \textbf{HCOO--H$_2$COO} & \textbf{HCOO--HCOOH} \\
\midrule
surf+CO$_2$+3H$_2$ & surf+CO$_2$+3H$_2$ & surf+CO$_2$+3H$_2$ & surf+CO$_2$+3H$_2$ \\
$\rightarrow$ (CO$_2$*+H*)+H*+2H$_2$ & $\rightarrow$ (CO$_2$*+H*)+H*+2H$_2$ & $\rightarrow$ (CO$_2$*+H*)+H*+2H$_2$ & $\rightarrow$ (CO$_2$*+H*)+H*+2H$_2$ \\
$\rightarrow$ TS01+H*+2H$_2$ & $\rightarrow$ TS01+H*+2H$_2$ & $\rightarrow$ TS01+H*+2H$_2$ & $\rightarrow$ TS01+H*+2H$_2$ \\
$\rightarrow$ (COOH*+H*)+2H$_2$ & $\rightarrow$ (COOH*+H*)+2H$_2$ & $\rightarrow$ (HCOO*+H*)+2H$_2$ & $\rightarrow$ (HCOO*+H*)+2H$_2$ \\
$\rightarrow$ TS02a+H*+2H$_2$ & $\rightarrow$ TS02+2H$_2$ & $\rightarrow$ TS02+2H$_2$ & $\rightarrow$ TS02+2H$_2$ \\
$\rightarrow$ (CO*+OH*)+H*+2H$_2$ & $\rightarrow$ (COHOH*+H*)+H*+H$_2$ & $\rightarrow$ (H$_2$COO*+H*)+H*+H$_2$ & $\rightarrow$ (HCOOH*+H*)+H*+H$_2$ \\
$\rightarrow$ TS02b+2H$_2$ & $\rightarrow$ TS03a+2H$_2$ & $\rightarrow$ TS03+H*+H$_2$ & $\rightarrow$ TS03+H*+H$_2$ \\
$\rightarrow$ CO*+H$_2$O*+2H$_2$ & $\rightarrow$ (COH*+OH*)+2H$_2$ & $\rightarrow$ (H$_2$COOH*)+H*+H$_2$ & $\rightarrow$ (H$_2$COOH*)+H*+H$_2$ \\
$\rightarrow$ (CO*+H*)+H*+H$_2$O+H$_2$ & $\rightarrow$ TS03b+H*+2H$_2$ & $\rightarrow$ TS04a+H*+H$_2$ & $\rightarrow$ TS04a+H*+H$_2$ \\
$\rightarrow$ TS03+H*+H$_2$O+H$_2$ & $\rightarrow$ (COH*+H$_2$O*)+H*+H$_2$ & $\rightarrow$ (H$_2$CO*+OH*+H*)+H$_2$ & $\rightarrow$ (H$_2$CO*+OH*+H*)+H$_2$ \\
$\rightarrow$ (CHO*+H*)+H$_2$O+H$_2$ & $\rightarrow$ (COH*+H*)+H$_2$O+H$_2$ & $\rightarrow$ TS04b+H*+H$_2$ & $\rightarrow$ TS04b+H*+H$_2$ \\
$\rightarrow$ TS04+H$_2$O+H$_2$ & $\rightarrow$ TS04+H$_2$O+H$_2$ & $\rightarrow$ (H$_2$CO*+H$_2$O*)+H$_2$ & $\rightarrow$ (H$_2$CO*+H$_2$O*)+H$_2$ \\
$\rightarrow$ (H$_2$CO*+H*)+H*+H$_2$O & $\rightarrow$ (HCOH*+H*)+H*+H$_2$O & $\rightarrow$ (H$_2$CO*+H*)+H*+H$_2$O & $\rightarrow$ (H$_2$CO*+H*)+H*+H$_2$O \\
$\rightarrow$ TS05+H*+H$_2$O & $\rightarrow$ TS05+H*+H$_2$O & $\rightarrow$ TS05+H*+H$_2$O & $\rightarrow$ TS05+H*+H$_2$O \\
$\rightarrow$ (H$_3$CO*+H*)+H$_2$O & $\rightarrow$ (H$_2$COH*+H*)+H$_2$O & $\rightarrow$ (H$_3$CO*+H*)+H$_2$O & $\rightarrow$ (H$_3$CO*+H*)+H$_2$O \\
$\rightarrow$ TS06+H$_2$O & $\rightarrow$ TS06+H$_2$O & $\rightarrow$ TS06+H$_2$O & $\rightarrow$ TS06+H$_2$O \\
$\rightarrow$ CH$_3$OH*+H$_2$O & $\rightarrow$ CH$_3$OH*+H$_2$O & $\rightarrow$ CH$_3$OH*+H$_2$O & $\rightarrow$ CH$_3$OH*+H$_2$O \\
\bottomrule
\end{tabularx}
\end{sidewaystable}
%\end{landscape}

\clearpage

\begin{figure}[H]
    \centering
    \includegraphics[width=0.64\textwidth]{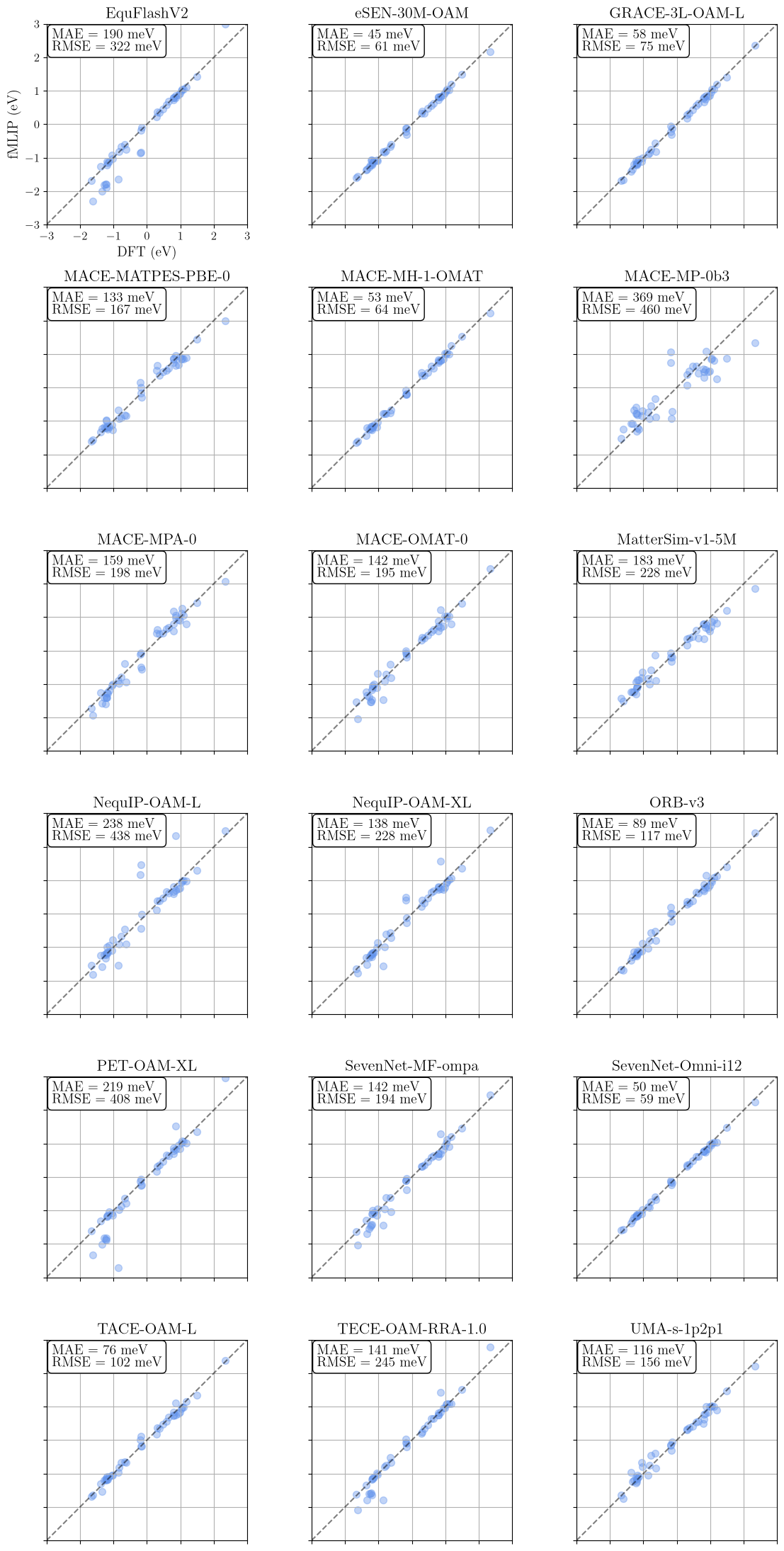}

    \caption{Parity plots comparing the fMLIPs' predicted reaction barriers in the hyd. of CO$_2$ on Cu(111) to reference PBE+D3 values.
    The dashed line represents perfect agreement between predicted and reference values. 
    For each model, the MAE and RMSE are reported at the top left.}

\end{figure}

\begin{figure}[H]
    \centering
    \includegraphics[width=1\textwidth]{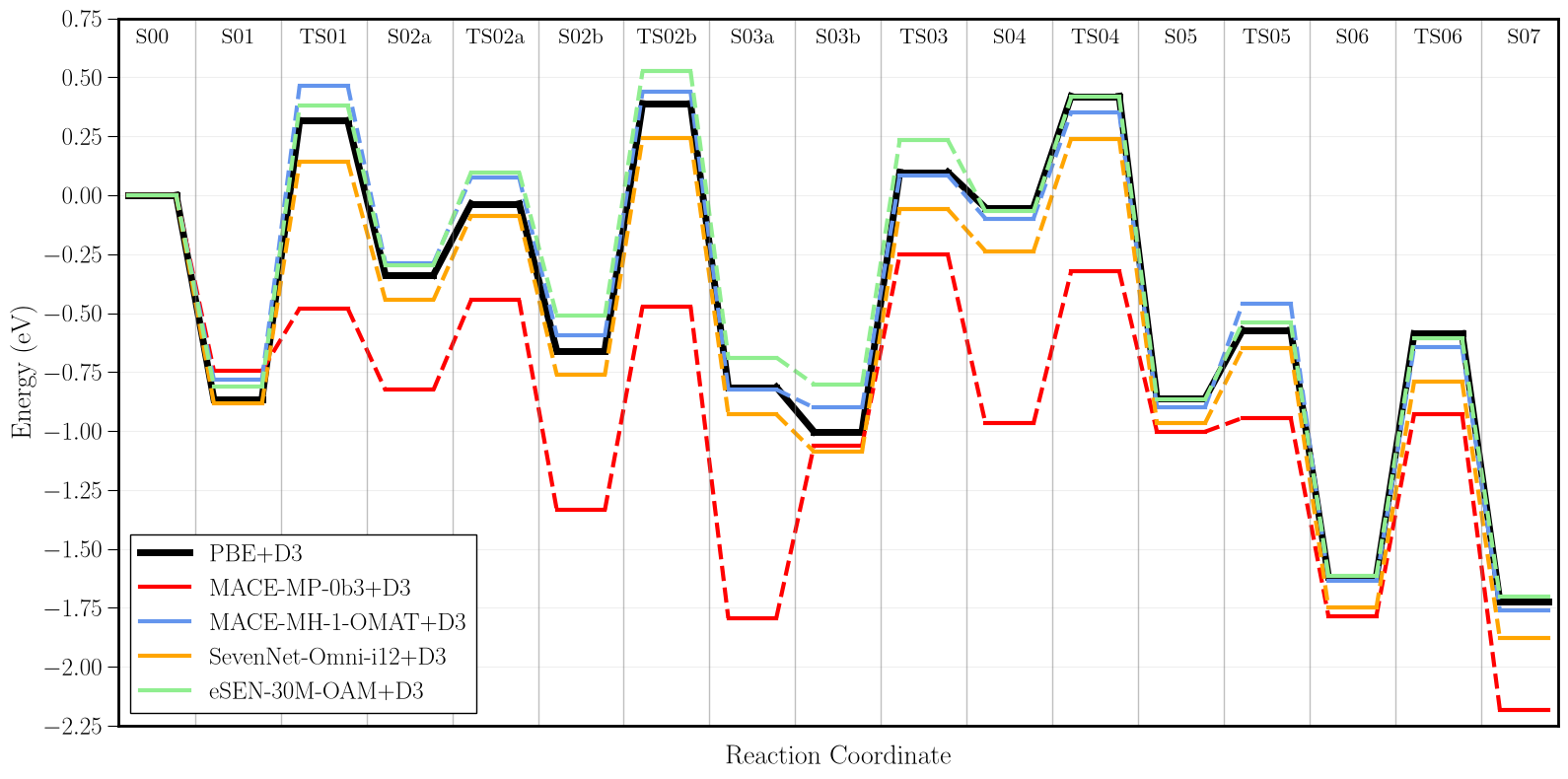}

    \caption{Comparison of PBE+D3 and three fMLIP-computed reaction pathways for CO$_2$ hydrogenation on Cu(111). 
    The data are identical to those in Figure 4 of the main text, but the initial state for each pathway is set to 0 eV.}

\end{figure}

\clearpage

\subsubsection{Reaction barriers (water--gas shift reaction on Cu)}

The water--gas shift reaction (WGSR) constitutes a second example catalytic reaction network in \textit{HetCat26}. 
The WGSR has five main subprocesses which include water activation, H$_2$ recombination, formate formation, and the redox and carboxyl mechanisms, as summarized in Table~\ref{tab:wgsr}.
The overall reaction is

\[
\mathrm{CO + H_2O \rightarrow CO_2 + H_2}
\]

\noindent and here we consider the full network on (100), (111), (211), and (874) Cu surfaces and at Cu single adatom, dimer, trimer and tetramer clusters attached to the (111) surface.

All periodic DFT calculations were performed using the Vienna Ab initio simulation package (VASP) code\cite{kresse1993ab,kresse1994ab,kresse1994norm,kresse1996efficiency,kresse1996efficient,kresse1999ultrasoft} employing the broadly used Perdew--Burke--Ernzerhof (PBE) exchange-correlation functional\cite{perdew1996} within the generalized gradient approach (GGA). The valence electron density was expanded using a plane wave basis set with an energy cutoff of 400 eV, while the contribution of core electrons to the valence region was accounted for using the projector augmented-wave (PAW) method, as implemented by Kresse and Joubert. Cu(111), Cu(100) surfaces were modeled using four-layer slabs in a (4$\times$4) supercell geometry, while the Cu(211) surface was represented by a four-layer slab in a (4$\times$1) supercell. The kinked Cu(874) surface was modeled using a (1$\times$1) supercell with a slab thickness equivalent to a four-layer (111) terrace. The supercell size and slab thickness of Cu clusters supported on Cu(111) were identical to those used for the clean Cu(111), ensuring negligible lateral interactions between periodic images of the supported clusters. All models are composed of 4 atomic layers, where the two topmost layers are fully relaxed, and the two bottommost are frozen at the bulk geometry. Numerical integrations in reciprocal space were performed using Monkhorst--Pack meshes. A (3$\times$3$\times$1) \textbf{k}-point grid was employed for Cu clusters supported on Cu(111) as well as for the clean Cu(111), Cu(100), and Cu(874) surfaces, while a (5$\times$4$\times$1) grid was used for Cu(211).

The ideal lattice parameter of bulk Cu was computed to be 3.63 \AA{} using a denser \textbf{k}-point mesh. The self-consistent solution of the Kohn--Sham equations was converged to an energy threshold of $10^{-5}$ eV, and structural optimizations were carried out until the residual atomic forces were below 0.02 eV·\AA{}$^{-1}$. Furthermore, the transition states (TS) were located using the climbing-image nudged elastic band (CI-NEB) method with 8 intermediate images. Each converged TS structure was further confirmed by vibrational frequency analysis.\\\\ 

\begin{table}[H]
\centering
\caption{Reactions steps within the Water Gas Shift Reaction (WGSR). 
These steps are considered on (100), (111), (211), and (874) Cu surfaces as well as at Cu single adatom, dimer, trimer and tetramer clusters attached to the (111) surface.}
\label{tab:wgsr}
\small
\setlength{\extrarowheight}{2.5pt} 
\begin{tabularx}{\linewidth}{>{\bfseries}l @{\hspace{2.5cm}} X}
\toprule
Reaction subpart & Elementary reaction steps \\
\midrule

Water activation & H$_2$O* $\rightarrow$ TS $\rightarrow$ OH* + H* \\ \addlinespace

H$_2$ recombination & H* + H* $\rightarrow$ TS $\rightarrow$ H$_2$ \\ \addlinespace

Carboxyl mechanism & CO* + OH* $\rightarrow$ TS $\rightarrow$ COOH* \\
& COOH* + OH* $\rightarrow$ TS $\rightarrow$ CO$_2$* + H$_2$O* \\ \addlinespace

Redox mechanism & OH* $\rightarrow$ TS $\rightarrow$ O* + H* \\
& OH* + OH* $\rightarrow$ TS $\rightarrow$ H$_2$O* + O* \\
& CO* + O* $\rightarrow$ TS $\rightarrow$ CO$_2$* \\ \addlinespace

Formate formation & COOH* $\rightarrow$ TS $\rightarrow$ HCOO* \\
& CO$_2$* + H* $\rightarrow$ TS $\rightarrow$ HCOO* \\

\bottomrule
\end{tabularx}
\end{table}

\clearpage

\begin{figure}[H]
    \centering
    \includegraphics[width=0.59\textwidth]{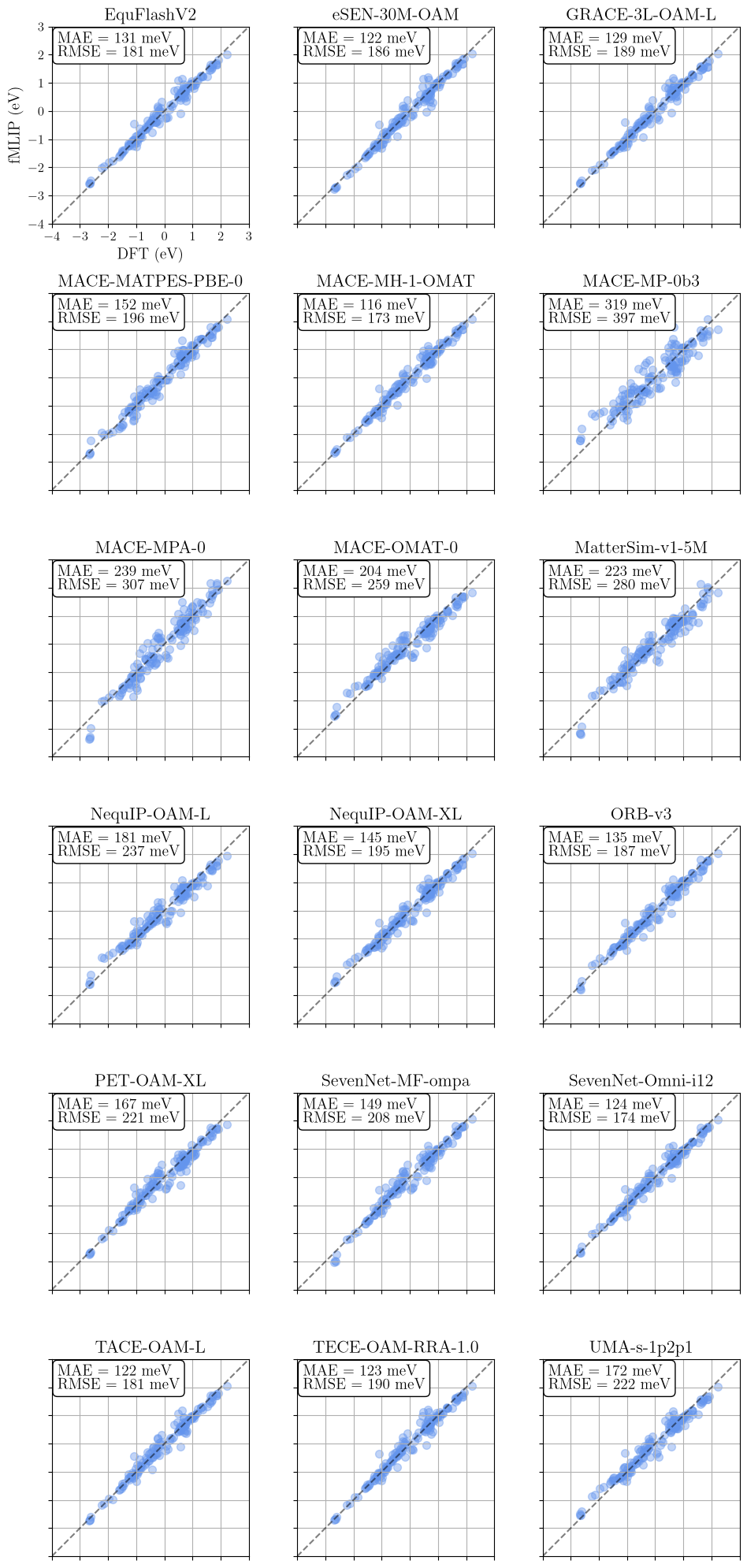}

    \caption{Parity plots comparing the fMLIPs' predicted reaction barriers in the WGSR on Cu to the reference PBE values. 
    The dashed line represents perfect agreement between predicted and reference values. 
    For each model, the MAE and RMSE are reported in the top left corner.}

\end{figure}

\section{Single point calculations set-up}

\subsection{General fMLIP calculation set-up}

All simulations were set up using the Atomic Simulation Environment (ASE).\cite{larsen2017} 
The fMLIPs considered in this work are available as calculators integrated in the ASE ecosystem, allowing for evaluation of energies from structures. 
Model-specific installation and ASE calculator implementations were used according to the documentation provided in the respective model repositories.
We note that because the different fMLIPs require incompatible versions of some underlying software libraries, a single unified environment cannot be used for all models. 
The dependency conflicts have been documented previously.\cite{loveday2026}
To avoid these problems, a dedicated environment was set up for each fMLIP family.

As stated in the main text, for all calculations but those pertaining to the final section of the paper, the reference structures were used as input for single point calculations with fMLIPs, with no additional relaxation.
From the individual total energy of structures, properties of interest are computed as detailed in the previous section.
Most reference calculations employ plain PBE\cite{perdew1996} but for calculations employing a D3 dispersion correction\cite{grimme2010}, the D3 contribution (with consistent damping choice) was evaluated separately and added to the raw fMLIP energy.

fMLIP inference calculations were performed on an NVIDIA GeForce RTX 5080 GPU.
Each model was run with its default settings (no optional acceleration), except for the changes needed for a consistent comparison described below.
This reflects the out-of-the-box performance a user would obtain, but is not the fastest configuration available for each model.
All models were run in single precision (float32).
Every call requested energies, forces, and stresses, so that all models performed the same task.
For each structure, an initial warm-up evaluation was followed by five timed evaluations.
The reported inference time is the sum over all structures of the median of these five runs.
CADS34 is excluded from timing as some models ran out of GPU memory for some structures and had to be evaluated on a CPU.

\subsection{Surface models}

All surface models in \textit{HetCat26} are constructed from DFT equilibrium bulk structures, so large deviations between the fMLIP and DFT equilibrium lattice parameters would introduce strain into the surface models when they are evaluated with the fMLIPs.
Given that the fMLIPs investigated in this work are trained primarily on bulk structures, they are expected to describe bulk equilibrium structures accurately.
Nevertheless, we verified this assumption explicitly.
To this end, the PBE equilibrium bulk structures used in the surface energy benchmark were relaxed with each fMLIP in ASE, optimizing atomic positions and cell vectors simultaneously at zero external pressure with the BFGS optimizer and the Fréchet cell filter.
Space-group symmetry was constrained throughout the relaxation using ASE's \texttt{FixSymmetry} constraint.
Convergence was defined by a maximum force threshold of 5~meV/\AA{}, applied to both the atomic forces and the generalized forces on the cell.
All relaxations converged within the limit of 1000 steps.
As reported in Table~S7, the mean absolute relative change in the lattice parameters is below 1\% for all models.
Constructing the surface models from the DFT equilibrium bulk structures is therefore justified.
Conversely, surface models constructed entirely with fMLIPs, starting from their own optimized bulk structures, would have lattice parameters within (on average) 1\% of the DFT-based models, so such fully fMLIP-based workflows are also sound in this respect.

\begin{table}[H]
\centering
\caption{Mean, standard deviation and maximum of the absolute percentage change in lattice parameters after relaxation of the bulk structures with each fMLIP.
For each structure, changes in independent lattice parameters are treated separately, each contributing to the statistics.}
\label{tab:lattice_change}
\begin{tabularx}{\textwidth}{
  >{\hsize=1.6\hsize\linewidth=\hsize\centering\arraybackslash}X
  >{\hsize=0.8\hsize\linewidth=\hsize\centering\arraybackslash}X
  >{\hsize=0.8\hsize\linewidth=\hsize\centering\arraybackslash}X
  >{\hsize=0.8\hsize\linewidth=\hsize\centering\arraybackslash}X}
\toprule
\textbf{fMLIP} & \textbf{Mean $\boldsymbol{|\Delta|}$ (\%)} & \textbf{Std.\ $\boldsymbol{|\Delta|}$ (\%)} & \textbf{Max $\boldsymbol{|\Delta|}$ (\%)} \\
\midrule
ORB-v3            & 0.123 & 0.148 & 0.671 \\
PET-OAM-XL        & 0.143 & 0.137 & 0.767 \\
EquFlashV2        & 0.144 & 0.142 & 0.793 \\
SevenNet-MF-ompa  & 0.160 & 0.219 & 1.000 \\
SevenNet-Omni-i12 & 0.180 & 0.173 & 0.741 \\
TECE-OAM-RRA-1.0  & 0.181 & 0.164 & 0.757 \\
eSEN-30M-OAM      & 0.237 & 0.251 & 1.015 \\
TACE-OAM-L        & 0.258 & 0.237 & 0.796 \\
GRACE-3L-OAM-L    & 0.261 & 0.254 & 0.752 \\
NequIP-OAM-XL     & 0.276 & 0.304 & 1.372 \\
NequIP-OAM-L      & 0.310 & 0.341 & 1.769 \\
MatterSim-v1-5M   & 0.358 & 0.457 & 2.064 \\
UMA-s-1p2p1       & 0.410 & 0.503 & 2.774 \\
MACE-MPA-0        & 0.426 & 0.380 & 1.417 \\
MACE-MH-1-OMAT    & 0.461 & 0.445 & 2.550 \\
MACE-MP-0b3       & 0.569 & 0.681 & 2.692 \\
MACE-OMAT-0       & 0.693 & 0.579 & 2.449 \\
MACE-MATPES-PBE-0 & 0.725 & 0.704 & 2.770 \\
\bottomrule
\end{tabularx}
\end{table}

\clearpage

\section{Beyond single point calculations}

\subsection{Structural optimizations with fMLIPs}

\subsubsection{Adsorption energies and structures (ADS41)}

To assess the performance of the fMLIPs beyond single point evaluations on the DFT structures, we relaxed all structures in the ADS41 dataset with each fMLIP before computing adsorption energies, excluding those associated with reactions involving oxygen-bearing adsorbates on Ni-, Co-containing surfaces.
This includes the gas-phase molecules, the clean slabs, and the adsorbate--slab systems.
Relaxations were performed with the BFGS optimizer as implemented in ASE, optimizing atomic positions until the maximum force on any mobile atom fell below 10~meV/\AA{}, with a limit of 1000 steps.
All relaxations converged within this limit.
Atoms fixed in the reference DFT calculations were kept fixed during the fMLIP relaxations.
The change in adsorption energy upon relaxation is reported as

\begin{equation}
    \Delta E_\mathrm{ads} = E_\mathrm{ads}^\mathrm{relaxed} - E_\mathrm{ads}^\text{single point}
\end{equation}

\noindent{where $E_\mathrm{ads}^\mathrm{relaxed}$ is the adsorption energy obtained with the fMLIP-relaxed structures and $E_\mathrm{ads}^\text{single point}$ is that obtained with the DFT reference structures.}
Structural changes are quantified by the root-mean-square deviation (RMSD) between the relaxed and the DFT atomic positions,

\begin{equation}
\mathrm{RMSD} = \sqrt{\tfrac{1}{N}\sum_{i=1}^{N} |\mathbf{r}_i^\mathrm{fMLIP} - \mathbf{r}_i^\mathrm{DFT}|^2},
\end{equation}

\noindent{where $N$ is the number of atoms considered and the displacements are evaluated using the minimum-image convention.}
For adsorbate--slab systems, the adsorbate atoms and the mobile slab atoms are treated separately; no alignment is applied, as the fixed layers define a common frame.
For clean slabs, all mobile atoms are included without alignment.
For gas-phase molecules, all atoms are included after optimal superposition with the Kabsch algorithm, which removes rigid translations and rotations that leave the energy unchanged.
In Figure~6 of the main text, mean RMSDs are reported per structure type (gas-phase molecules, clean slabs, slab atoms in adsorbate--slab systems, and adsorbate atoms), counting each unique structure once.
Figures~S13 and S14 provide a breakdown of the results on a model-system pair basis.

\begin{figure}[H]
    \centering
    \includegraphics[width=0.92\textwidth]{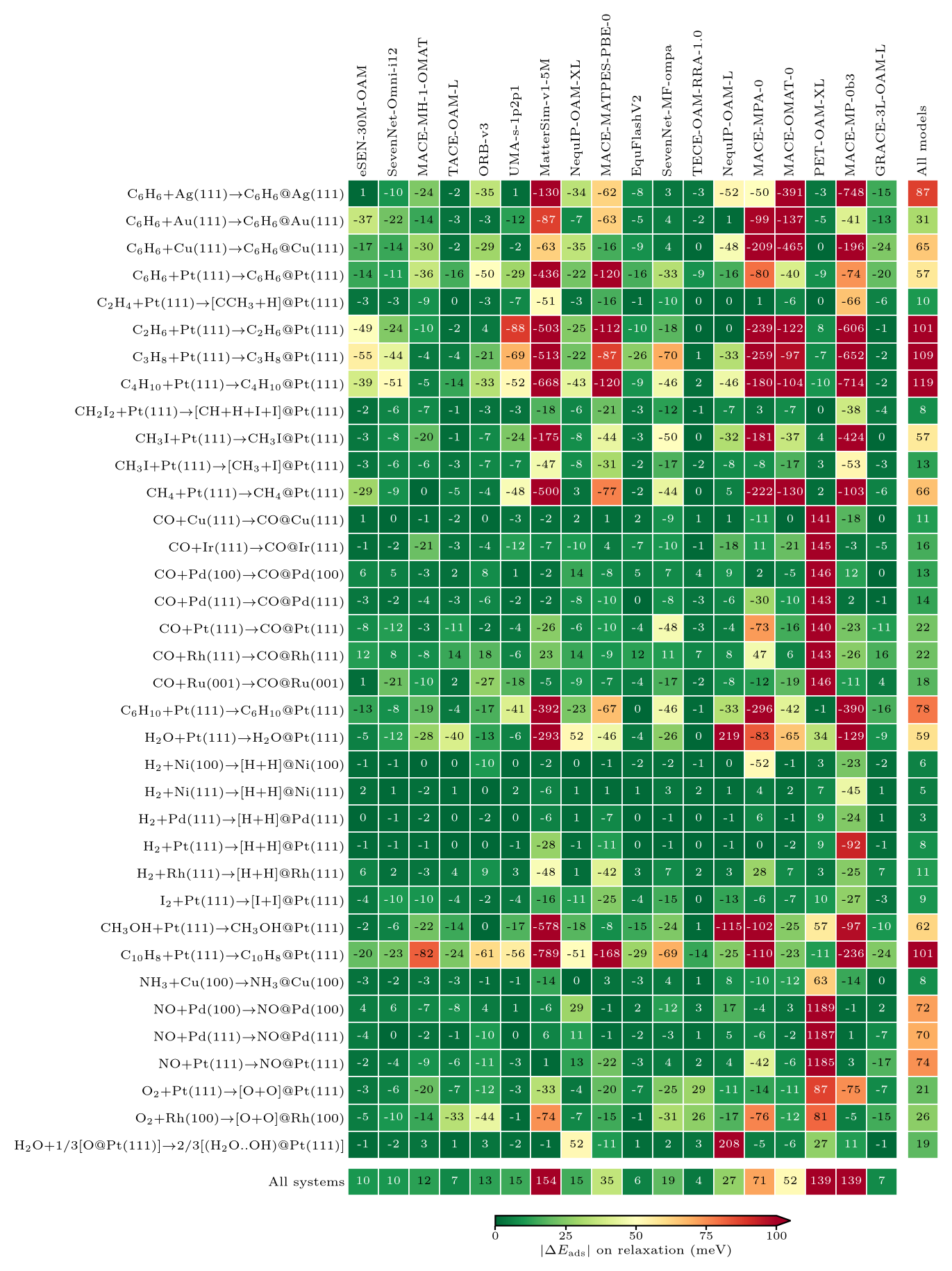}

    \caption{Changes in adsorption energies upon relaxation, with each fMLIP, of all structures in the ADS41 dataset, in meV.
    The last column gives the mean $\lvert\Delta E_\mathrm{ads}\rvert$ over models for each reaction, and the last row the mean over reactions for each model.}

\end{figure}

\begin{figure}[H]
    \centering
    \includegraphics[width=0.92\textwidth]{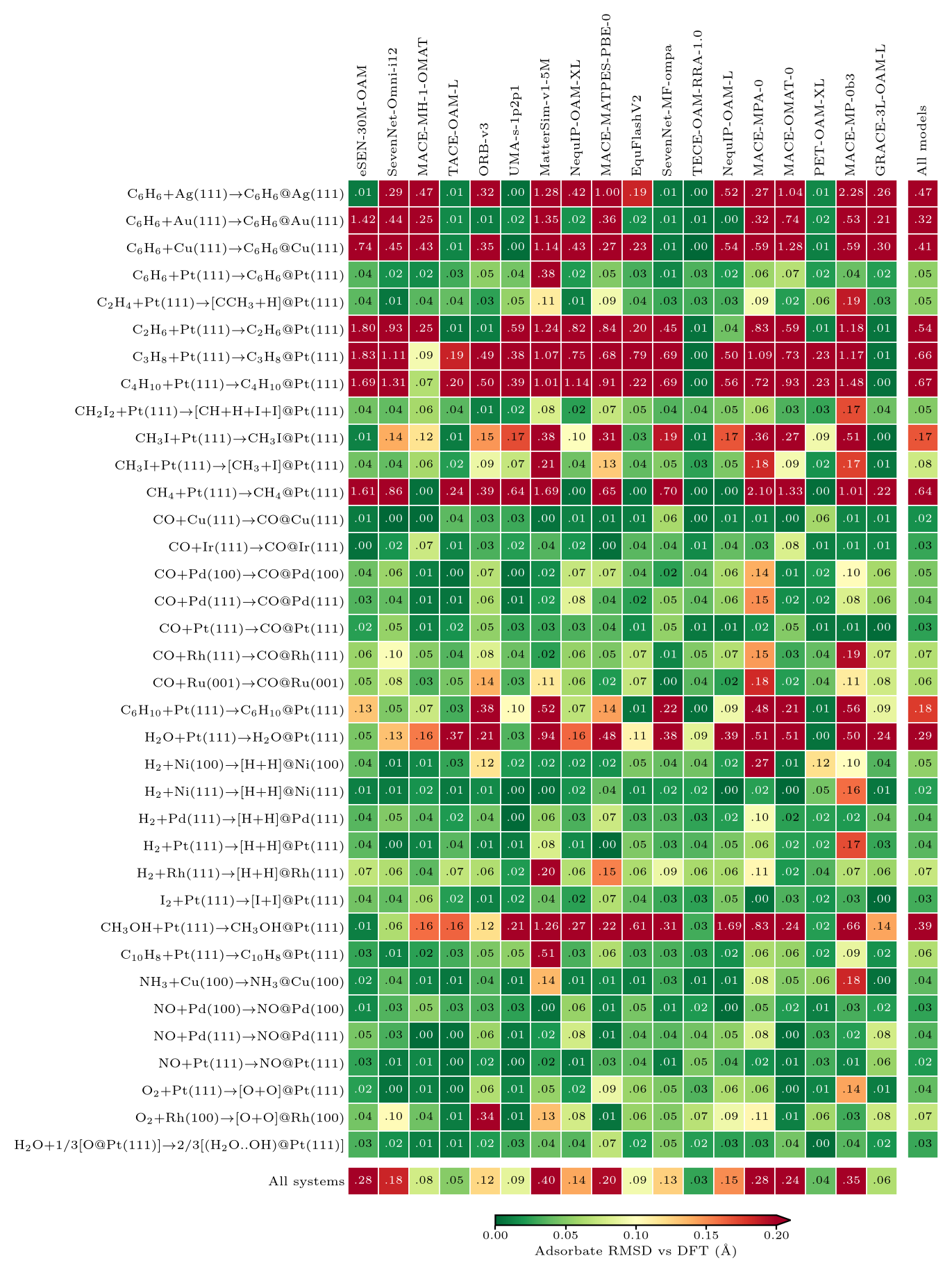}

    \caption{RMSD of the adsorbate atoms after relaxation with each fMLIP, relative to the DFT structure, in \AA{}.
    For reactions with more than one adsorbed product, the largest product RMSD is shown.
    The last column gives the mean over models for each reaction, and the last row the mean over reactions for each model.}

\end{figure}

\subsubsection{Transition state energies and structures (SBH17)}

We also investigated the effect of refining each transition state (TS) of the dissociative chemisorption dataset (SBH17) with the fMLIPs.
Each DFT TS was used as the starting structure; only the molecule (containing H, C, and N atoms) was allowed to move, as in the reference calculations.
The TS was refined with the dimer method as implemented in ASE (\texttt{MinModeTranslate}), using the ASE default settings with the following exceptions: the initial dimer direction was taken as the eigenvector of the lowest eigenvalue of the fMLIP Hessian at the DFT TS, the dimer separation was 0.01~\AA{}, and the maximum step was 0.03~\AA{}, with up to five rotations per step.
Refinement stopped when the maximum true force on the molecule was below 1~meV/\AA{} and the curvature along the dimer was negative, or after 1000 steps.
Harmonic frequencies of the refined structure were computed with \texttt{ase.vibrations.Vibrations}, displacing only the atoms of the molecule (four-point central differences, 0.01~\AA{}), and all imaginary frequencies were counted, without a threshold for small values.
If the refined structure had more than one imaginary frequency, the reaction mode was taken as the imaginary mode closest to the final dimer direction.
The structure was displaced by $\pm 0.05$~\AA{} along the largest other imaginary mode and refined again starting along the reaction mode, and the lower-energy converged result was kept.
A refinement was considered successful if it converged to a structure with exactly one imaginary frequency.

Single-precision GPU calculations are not bitwise reproducible, so each refinement was performed twice.
A run was considered to have reached a TS if it converged to a structure with exactly one imaginary frequency in which the molecule remained on the surface (within 4~\AA{} of a metal atom).
A TS was considered located if both runs reached a TS and their results agreed, with energies within 5~meV and RMSDs from the DFT TS within 0.1~\AA{} of each other.
For each located TS, we calculate the energy change

\begin{equation}
  \Delta E_\mathrm{TS} = E_\mathrm{model}(\mathbf{R}_\mathrm{TS}^\mathrm{model}) - E_\mathrm{model}(\mathbf{R}_\mathrm{TS}^\mathrm{DFT})
\end{equation}

\noindent{where $E_\mathrm{model}(\mathbf{R}_\mathrm{TS}^\mathrm{DFT})$ is the fMLIP energy of the DFT TS and $E_\mathrm{model}(\mathbf{R}_\mathrm{TS}^\mathrm{model})$ that of the fMLIP-refined TS.}
The RMSD of the molecule from the DFT TS is also calculated to quantify the structural changes.

\begin{figure}[H]
    \centering
    \includegraphics[width=0.82\textwidth]{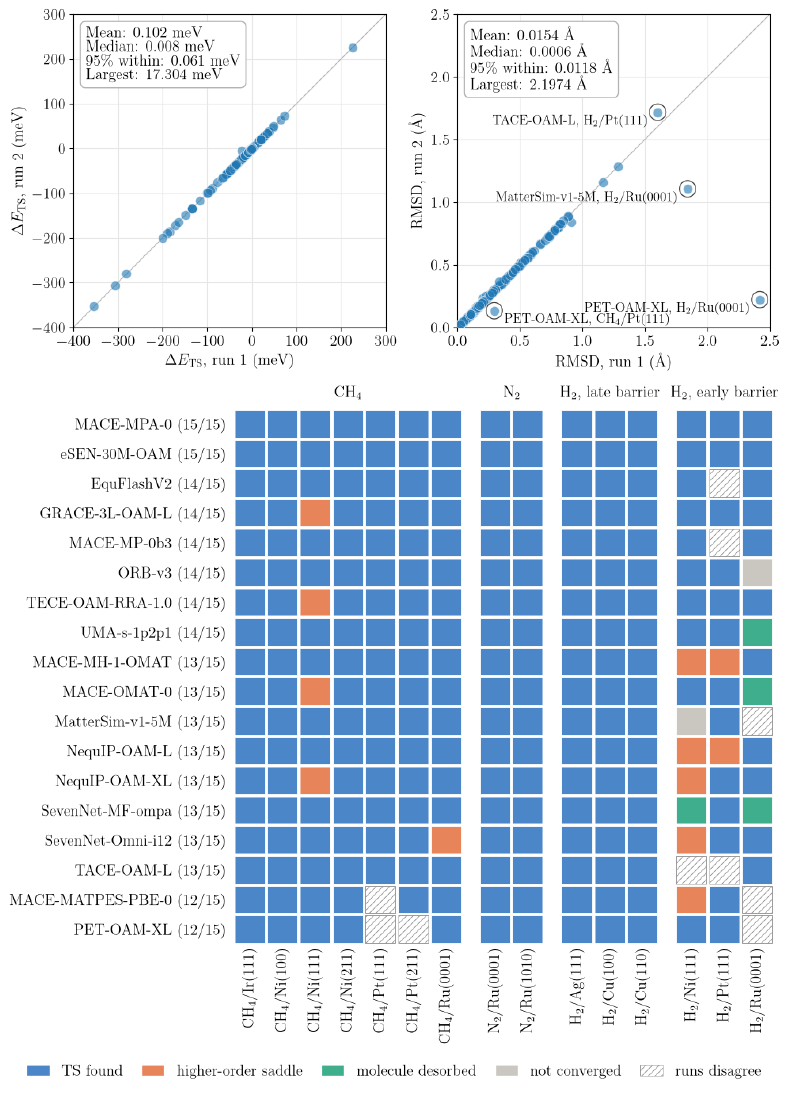}

    \caption{Reproducibility and outcome of the transition state refinements.
    Each DFT TS was refined twice with identical settings.
    At the top, $\Delta E_\mathrm{TS}$ (left) and the RMSD of the molecule from the DFT TS (right) of run 2 against run 1, for all model/system pairs in which both runs reached a TS are reported.
    The four ringed points (right) are pairs whose energies differ by more than 5~meV and/or whose RMSDs differ by more than 0.1~\AA{}.
    At the bottom, the outcomes for each model and system are summarized.
    Blue: both runs reach the same TS.
    Orange: both runs reach a higher-order saddle point.
    Green: the molecule desorbed in both runs.
    Grey: neither run converged.
    Hatched: the runs disagree.}

\end{figure}

\subsection{Free-energy calculations}

To assess the performance of the models in dynamical simulations, we computed two-dimensional free energy surfaces for CO adsorption on Pt(111) and Cu(111) with eSEN-30M-OAM and MACE-MH-1-OMAT.
We used a $3\times3$ surface cell with six layers (54 metal atoms), in which the bottom two layers were fixed and the upper four layers and a single CO molecule were free to move.
A 20~\AA{} vacuum region was added between slabs.
The equilibrium lattice constants predicted by eSEN-30M-OAM, 3.976~\AA{} for Pt and 3.626~\AA{} for Cu, were used for both models.

MD simulations were performed in the NVT ensemble at 300~K using a Bussi stochastic velocity-rescaling thermostat\cite{bussi2007} with a coupling time of 50~fs, an integration time step of 1~fs, and a sampling interval of 20~fs.
Two-dimensional umbrella sampling was performed with PLUMED,\cite{plumed2019} using 36 windows arranged on a $6\times6$ grid spanning one primitive surface unit cell.
Each window was simulated for 100~ps, of which the first 20~ps were discarded as equilibration.
The collective variables were the two periodic fractional lattice coordinates $a$ and $b$, which describe the lateral position of the carbon atom relative to the top layer.
The bias potential was

\begin{equation}
U_{\mathrm{bias}} = 
\frac{k}{4} \left[(\Delta a)^2 + (\Delta b)^2
+ \left(\operatorname{wrap}(\Delta a + \Delta b)\right)^2 \right],
\end{equation}

\noindent{where $\Delta a = \operatorname{wrap}(a - a_0)$, $\Delta b = \operatorname{wrap}(b - b_0)$, and $(a_0, b_0)$ denotes the umbrella center.}
The wrapping function maps fractional coordinate displacements into the interval $[-0.5, 0.5)$ to account for periodicity, and the final term accounts for the non-orthogonality of the primitive lattice vectors.
A force constant $k = 40$~eV per squared fractional coordinate unit was used.

The free-energy surfaces, i.e., the potentials of mean force (PMFs), were constructed using two-dimensional Gaussian-process regression (GPR).\cite{stecher2014}
The free-energy gradients at the 36 umbrella centers were estimated by averaging the forces exerted by the umbrella restraints over the final 80~ps of each trajectory.
GPR was then used to reconstruct the PMFs on a $120\times120$ grid covering the primitive surface cell.
The statistical covariance of the gradient estimates was determined from eight 10-ps blocks per umbrella window and incorporated into the GPR analysis.
An anisotropic periodic squared-exponential kernel was used, with separate length-scale parameters for $a$ and $b$ optimized independently for each system.\\

\begin{figure}[H]
    \centering
    \includegraphics[width=1\textwidth]{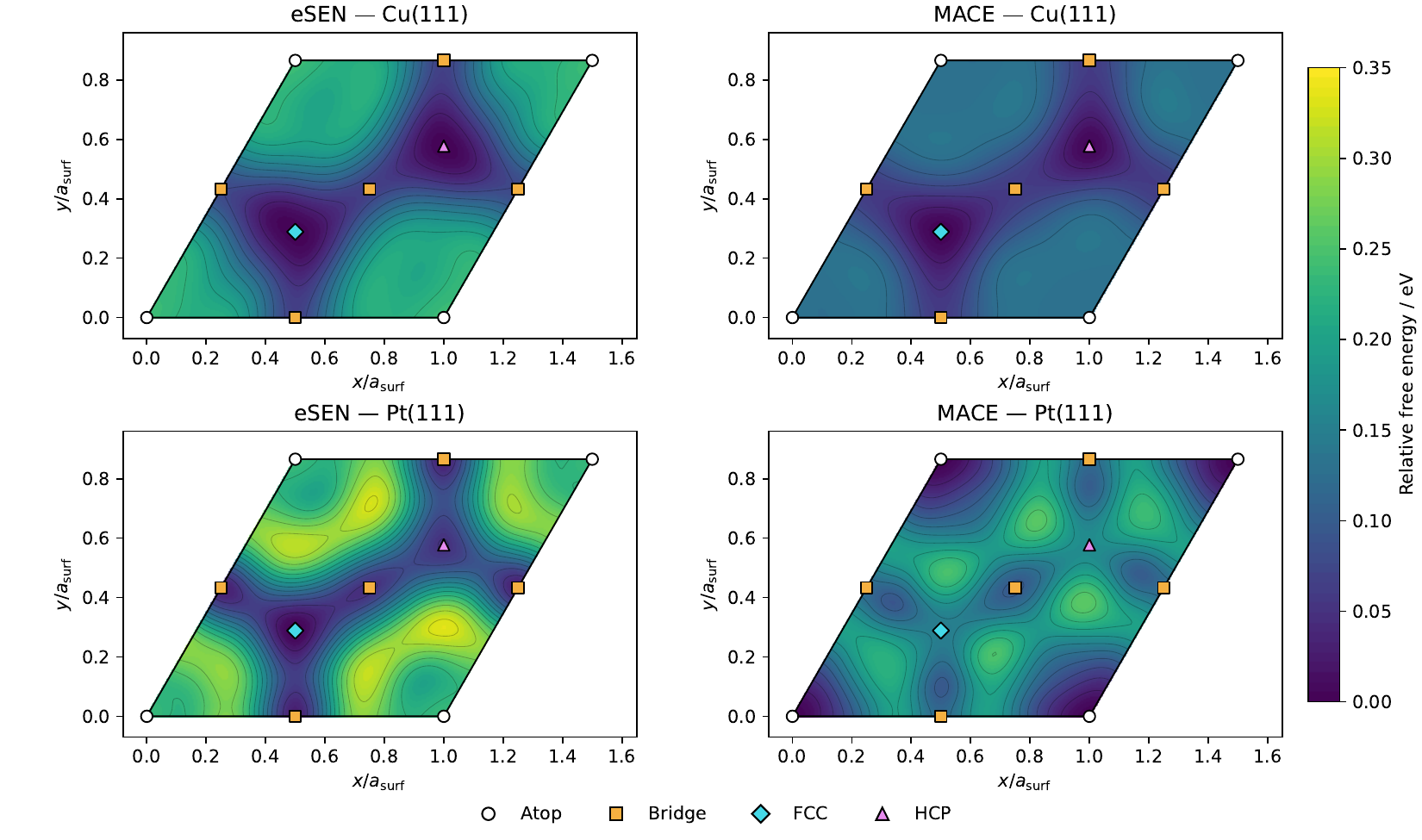}
    \caption{Two-dimensional potentials of mean force (PMFs) for CO adsorption on Cu(111) (top) and Pt(111) (bottom), calculated using eSEN-30M-OAM (left) and MACE-MH-1-OMAT (right) at 300~K. Each PMF is referenced to its minimum free energy, with a common color scale across all four panels. Symbols indicate the ideal atop (circle), bridge (square), fcc hollow (diamond), and hcp hollow (triangle) adsorption sites.}
    \label{fig:pmf}
\end{figure}

Figure~\ref{fig:pmf} shows the free-energy surfaces of CO on Cu(111) and Pt(111) at 300~K obtained with eSEN-30M-OAM and MACE-MH-1-OMAT, following the protocol described above.
Both models produce stable MD simulations out of the box, without system-specific fine-tuning.
On NVIDIA GH200 Grace Hopper Superchips (H100 GPU with 96 GB HBM3, 72-core Grace CPU) of the JUPITER supercomputer at Forschungszentrum Jülich (Germany), with one GPU per umbrella window, a 100-ps run took on average 3.80~h for CO on Cu(111) and 3.68~h for CO on Pt(111) with eSEN-30M-OAM, and 1.21~h for both systems with MACE-MH-1-OMAT.
At these costs, such simulations are far more affordable than with AIMD, for which sampling of this extent (3.6~ns in total per free-energy surface) is rarely feasible.

For CO on Cu(111), both models place the free-energy minima at the fcc and hcp hollow sites.
eSEN-30M-OAM predicts a more corrugated free-energy surface, with the atop site around 0.20~eV above the hollow sites, whereas MACE-MH-1-OMAT predicts a much flatter landscape, with all other sites within around 0.12~eV of the hollow sites.

On Pt(111), the two models predict opposite site preferences.
With eSEN-30M-OAM, the free-energy minima lie at the hollow sites and are connected through the bridge sites, with the atop site around 0.20~eV higher.
MACE-MH-1-OMAT instead predicts a clear preference for the atop site, with the bridge and hollow sites around 0.10 and 0.15--0.20~eV higher, respectively.
Recent PBE-based MLIP simulations at a lower coverage (1/36~ML, compared with 1/9~ML here) agree qualitatively with eSEN-30M-OAM.\cite{wei2025}
Since static PBE calculations favor the hollow sites on Pt(111), the atop preference of MACE-MH-1-OMAT, although in line with experiment, reflects a deviation from its PBE reference level rather than a more accurate description.

eSEN-30M-OAM and MACE-MH-1-OMAT thus provide stable and inexpensive free-energy simulations out of the box.
In the absence of extensive AIMD reference data or validation of the foundation models against DFT energies and forces of representative snapshots from these simulations, however, the accuracy of the resulting free-energy surfaces remains an open question.
Nonetheless, these models are promising starting points for active-learning workflows, which could yield free-energy surfaces of DFT quality at a fraction of the cost of AIMD.

\clearpage

\bibliography{refs}